\documentclass[11pt]{article}

\usepackage{amssymb}
\usepackage{amsfonts}
\usepackage{mathrsfs}
\usepackage{graphicx}
\IfFileExists{epstopdf.sty}{\usepackage{epstopdf}}{}
\graphicspath{{figures/}}
\usepackage{amsmath}
\usepackage{color}
\usepackage{amsthm}
\usepackage{pifont,bm}
\usepackage{enumerate}
\IfFileExists{subcaption.sty}{%
  \usepackage{subcaption}%
}{%
  \newcounter{legacysubfigure}[figure]
  \newenvironment{subfigure}[2][b]{%
    \begin{minipage}[#1]{#2}%
    \renewcommand{\caption}[1]{\par\refstepcounter{legacysubfigure}%
      {\centering\small(\alph{legacysubfigure})\space ##1\par}}%
  }{\end{minipage}}%
}
\usepackage{tikz}
\newtheorem{thm}{Theorem}[section]

\theoremstyle{definition}

\makeatletter
\def\@biblabel#1{[#1]}
\makeatother

\makeatletter \@addtoreset{equation}{section}
\makeatother

\makeatletter
\let\OriginalIncludeGraphics\includegraphics
\newif\ifNLSimagefound
\renewcommand{\includegraphics}[2][]{%
  \begingroup
  \NLSimagefoundfalse
  \IfFileExists{#2}{\NLSimagefoundtrue}{}%
  \IfFileExists{figures/#2}{\NLSimagefoundtrue}{}%
  \@for\NLSext:=\Gin@extensions\do{%
    \IfFileExists{#2\NLSext}{\NLSimagefoundtrue}{}%
    \IfFileExists{figures/#2\NLSext}{\NLSimagefoundtrue}{}%
  }%
  \ifNLSimagefound
    \OriginalIncludeGraphics[#1]{#2}%
  \else
    \PackageWarning{Manuscript}{Image `#2' not found; placeholder used}%
    \fbox{\parbox[c][3cm][c]{0.85\linewidth}{\centering\footnotesize
      Missing image:\\ \texttt{\detokenize{#2}}}}%
  \fi
  \endgroup
}
\makeatother

\begin{document}

\title{\bf{Vector breathers and exact spectra analysis of the coupled nonlinear Schr\"{o}dinger equation in the focusing regime}}
\author{Xue-Jiao Cheng$^{1}$, Xiu-Bin Wang$^{1}$\footnote{
E-mail addresses: xbwang@cumt.edu.cn, xiubinwang@163.com.}, Haie Long$^{2}$, Xing-Jie Yan$^{1}$\\
\small \emph{$^{1}$School of Mathematics and  Jiangsu Center for Applied Mathematics, China University of Mining and Technology,}\\
\small \emph{Xuzhou 221116,People's Republic of China}\\
\small \emph{$^{2}$MSU-BIT-SMBU Joint Research Center of Applied Mathematics, Shenzhen MSU-BIT University, Shenzhen,}\\
\small \emph{Guangdong, People's Republic of China}\\
\date{}}
\maketitle
\thispagestyle{empty}

\vspace{-0.5cm}
\begin{center}
\rule{15cm}{1pt}\vspace{0.3cm}

\parbox{15cm}{\small
{\bf Abstract}\\
\hspace{0.5cm}
In this work, we construct explicit breather solutions and analyze their Fourier spectra for the coupled nonlinear Schr\"odinger system with pair-transition terms. First, the Darboux transformation is used to derive first- and second-order breather building blocks in the symmetric sector. Then by introducing two normal-mode reconstructions, we find that the Type-I configuration combines unshifted symmetric and antisymmetric normal-mode contributions, whereas the Type-II configuration includes a relative spatiotemporal displacement.
These reconstructions generate equal-order and mixed-order component patterns with distinct overlap, separation, and walk-off-like features.
Furthermore, as a typical example, we derive a closed-form componentwise Fourier decomposition of the first-order Type-I Kuznetsov--Ma (KM) configuration by applying the residue theorem. The background-subtracted localized parts of the two components have an identical continuous spectrum, whereas the component retaining a plane-wave background contains an additional zero-wavenumber Dirac-$\delta$ contribution. Finally, direct numerical integration and fast Fourier transforms are used to verify the analytical solutions and physical spectra.
}
\end{center}
\vspace{0.3cm} \rule{15cm}{1pt} \vspace{0.2cm}

\section{Introduction}\label{sec.1}
The nonlinear Schr\"odinger (NLS) equation
\begin{equation*}
\mathrm{i}\psi_t+\frac{1}{2}\psi_{xx}
+\sigma|\psi|^2\psi=0,
\end{equation*}
where $\sigma=\pm1$ denote the focusing and defocusing cases, respectively, is a standard integrable model for nonlinear waves \cite{ZVABS-1972}. It describes optical pulses in fibers \cite{HT-1973,H-2022,A-2023}, matter waves in Bose--Einstein condensates \cite{S-2021,WZL-2022}, and surface waves in deep water \cite{KDD-2020}. However, the scalar NLS equation contains only one wave component and cannot describe distinct dynamics in multicomponent systems. Examples include polarization modes, fiber modes, and different internal atomic states \cite{M-2022,SSRL-2021,SPAA-2024}, which are commonly modeled by coupled nonlinear Schr\"odinger (CNLS) equations.
Among the integrable CNLS models, the Manakov system is one of the most fundamental examples \cite{MSV-1974}. Its nonlinear terms contain self-phase modulation and cross-phase modulation, but no explicit phase-sensitive mixing term. Consequently, the norm of each component is conserved separately.
This situation changes when four-wave mixing (FWM) becomes significant. FWM introduces phase-sensitive nonlinear terms \cite{WLZ-2021,E-2022,SMT-2020}. In the physical-component basis, these terms can lead to a redistribution of population or power between the components. These phase-sensitive effects motivate the study of localized wave structures and their component-dependent behavior.

Breathers are important nonlinear wave structures in focusing NLS-type systems. They describe localized and recurrent wave dynamics on finite backgrounds. The fundamental breather family includes the Kuznetsov--Ma (KM) breather \cite{Kuznetsov1977}, the Akhmediev breather \cite{AkhmedievKorneev1986}, and the limiting Peregrine state \cite{P-1983}. These structures are closely related to modulation instability and rogue-wave formation \cite{DudleyEtAl2014,HJJSS2013}. In multicomponent systems, breather dynamics can become richer. Different wave components may exhibit distinct localization, interference, and spectral features. Therefore, breather solutions and their physical spectra provide a useful framework for studying localized nonlinear-wave dynamics in coupled systems \cite{CheChenLiu2024,YANZYJNS,WLEEPRA,QZYUNNSAM,CYYY1,HJSSS2020,RJGHJS2023,LLLBBB2014,MAWWW1,MAWWW2,MAWWW3,WDSSYZY2,WANGXB1,WANGXB2,WANGXB3,WANGXB4}.

An suitable model for studying such breather dynamics in the presence of phase-sensitive coupling is the integrable CNLS system with pair-transition terms (CNLS-p) \cite{LLLZ-2015}
\begin{equation}\label{qr1}
\begin{cases}
\displaystyle
\mathrm{i}p_t+\frac{1}{2}p_{xx}
+\sigma\left(|p|^2+2|q|^2\right)p
+\sigma q^2p^*=0,\\[8pt]
\displaystyle
\mathrm{i}q_t+\frac{1}{2}q_{xx}
+\sigma\left(2|p|^2+|q|^2\right)q
+\sigma p^2q^*=0.
\end{cases}
\end{equation}
Here, $p^*$ and $q^*$ denote the complex conjugates of $p$ and $q$, respectively. In the physical-component basis $(p,q)$, the terms $\sigma q^2p^*$ and $\sigma p^2q^*$ appear as phase-sensitive exchange terms.
In addition, Eq.\eqref{qr1} also has a special normal-mode structure.
By introducing the transformation $u_{+}=p+q$, $u_{-}=p-q$,
we have
\begin{equation}\label{inverse-normalmodes}
p=\frac{u_{+}+u_{-}}{2},\qquad
q=\frac{u_{+}-u_{-}}{2}.
\end{equation}
Then we check that two normal modes satisfy the scalar NLS equations
\begin{equation}\label{decoupledNLS}
\mathrm{i}u_{\pm,t}
+\frac{1}{2}u_{\pm,xx}
+\sigma|u_{\pm}|^2u_{\pm}=0.
\end{equation}
Thus, the CNLS-p system is exactly diagonalizable, and the two normal modes do not interact directly. Different structures in the $p$ and $q$ components arise when the normal modes are reconstructed in the physical-component basis. They therefore result from normal-mode reconstruction and interference.

The decoupled normal-mode structure provides a useful perspective on previous studies of the CNLS-p system, which have addressed both integrability and localized nonlinear waves. Early studies examined the integrability of related coherently coupled models \cite{PS-1999}. Zhao et al. obtained bright solitons, KM and Akhmediev breathers, and rogue waves through the Darboux transformation \cite{ZLYL-2015}. More directly related to the present work, Ling and Zhao developed two forms of the Darboux transformation for the CNLS-p system and showed that its solutions can be expressed as linear combinations of scalar NLS solutions \cite{LLLZ-2015}. Related superposition and interference mechanisms were investigated by Mareeswaran and Kanna in a coherently coupled Gross--Pitaevskii system \cite{MareeswaranKanna2016} and by Qin et al. in a pair-transition-coupled system \cite{QinZhaoYang2018}. Further studies addressed modulational instability, higher-order rogue waves, asymmetric breathers, and rational wave structures \cite{ZGY-2017,LWH-2019,WH-2020a}. Riemann--Hilbert formulations, soliton solutions, and long-time asymptotic behavior were also investigated \cite{WXBH-2019,PTZ-2020,WGYL-2023}.

Despite these advances, a unified comparison of equal-order and mixed-order normal-mode reconstructions involving first- and second-order breather building blocks remains limited for the CNLS-p system. In particular, the combined effects of Darboux order, spectral mismatch, and relative spatiotemporal displacement on background cancellation, peak separation, and component interference have not been examined systematically.

Beyond their spatiotemporal structures, breathers can also be characterized by their wavenumber-domain properties. Exact spatial Fourier spectra of scalar NLS breathers, including the KM, Akhmediev, and Peregrine families, have been derived previously \cite{Karjanto2023}. Exact physical spectra of nondegenerate vector KM solitons have also been studied for the Manakov system \cite{CheChenLiu2022}. By contrast, Wang et al. investigated the nonlinear inverse-scattering spectrum of vector rogue waves in a related CNLS model with coherent coupling \cite{WZLL-2023}. This spectrum, associated with the Lax operator, differs from the spatial Fourier spectrum of the physical wave profile considered here. The present work therefore focuses on the componentwise Fourier decomposition induced by the Type-I normal-mode reconstruction of the CNLS-p system. In particular, we distinguish the common continuous spectrum of the localized component fields from the zero-wavenumber Dirac contribution associated with the retained plane-wave background.

Throughout this work, we study the focusing CNLS-p system with $\sigma=1$. We first use the Darboux transformation to construct first- and second-order breather building blocks in the symmetric sector. We then choose two normal-mode solutions independently and reconstruct the physical fields.
For convenience, we consider two configurations. The Type-I configuration contains no relative spatiotemporal shift and includes equal-order and mixed-order normal-mode combinations. The Type-II configuration introduces a relative spatiotemporal shift between the two normal modes.
The reconstructed fields display component-dependent overlap, separation, and walk-off-like patterns. These patterns result from normal-mode interference rather than direct nonlinear scattering between the normal modes. We also analyze the Type-I KM configuration. After separating the nonzero background from the localized perturbation, we derive a closed-form componentwise Fourier decomposition by applying the residue theorem.
For this family, the localized parts of the two components have identical continuous Fourier spectra. The full spectrum of the component that retains a nonzero background contains an additional zero-wavenumber Dirac-$\delta$ term. The Type-I KM solution is compared with direct numerical integration based on the split-step Fourier method \cite{TA-1984}. The analytical spectrum is also compared with a numerical spectrum obtained using a fast Fourier transform.

The remainder of this paper is organized as follows. Section~2 presents the Lax pair and the $N$-fold Darboux transformation. Section~3 constructs the first- and second-order breather building blocks. Section~4 studies the unshifted Type-I configurations. Section~5 considers the shifted Type-II configurations. Section~6 derives the Fourier spectrum of the Type-I KM configuration. Section~7 presents the numerical verification. Section~8 summarizes the main conclusions.

\section{Lax pair and Darboux transformation}

We consider the CNLS-p system \eqref{qr1} in the focusing regime, i.e., $\sigma=1$. The system admits the following $4\times4$ Lax pair
\begin{equation}\label{qr2}
\begin{aligned}
\Phi_x=U\Phi,\qquad
\Phi_t=V\Phi,
\end{aligned}
\end{equation}
where $\Phi(x,t,\lambda)$ is a $4\times1$ column-vector eigenfunction and $\lambda$ is the spectral parameter. The spectral matrices are
\begin{equation}\label{UV}
\begin{aligned}
U=\mathrm{i}(\lambda\sigma_3+Q),\qquad
V=\mathrm{i}(\lambda^2\sigma_3+\lambda Q)
+\frac{1}{2}\sigma_3(Q_x-\mathrm{i}Q^2),
\end{aligned}
\end{equation}
where $\sigma_3=\operatorname{diag}(1,1,-1,-1)$ and
\begin{equation*}
Q=
\begin{pmatrix}
0&W^\dagger\\
W&0
\end{pmatrix},
\qquad
W=
\begin{pmatrix}
p&q\\
q&p
\end{pmatrix}.
\end{equation*}
Here, $\dagger$ denotes Hermitian conjugation. The compatibility condition $\Phi_{xt}=\Phi_{tx}$ yields the zero-curvature equation
\begin{equation*}
U_t-V_x+[U,V]=0.
\end{equation*}
Substituting \eqref{UV} into this equation reproduces the focusing CNLS-p system.

The block structure of $Q$ gives the Lax pair two useful symmetries. The first follows from $Q^\dagger=Q$:
\begin{equation}\label{sym1}
U^\dagger(\lambda)=-U(\lambda^*),\qquad
V^\dagger(\lambda)=-V(\lambda^*).
\end{equation}
For the second symmetry, note that $W=pI_2+q\sigma_1$ and hence $\sigma_1W\sigma_1=W$. Therefore,
\begin{equation*}\label{sym2}
\Sigma U(\lambda)\Sigma=U(\lambda),\qquad
\Sigma V(\lambda)\Sigma=V(\lambda),
\end{equation*}
where
\begin{equation}\label{SigmaDef}
\Sigma=
\begin{pmatrix}
\sigma_1&0\\[2pt]
0&\sigma_1
\end{pmatrix},
\qquad
\sigma_1=
\begin{pmatrix}
0&1\\[2pt]
1&0
\end{pmatrix}.
\end{equation}
These symmetries constrain the Darboux matrix and ensure that the transformed potential preserves the algebraic structure of the original system.

To construct the Darboux transformation, we introduce the gauge transformation $\Phi[1]=T[1]\Phi$ and require the transformed eigenfunction to satisfy a Lax pair of the same form
\begin{equation*}
\Phi[1]_x=U[1]\Phi[1],\qquad
\Phi[1]_t=V[1]\Phi[1].
\end{equation*}
Substituting $\Phi[1]=T[1]\Phi$ gives the compatibility conditions
\begin{equation}\label{Tcond}
\begin{aligned}
T[1]_x+T[1]U=U[1]T[1],\qquad
T[1]_t+T[1]V=V[1]T[1].
\end{aligned}
\end{equation}

Let $\lambda_1\in\mathbb{C}\setminus\mathbb{R}$ be an admissible spectral parameter, and let $y_1$ be a column-vector solution of the Lax pair at $\lambda=\lambda_1$. The Hermitian reduction \eqref{sym1} leads to the rank-one Darboux matrix
\begin{equation}\label{T1}
T[1](\lambda)
=I_4+
\frac{\lambda_1^*-\lambda_1}{\lambda-\lambda_1^*}
\frac{y_1y_1^\dagger}{y_1^\dagger y_1}.
\end{equation}
Because $U$ and $V$ commute with $\Sigma$, the eigenfunction can be chosen to have definite parity,
\begin{equation*}\label{yparity}
\Sigma y_1=\varepsilon_1y_1,
\qquad \varepsilon_1=\pm1.
\end{equation*}
This choice ensures that $\Sigma T[1](\lambda)\Sigma=T[1](\lambda)$ and that the transformed potential retains the block-symmetric form of $Q$.

For admissible spectral data $\{(\lambda_j,y_j)\}_{j=1}^N$, repeated application of \eqref{T1} yields the $N$-fold Darboux matrix
\begin{equation*}\label{TN}
T[N](\lambda)
=I_4-\mathcal{Y}M^{-1}(\lambda I_N-S)^{-1}\mathcal{Y}^\dagger,
\end{equation*}
where
\begin{equation*}\label{defs}
\begin{aligned}
&S=\operatorname{diag}(\lambda_1^*,\ldots,\lambda_N^*),\qquad
\mathcal{Y}=(y_1,\ldots,y_N),\\
&M_{ij}=\frac{y_i^\dagger y_j}{\lambda_j-\lambda_i^*},
\qquad i,j=1,\ldots,N.
\end{aligned}
\end{equation*}
Here, $\mathcal{Y}$ is a $4\times N$ matrix, whereas $M$ and $S$ are $N\times N$ matrices. The spectral data are assumed to satisfy $\lambda_j\neq\lambda_i^*$ and $\det M\neq0$.

To extract the transformed potentials, we use the expansion
\begin{equation*}\label{asymp}
T[N](\lambda)
=I_4-\frac{K}{\lambda}+\mathcal{O}(\lambda^{-2}),
\qquad
K=\mathcal{Y}M^{-1}\mathcal{Y}^\dagger,
\qquad \lambda\to\infty.
\end{equation*}
Substituting this expansion into \eqref{Tcond} and comparing equal powers of $\lambda$ gives
\begin{equation}\label{comm}
Q[N]=Q+[\sigma_3,K].
\end{equation}
Since the lower-left block of $Q$ is $W$, equation \eqref{comm} gives
\begin{equation*}\label{pqK}
p[N]=p-2K_{31},\qquad
q[N]=q-2K_{41}.
\end{equation*}
Let $\mathcal{Y}_j$ denote the $j$th row of $\mathcal{Y}$. Applying the Schur-complement identity yields
\begin{equation}\label{pN}
p[N]=p+2\,
\frac{\det\!\begin{pmatrix}
M&\mathcal{Y}_1^\dagger\\[2pt]
\mathcal{Y}_3&0
\end{pmatrix}}
{\det M},
\end{equation}
and
\begin{equation}\label{qN}
q[N]=q+2\,
\frac{\det\!\begin{pmatrix}
M&\mathcal{Y}_1^\dagger\\[2pt]
\mathcal{Y}_4&0
\end{pmatrix}}
{\det M}.
\end{equation}
Equations \eqref{pN} and \eqref{qN} are the main algebraic formulas used to construct the breather solutions in this paper. Once the seed solution and the $N$ eigenvectors are specified, the transformed fields follow from these determinants.

\section{First- and second-order breather building blocks}

We begin with the following plane-wave solution as a seed solution
\begin{equation*}\label{seed-general}
p[0]=q[0]=c_0e^{4\mathrm{i}\sigma c_0^2t},
\end{equation*}
where $c_0>0$ is the background amplitude.

A standard linear-stability analysis shows that the plane-wave background is modulationally unstable for perturbation wavenumbers
\begin{equation*}
0<|k|<4c_0.
\end{equation*}
This modulational-instability band is associated with the formation of breathers and rogue waves.
Without loss of generality, we set $c_0=1/2$ throughout this paper, so that
\begin{equation}\label{seed-fixed}
p[0]=q[0]=\frac{1}{2}e^{\mathrm{i}t}.
\end{equation}

Because the seed \eqref{seed-fixed} contains the phase factor $e^{\mathrm{i}t}$, the original Lax pair contains the factors $e^{\pm\mathrm{i}t}$. We remove these factors by introducing the gauge transformation
\begin{equation}\label{gauge2}
\Phi=C\Psi,
\qquad
C=\operatorname{diag}\left(1,1,e^{\mathrm{i}t},e^{\mathrm{i}t}\right).
\end{equation}
Substitution \eqref{gauge2} into \eqref{qr2} gives
\begin{equation}\label{qr5}
\begin{aligned}
\Psi_x=A\Psi,\qquad
\Psi_t=B\Psi,
\end{aligned}
\end{equation}
where $A=C^{-1}UC$ and $B=C^{-1}VC-C^{-1}C_t$. For the seed solution \eqref{seed-fixed}, both matrices are independent of $x$ and $t$.

The matrices $A$ and $B$ commute with the symmetry matrix $\Sigma$ in \eqref{SigmaDef}. The spectral problem therefore separates into symmetric and antisymmetric subspaces. The eigenfunctions used below belong to the symmetric subspace, $\Sigma\Psi=\Psi$. In this subspace, the eigenvalues of the $x$-part are $\pm\mu$, where
\begin{equation*}\label{mu}
\mu=\sqrt{-\lambda^2-1}.
\end{equation*}
Throughout this paper, the branch of the square root is chosen such that $\operatorname{Re}(\mu)\geq0$. When $\operatorname{Re}(\mu)=0$, we take $\operatorname{Im}(\mu)\geq0$. The square roots entering $\kappa_1$ and $\kappa_2$ are chosen consistently with this convention.

We introduce the phase
\begin{equation*}\label{theta}
\Theta=\mu(x+\lambda t)+\gamma+\mathrm{i}\theta,
\end{equation*}
where $\gamma,\theta\in\mathbb{R}$ control the spatiotemporal position of the breather.

For convenience, define
\begin{equation*}
h=-\mathrm{i}\lambda
\end{equation*}
and
\begin{equation*}\label{kappa12}
\kappa_1=\frac{1}{\sqrt{h^2-1}\sqrt{h-\sqrt{h^2-1}}},
\qquad
\kappa_2=\frac{1}{\sqrt{h^2-1}\sqrt{h+\sqrt{h^2-1}}}.
\end{equation*}
A fundamental eigenvector of the constant-coefficient system \eqref{qr5} is then
\begin{equation*}\label{Psi1}
\Psi_1=
\begin{pmatrix}
-\kappa_2e^{\Theta}+\kappa_1e^{-\Theta}\\
-\kappa_2e^{\Theta}+\kappa_1e^{-\Theta}\\
\mathrm{i}\kappa_1e^{\Theta}-\mathrm{i}\kappa_2e^{-\Theta}\\
\mathrm{i}\kappa_1e^{\Theta}-\mathrm{i}\kappa_2e^{-\Theta}
\end{pmatrix}.
\end{equation*}
Returning to the original variables gives
\begin{equation*}\label{qr6}
y_1=C\Psi_1.
\end{equation*}
The equal first two entries and equal last two entries imply $\Sigma y_1=y_1$. Thus, this eigenvector generates Darboux solutions in the symmetric sector.

For the first-order case (i.e., $N=1$), substituting $y_1$ into \eqref{pN} and \eqref{qN} gives
\begin{equation}\label{eq:3.4}
p[1]=q[1]=\frac{1}{2}e^{\mathrm{i}t}(1+H_1),
\end{equation}
where
\begin{align*}
H_1&=\frac{\mathrm{i}(\lambda^*-\lambda)\mathcal{A}}{\mathcal{D}},\\
\mathcal{C}&=\cosh(2\mu x+2\gamma)
+\mathrm{i}\lambda\cosh(2\lambda\mu t+2\mathrm{i}\theta),\\
\mathcal{A}&=\mathcal{C}
-\mu\sinh(2\lambda\mu t+2\mathrm{i}\theta),\\
\mathcal{D}&=\mathrm{i}\lambda\cosh(2\mu x+2\gamma)
+\cosh(2\lambda\mu t+2\mathrm{i}\theta).
\end{align*}

For the second-order case (i.e., $N=2$), we use two distinct spectral parameters $\lambda_1$ and $\lambda_2$. The resulting two-fold solution is
\begin{equation}\label{eq:3.5}
p[2]=q[2]=\frac{1}{2}e^{\mathrm{i}t}(1+H_2),
\end{equation}
where
\begin{equation}\label{H2-general}
H_2=
\frac{2\mathrm{i}(\lambda_1-\lambda_2^*)(\lambda_1-\lambda_2)
[\lambda_2\mathcal{A}_2\mathcal{D}_1-\lambda_1\mathcal{A}_1\mathcal{D}_2]}
{\mathcal{D}_1\mathcal{D}_2(\lambda_1-\lambda_2^*)^2
-2\lambda_1\lambda_2\mathcal{B}}.
\end{equation}
The functions $\mathcal{C}_j$, $\mathcal{D}_j$, and $\mathcal{A}_j$ retain the first-order form, with $(\lambda,\mu,\gamma,\theta)$ replaced by $(\lambda_j,\mu_j,\gamma_j,\theta_j)$ for $j=1,2$. The mixed term is
\begin{align*}
\mathcal{B}
={}&\mathcal{D}_1\mathcal{D}_2+\mathcal{C}_1\mathcal{C}_2\\
&+\mu_1\mu_2\sinh(2\mu_1x+2\gamma_1)
\sinh(2\mu_2x+2\gamma_2)\\
&-\mu_1\mu_2\sinh(2\lambda_1\mu_1t+2\mathrm{i}\theta_1)
\sinh(2\lambda_2\mu_2t+2\mathrm{i}\theta_2).
\end{align*}
The relative positions of the two spectral points, together with the phase parameters, control the overlap, separation, and internal multipeak structure of the two-fold solution. Formulas \eqref{eq:3.5} and \eqref{H2-general} assume $\lambda_1\neq\lambda_2$, because a repeated spectral point requires a confluent Darboux transformation.

The relation $p[N]=q[N]$ is not an intrinsic degeneracy of the CNLS-p system. It follows from the symmetric seed and the symmetric eigenvectors used in this section. In terms of the normal modes, each building block satisfies $u_+^{(N)}=2p[N]$ and $u_-^{(N)}=0$. In the next two sections, two symmetric-sector building blocks are assigned independently to the half-normal-mode fields and then reconstructed in the physical-component basis.

\section{Type-I breather solutions}

We now reconstruct the physical fields from two symmetric-sector breather building blocks. Let
\begin{equation*}\label{gMN}
g_N=p[N]=q[N],\qquad
g_M=p[M]=q[M],
\end{equation*}
where $g_0=(1/2)e^{\mathrm{i}t}$. The unshifted Type-I configuration is defined by
\begin{equation}\label{qr7}
\begin{aligned}
p_{\mathrm I}^{[M,N]}=g_N+g_M,\qquad
q_{\mathrm I}^{[M,N]}=g_N-g_M.
\end{aligned}
\end{equation}
We use the combined amplitude
\begin{equation*}\label{rhoI}
\rho_{\mathrm I}=\left(|p_{\mathrm I}|^2+|q_{\mathrm I}|^2\right)^{1/2}
\end{equation*}
to display the joint component structure.

To parametrize the breather solutions, we introduce the Joukowsky transformation
\begin{equation*}\label{Joukowsky}
\lambda=\frac{\mathrm{i}}{2}\left(\xi+\xi^{-1}\right),\qquad
\mu=\frac{1}{2}\left(\xi-\xi^{-1}\right),\qquad
\xi=e^{z+\mathrm{i}\alpha},\qquad
R=e^z.
\end{equation*}
It gives
\begin{equation*}\label{lambda-mu-J}
\lambda=\mathrm{i}\cosh z\cos\alpha-\sinh z\sin\alpha,
\qquad
\mu=\sinh z\cos\alpha+\mathrm{i}\cosh z\sin\alpha.
\end{equation*}
For compactness, define
\begin{equation*}\label{J-parameters}
\begin{aligned}
\mathcal{U}&=2\sinh z\cos\alpha,
&\mathcal{R}&=2\cosh z\sin\alpha,\\
\beta&=\cosh(2z)\sin(2\alpha),
&\kappa&=\sinh(2z)\cos(2\alpha),
\end{aligned}
\end{equation*}
and
\begin{equation*}\label{cs-def}
c=\cosh z\cos\alpha+\mathrm{i}\sinh z\sin\alpha,
\qquad
s=\sinh z\cos\alpha+\mathrm{i}\cosh z\sin\alpha.
\end{equation*}
The phase variables are
\begin{equation*}\label{Theta-def}
\Theta(x)=x(\mathcal{U}+\mathrm{i}\mathcal{R})+2\gamma,
\qquad
\Theta(t)=t(-\beta+\mathrm{i}\kappa)+2\mathrm{i}\theta.
\end{equation*}
The auxiliary functions are
\begin{equation*}\label{eq:4.7}
\begin{aligned}
\mathcal{C}&=\cosh\Theta(x)-c\cosh\Theta(t),\qquad
\mathcal{A}&=\mathcal{C}-s\sinh\Theta(t),\\
\mathcal{D}=\cosh\Theta(t)-c\cosh\Theta(x).
\end{aligned}
\end{equation*}
Consequently,
\begin{equation*}\label{H1-J}
H_1=\frac{2\cosh z\cos\alpha\,\mathcal{A}}{\mathcal{D}},
\qquad
p[1]=q[1]=\frac12e^{\mathrm{i}t}(1+H_1).
\end{equation*}

For the two-fold solution, we choose two distinct spectral points
\begin{equation*}
\xi_j=e^{z_j+\mathrm{i}\alpha_j},\qquad
R_j=e^{z_j},\qquad j=1,2.
\end{equation*}
Then
\begin{equation*}
p[2]=q[2]=\frac12e^{\mathrm{i}t}(1+H_2),
\end{equation*}
where
\begin{equation*}\label{eq:H2}
H_2=
\frac{2\mathrm{i}(\lambda_1-\lambda_2^*)(\lambda_1-\lambda_2)
[\lambda_2\mathcal{A}_2\mathcal{D}_1-\lambda_1\mathcal{A}_1\mathcal{D}_2]}
{\mathcal{D}_1\mathcal{D}_2(\lambda_1-\lambda_2^*)^2
-2\lambda_1\lambda_2\mathcal{B}}.
\end{equation*}
Here, $\mathcal{A}_j$, $\mathcal{C}_j$, and $\mathcal{D}_j$ follow the first-order definitions with the subscript $j$. The mixed term is
\begin{equation*}\label{B12}
\begin{aligned}
\mathcal{B}
={}&\mathcal{D}_1\mathcal{D}_2+\mathcal{C}_1\mathcal{C}_2\\
&+\mu_1\mu_2\sinh\Theta(x)_{1}\sinh\Theta(x)_{2}
-\mu_1\mu_2\sinh\Theta(t)_{1}\sinh\Theta(t)_{2}.
\end{aligned}
\end{equation*}
Unless otherwise stated, the phase parameters are set to zero so that the main structure is centered at $(x,t)=(0,0)$. We consider three representative order combinations.

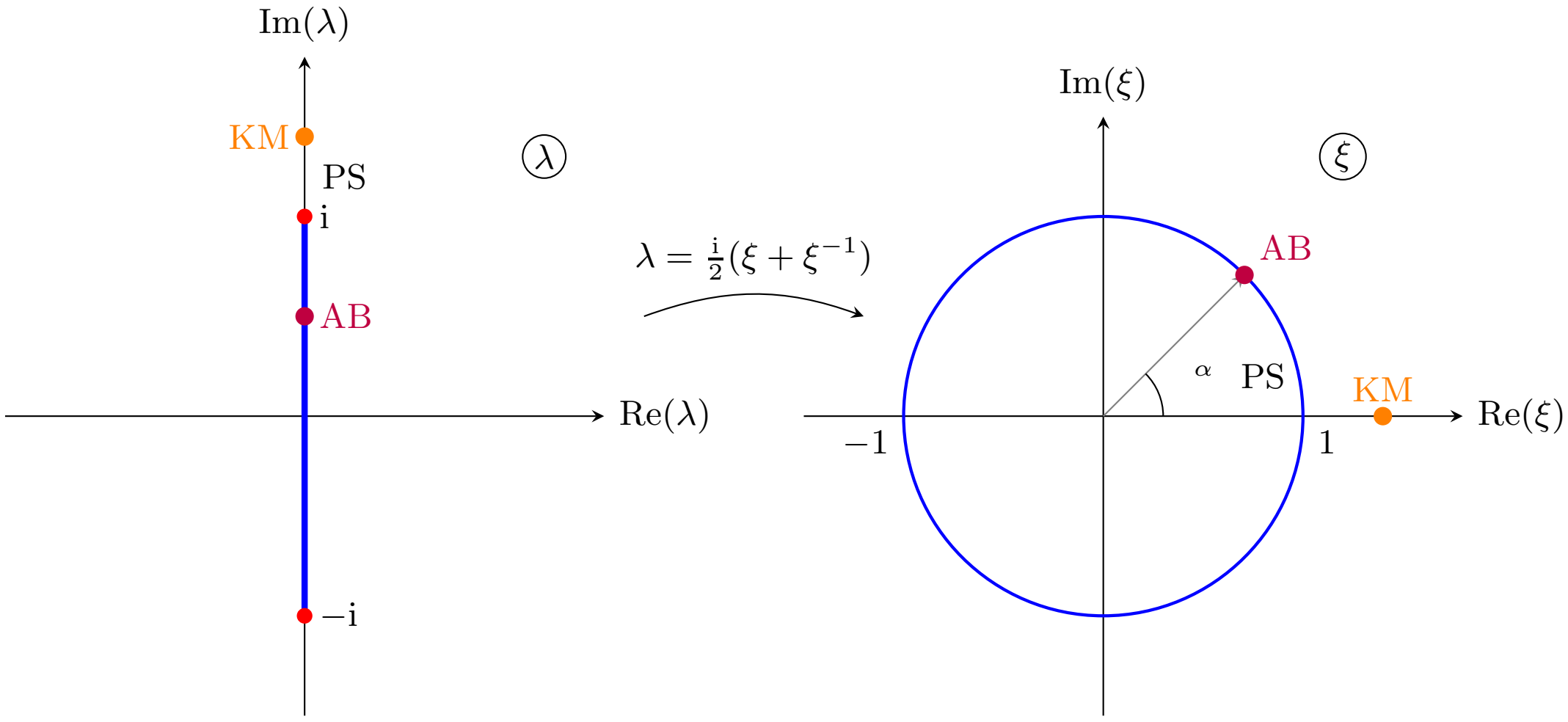
\begin{figure}[!tbh]
    \centering
    \begin{tikzpicture}[scale=1.8, >=stealth]
        \begin{scope}[shift={(-2.5,0)}]
            \draw[->] (-1.5,0) -- (1.5,0) node[right] {\footnotesize Re($\lambda$)};
            \draw[->] (0,-1.5) -- (0,1.8) node[above] {\footnotesize Im($\lambda$)};
            \node at (1.2,1.3) [circle, draw, inner sep=1pt] {\footnotesize $\lambda$};

            \draw[blue, ultra thick] (0,-1) -- (0,1);
            \filldraw[red] (0,1) circle (1pt) node[right, black] {\footnotesize $\mathrm{i}$};
            \filldraw[red] (0,-1) circle (1pt) node[right, black] {\footnotesize $-\mathrm{i}$};

            \filldraw[orange] (0,1.4) circle (1.2pt) node[left] {\footnotesize KM};
            \filldraw[purple] (0,0.5) circle (1.2pt) node[right] {\footnotesize AB};
            \node at (0.2,1.2) {\footnotesize PS};
        \end{scope}

        \draw[->, bend left=20] (-0.8, 0.5) to node[above] {\footnotesize $\lambda = \frac{\mathrm{i}}{2}(\xi + \xi^{-1})$} (0.3, 0.5);

        \begin{scope}[shift={(1.5,0)}]
            \draw[->] (-1.5,0) -- (1.8,0) node[right] {\footnotesize Re($\xi$)};
            \draw[->] (0,-1.5) -- (0,1.5) node[above] {\footnotesize Im($\xi$)};
            \node at (1.2,1.3) [circle, draw, inner sep=1pt] {\footnotesize $\xi$};

            \draw[blue, thick] (0,0) circle (1);
            \node[below right] at (1,0) {\footnotesize $1$};
            \node[below left] at (-1,0) {\footnotesize $-1$};

            \filldraw[orange] (1.4,0) circle (1.2pt) node[above] {\footnotesize KM};
            \draw[->, gray, thin] (0,0) -- (45:1);
            \filldraw[purple] (45:1) circle (1.2pt) node[above right] {\footnotesize AB};
            \draw (0.3,0) arc (0:45:0.3) node[pos=0.5, above right, xshift=5.2, yshift=0.6] {\tiny $\alpha$};;
            \node at (0.8,0.2) {\footnotesize PS};
        \end{scope}
    \end{tikzpicture}
    \caption{Uniformization of the spectral plane by the Joukowsky transformation. The branch cut on the imaginary $\lambda$ axis is mapped onto the unit circle in the $\xi$ plane.}
\end{figure}

\subsection{The $(0,1)$-order configuration}

For $M=0$ and $N=1$, equation \eqref{qr7} gives
\begin{equation}\label{TypeI01}
\begin{aligned}
p_{\mathrm I}^{[0,1]}=e^{\mathrm{i}t}\left(1+\frac12H_1\right),\qquad
q_{\mathrm I}^{[0,1]}=\frac12e^{\mathrm{i}t}H_1.
\end{aligned}
\end{equation}
The principal spectral regimes are shown in Figure 1:
\begin{itemize}
    \item \textbf{KM regime:} $\alpha=0$ and $z>0$, for which the spectral point lies on the positive real axis of the $\xi$ plane outside the unit circle;
    \item \textbf{Akhmediev regime:} $z=0$ and $\alpha\neq0$, for which the spectral point lies on the unit circle $|\xi|=1$;
    \item \textbf{Peregrine limit:} $(z,\alpha)\to(0,0)$, corresponding to the branch point $\xi=1$.
\end{itemize}


For $\alpha=0$ and $z>0$, we have $\lambda=\mathrm{i}\cosh z$, and the solution reduces to
\begin{equation}
\begin{aligned}
    p_{\textbf{I},\textbf{KM}}^{[0,1]} &=  e^{\mathrm{i}t} \sinh z \frac{ \sinh z \cos(t \sinh(2z)) + \mathrm{i} \cosh z \sin(t \sinh(2z)) }{ \cosh z \cosh(2x \sinh z) - \cos(t \sinh(2z))},\\
    q_{\textbf{I},\textbf{KM}}^{[0,1]} &= - e^{\mathrm{i}t} \cosh z \frac{ \cosh(2x \sinh z) - \cosh z \cos(t \sinh(2z)) - \mathrm{i} \sinh z \sin(t \sinh(2z)) }{ \cosh z \cosh(2x \sinh z) - \cos(t \sinh(2z)) }.
\end{aligned}
\end{equation}
This solution is localized in $x$ and periodic in $t$, with period
\begin{equation*}\label{KM-period}
T_{\textbf{KM}}=\frac{2\pi}{\sinh(2z)}.
\end{equation*}

The KM breather is a canonical space-localized and time-periodic solution of the focusing NLS equation and has been observed experimentally in optical fibre \cite{KiblerEtAl2012}. Its dynamical mechanism has also been investigated in relation to modulation instability and nonlinear-wave interference \cite{ZhaoLingYang2018}.

As shown in Figure 2, $p_{\mathrm I}$ is localized on a zero background, whereas $q_{\mathrm I}$ approaches the finite background $|q_{\mathrm I}|=1$. The combined amplitude $\rho_{\mathrm I}$ exhibits periodic compression and relaxation near the origin.

\begin{figure}[!tbh]
    \centering
    \begin{subfigure}[b]{0.32\textwidth}
        \centering
        \includegraphics[height=3.2cm]{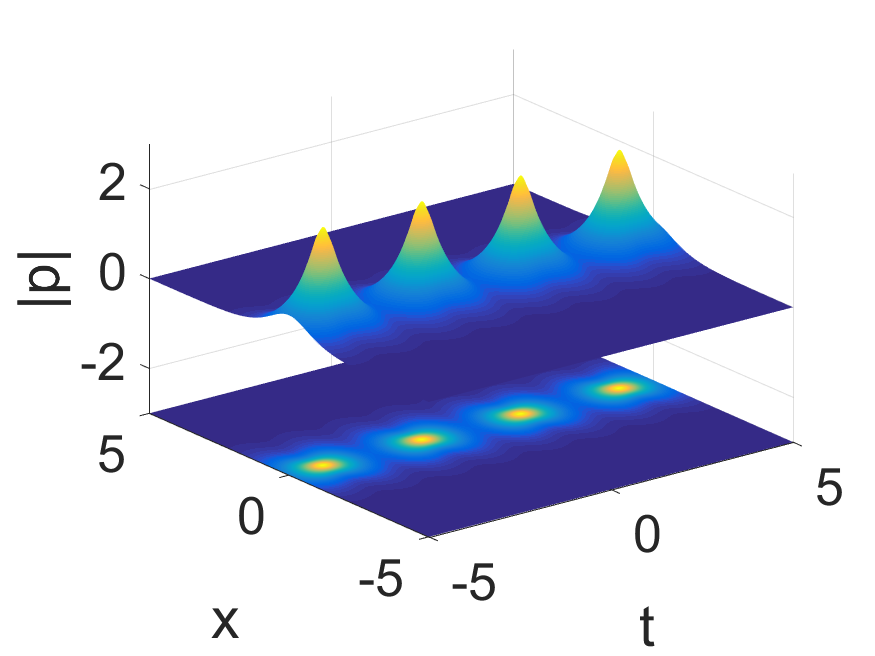}
        \caption{}
    \end{subfigure}
    \hfill
    \begin{subfigure}[b]{0.32\textwidth}
        \centering
        \includegraphics[height=3.2cm]{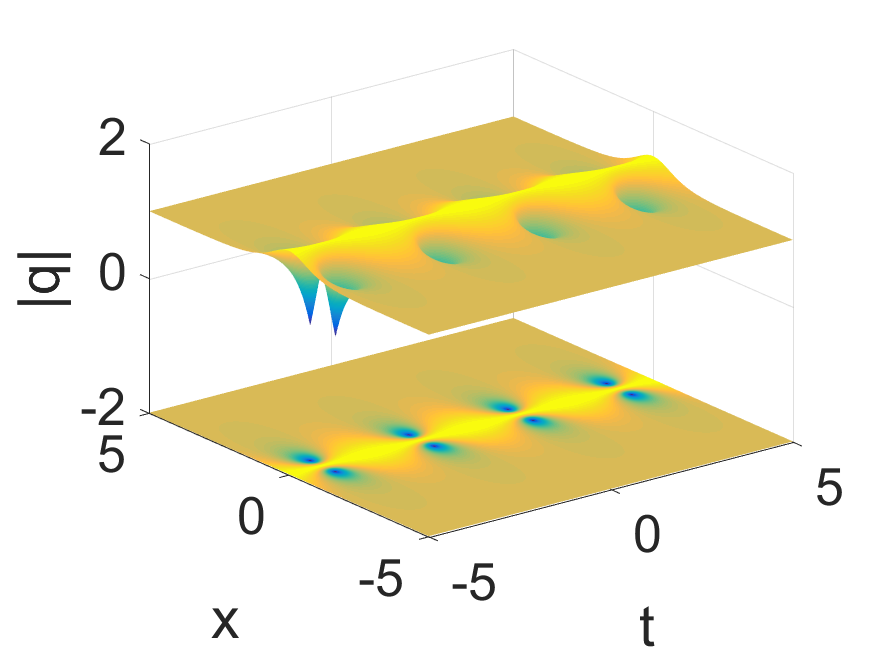}
        \caption{}
    \end{subfigure}
    \hfill
    \begin{subfigure}[b]{0.32\textwidth}
        \centering
        \includegraphics[height=3.2cm]{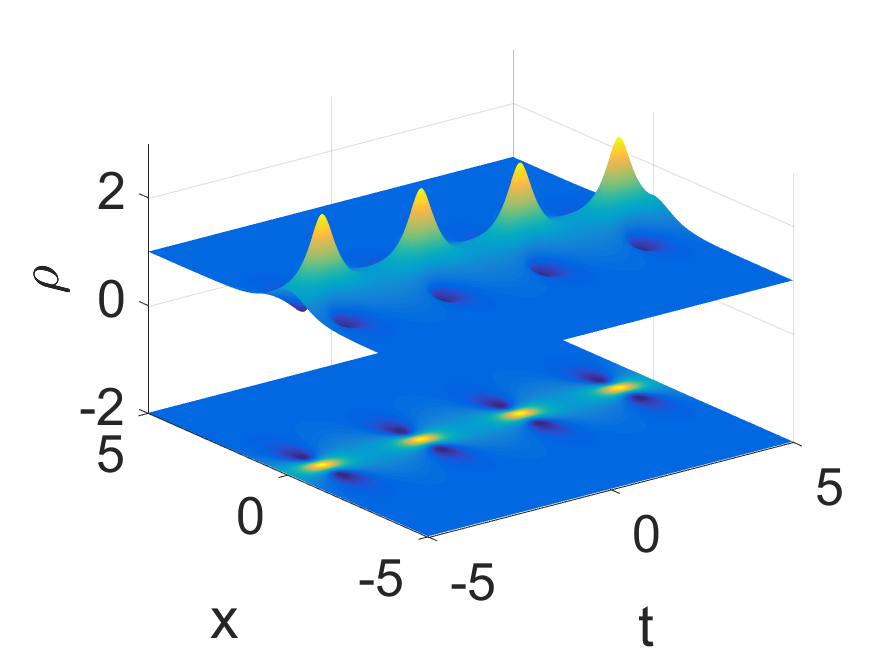}
        \caption{}
    \end{subfigure}
    \caption{Type-I KM configuration. (a) $|p_{\mathrm I}|$; (b) $|q_{\mathrm I}|$; (c) combined amplitude $\rho_{\mathrm I}=(|p_{\mathrm I}|^2+|q_{\mathrm I}|^2)^{1/2}$.}
\end{figure}

For $z=0$ and $\alpha\neq0$, the spectral point lies on the unit circle in the $\xi$ plane. The explicit solution is
\begin{equation}
\begin{aligned}
   p_{\textbf{I},\textbf{AB}}^{[0,1]} &=  e^{\mathrm{i}t} \sin\alpha \frac{ \sin\alpha \cosh(\sin(2\alpha) t) + \mathrm{i}\cos\alpha \sinh(\sin(2\alpha) t) }{\cosh(\sin(2\alpha) t) - \cos\alpha \cos(2\sin\alpha x)},\\
   q_{\textbf{I},\textbf{AB}}^{[0,1]} &= - e^{\mathrm{i}t} \cos\alpha \frac{ \cos\alpha \cosh(\sin(2\alpha) t) - \cos(2\sin\alpha x) - \mathrm{i}\sin\alpha \sinh(\sin(2\alpha) t) }{\cosh(\sin(2\alpha) t) - \cos\alpha \cos(2\sin\alpha x)}.
\end{aligned}
\end{equation}

The Akhmediev configuration is periodic in $x$, with period
\begin{equation*}\label{AB-period}
L_{\textbf{AB}}=\frac{\pi}{|\sin\alpha|},
\end{equation*}
and localized in $t$. As shown in Figure 3, the solution forms a periodic array of spatial peaks and undergoes a single amplification event in time. As $|\alpha|$ decreases, the spatial period increases and the peak array becomes sparser. The combined amplitude $\rho_{\mathrm I}$ displays the same periodic spatial organization.
\begin{figure}[!tbh]
    \centering
    \begin{subfigure}[b]{0.32\textwidth}
        \centering
        \includegraphics[height=3.2cm]{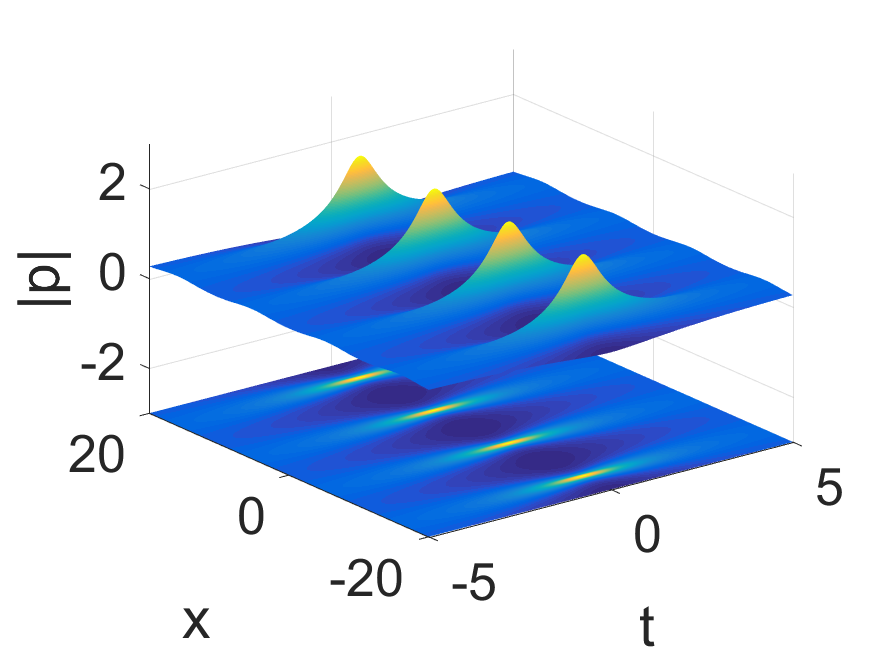}
        \caption{}
    \end{subfigure}
    \hfill
    \begin{subfigure}[b]{0.32\textwidth}
        \centering
        \includegraphics[height=3.2cm]{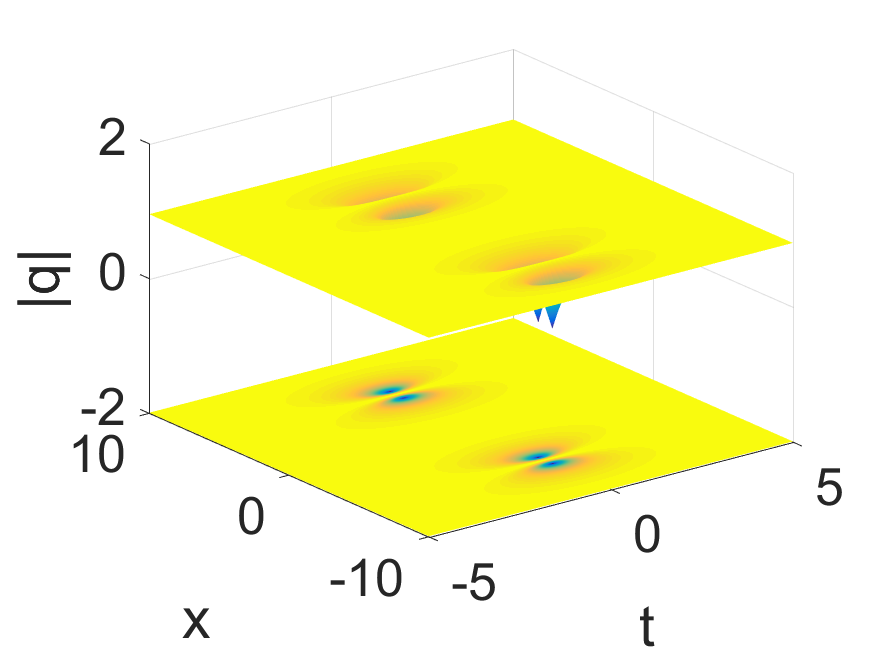}
        \caption{}
    \end{subfigure}
    \hfill
    \begin{subfigure}[b]{0.32\textwidth}
        \centering
        \includegraphics[height=3.2cm]{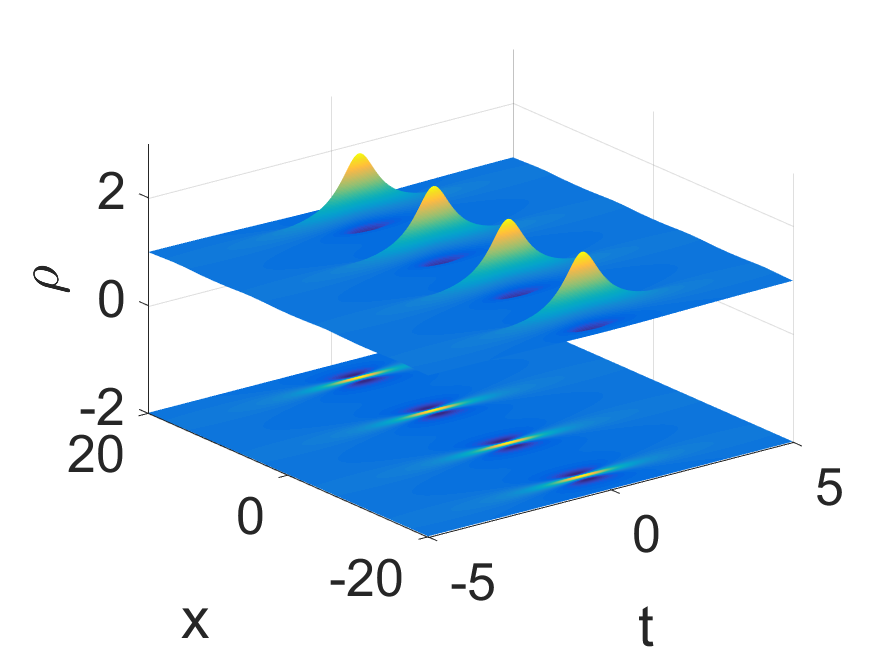}
        \caption{}
    \end{subfigure}
    \caption{Type-I Akhmediev configuration. (a) $|p_{\mathrm I}|$; (b) $|q_{\mathrm I}|$; (c) combined amplitude $\rho_{\mathrm  I}=(|p_{\mathrm I}|^2+|q_{\mathrm I}|^2)^{1/2}$.}
\end{figure}

As $(z,\alpha)\to(0,0)$, the spectral point approaches the branch point $\xi=1$, and the first-order solution reduces to the doubly localized Peregrine form, consistently with the standard KM/AB-to-Peregrine limiting relation of the focusing scalar NLS equation \cite{Karjanto2021}.
\begin{equation}
\begin{aligned}
    p_{\textbf{I},\textbf{PS}}^{[0,1]} = e^{\mathrm{i}t} \left( \frac{2+4\mathrm{i}t}{1+4x^2+4t^2} \right),\qquad
    q_{\textbf{I},\textbf{PS}}^{[0,1]} = - e^{\mathrm{i}t} \left( 1 - \frac{2+4\mathrm{i}t}{1+4x^2+4t^2} \right).
\end{aligned}
\end{equation}

\begin{figure}[!tbh]
    \centering
    \begin{subfigure}[b]{0.32\textwidth}
        \centering
        \includegraphics[height=3.2cm]{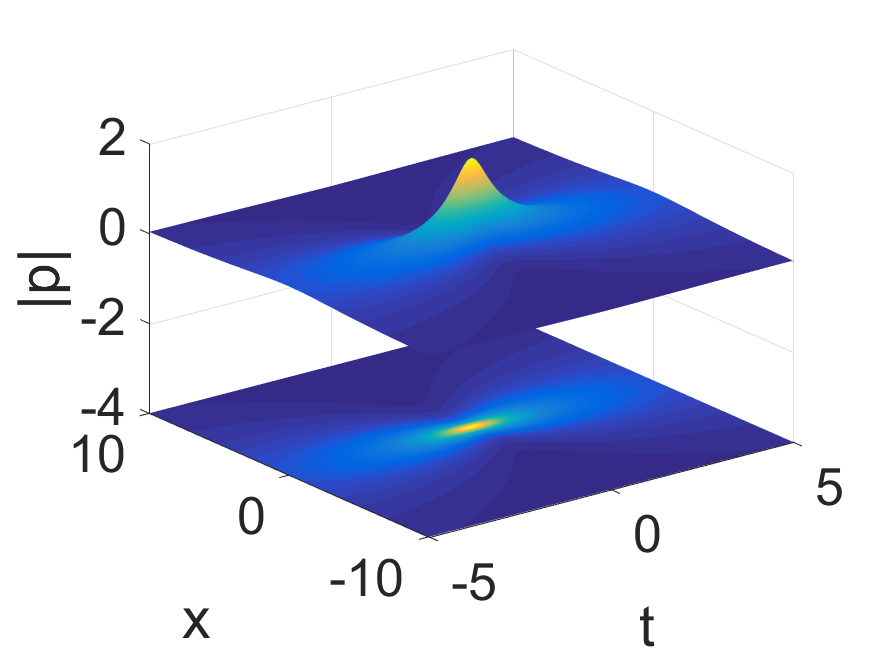}
        \caption{}
    \end{subfigure}
    \hfill
    \begin{subfigure}[b]{0.32\textwidth}
        \centering
        \includegraphics[height=3.2cm]{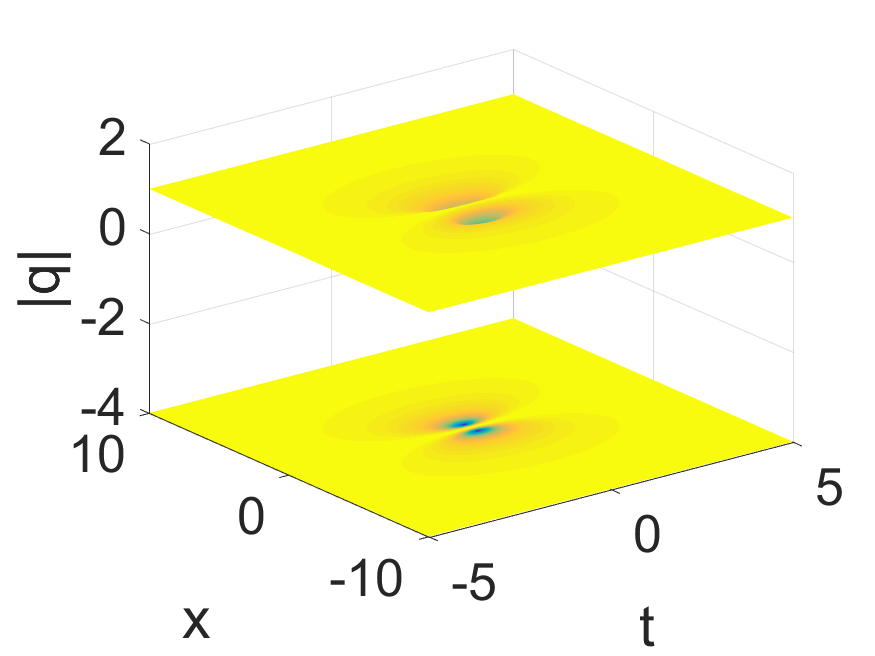}
        \caption{}
    \end{subfigure}
    \hfill
    \begin{subfigure}[b]{0.32\textwidth}
        \centering
        \includegraphics[height=3.2cm]{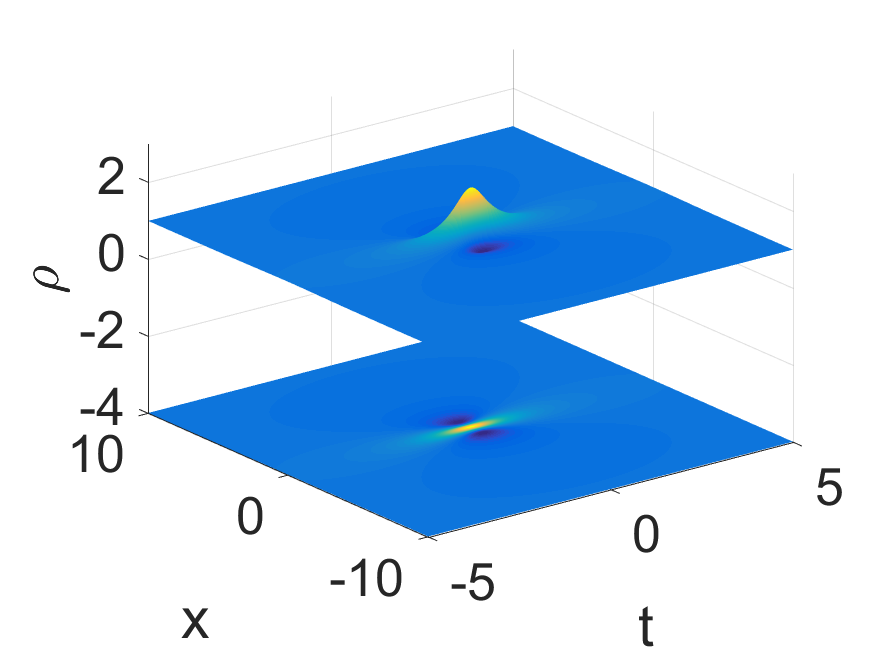}
        \caption{}
    \end{subfigure}
    \caption{Type-I Peregrine configuration. (a) $|p_{\mathrm I}|$; (b) $|q_{\mathrm I}|$; (c) combined amplitude $\rho_{\mathrm I}=(|p_{\mathrm I}|^2+|q_{\mathrm I}|^2)^{1/2}$.}
\end{figure}

Figure~4 shows a doubly localized Type-I Peregrine configuration. The $p_{\mathrm I}$ component is localized on a zero background, whereas $q_{\mathrm I}$ approaches the finite background $|q_{\mathrm I}|=1$. The combined amplitude $\rho_{\mathrm I}$ reaches its maximum near $(x,t)=(0,0)$.

\begin{figure}[!tbh]
    \centering
    \begin{subfigure}[b]{0.48\textwidth}
        \centering
        \includegraphics[width=\textwidth, height=4.2cm]{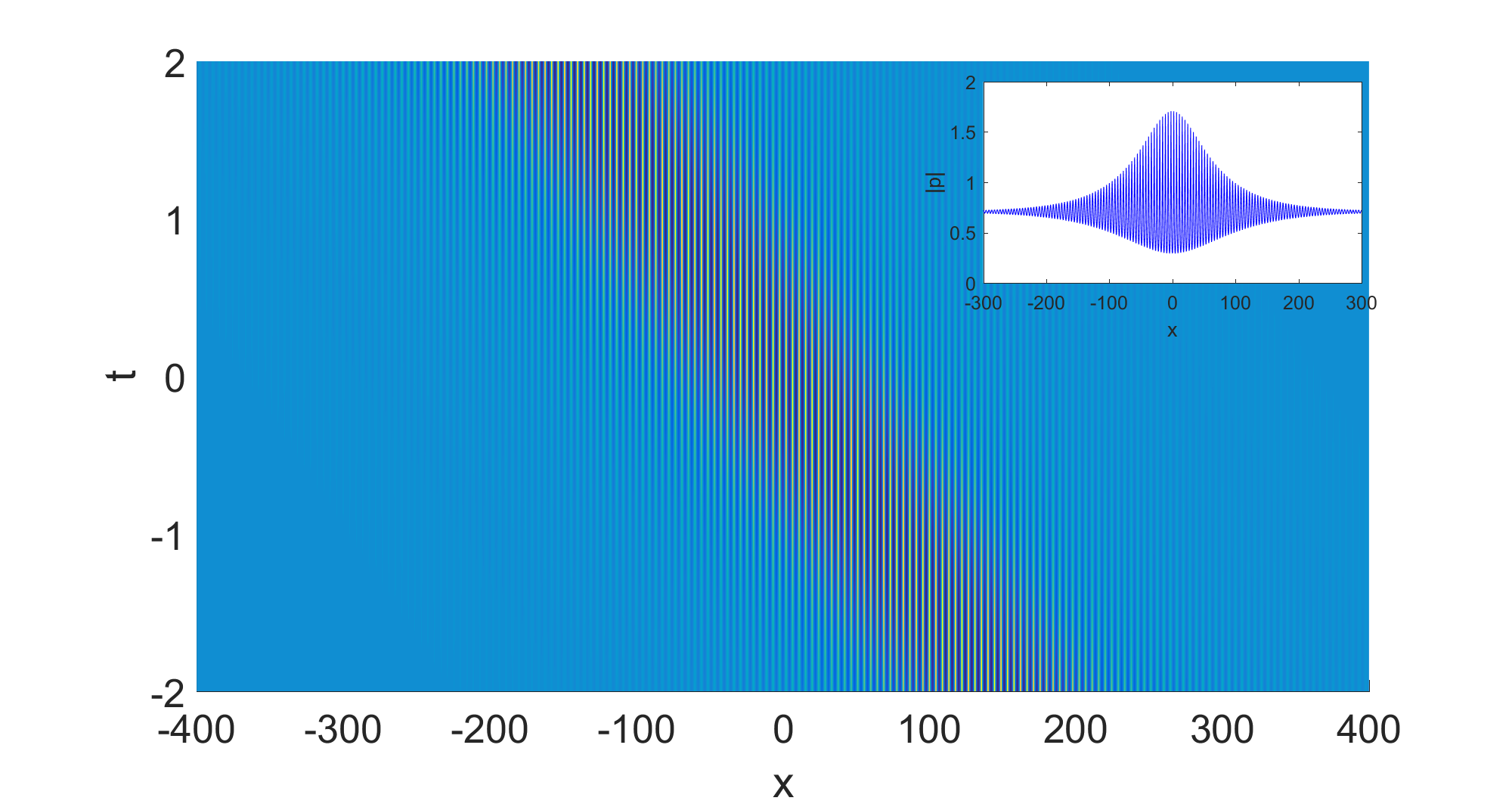}
        \caption{}
    \end{subfigure}
    \hfill
    \begin{subfigure}[b]{0.48\textwidth}
        \centering
        \includegraphics[width=\textwidth, height=4.2cm]{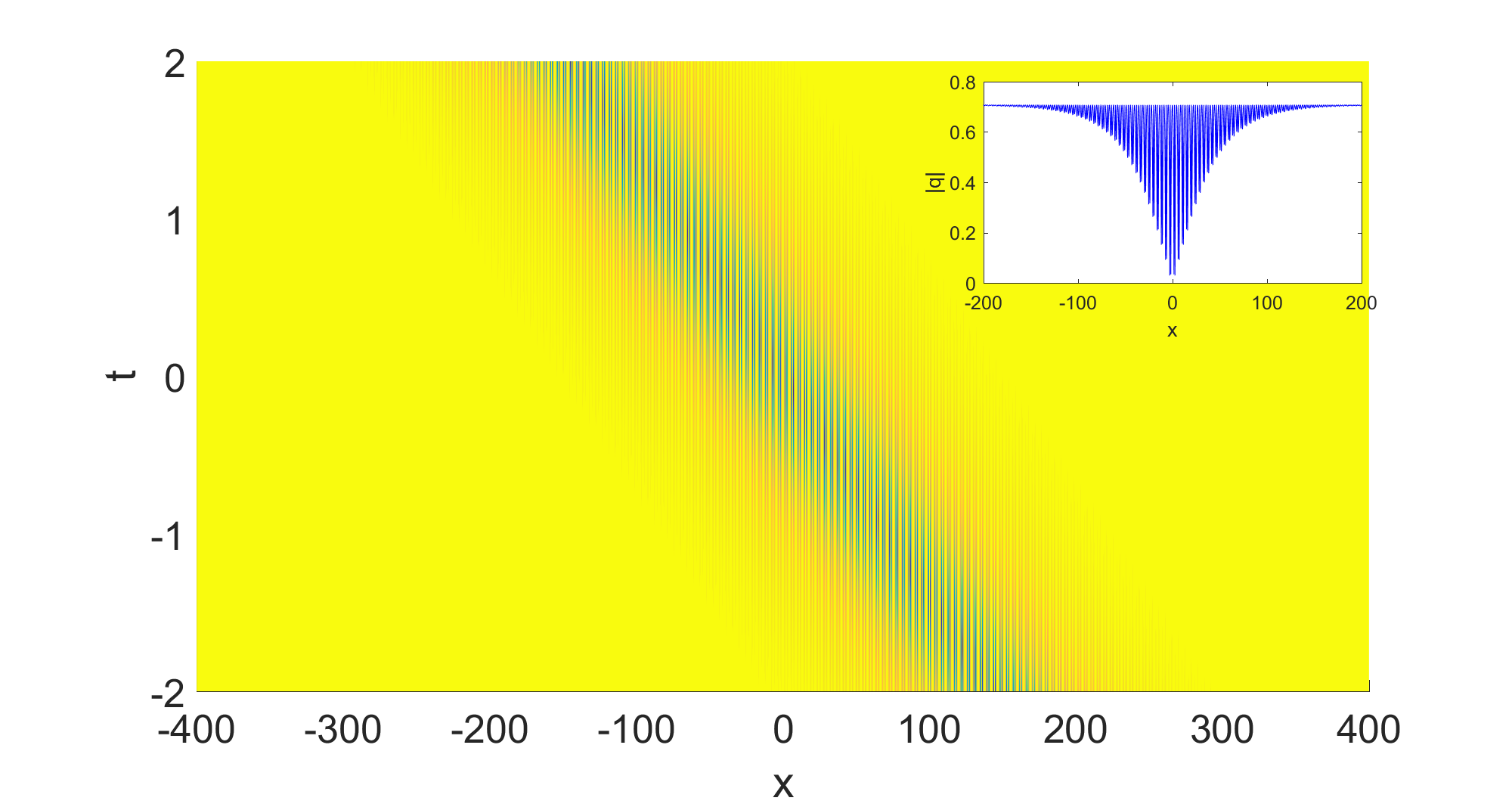}
        \caption{}
    \end{subfigure}

    \vspace{6pt}

    \begin{subfigure}[b]{0.48\textwidth}
        \centering
        \includegraphics[width=\textwidth, height=4.2cm]{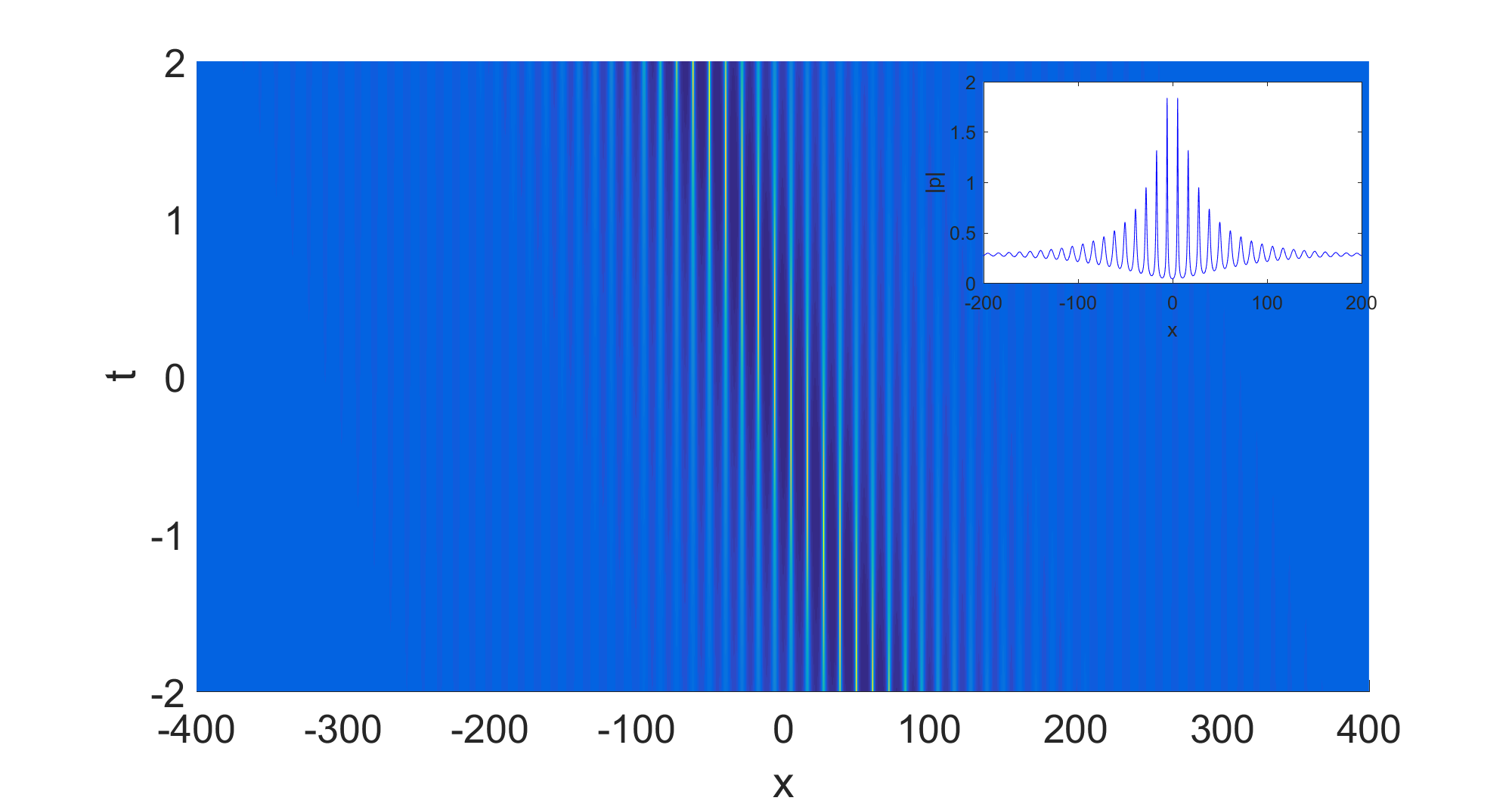}
        \caption{}
    \end{subfigure}
    \hfill
    \begin{subfigure}[b]{0.48\textwidth}
        \centering
        \includegraphics[width=\textwidth, height=4.2cm]{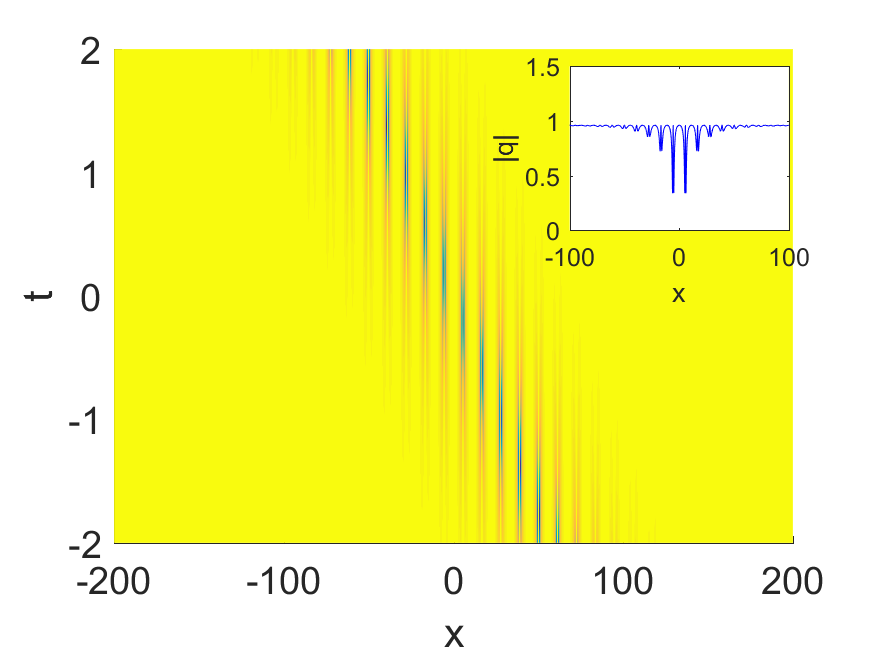}
        \caption{}
    \end{subfigure}

    \vspace{6pt}

    \begin{subfigure}[b]{0.48\textwidth}
        \centering
        \includegraphics[width=\textwidth, height=4.2cm]{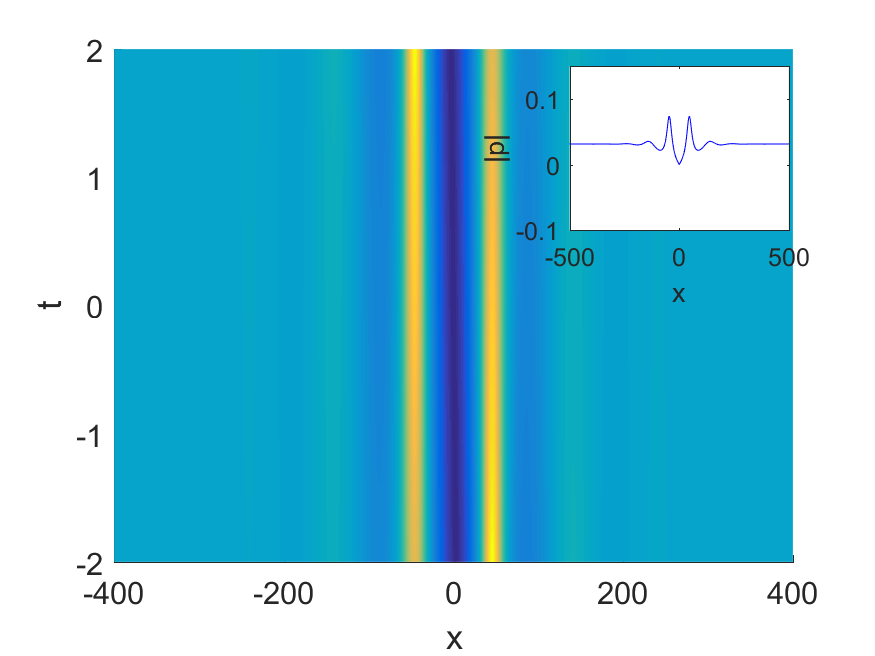}
        \caption{}
    \end{subfigure}
    \hfill
    \begin{subfigure}[b]{0.48\textwidth}
        \centering
        \includegraphics[width=\textwidth, height=4.2cm]{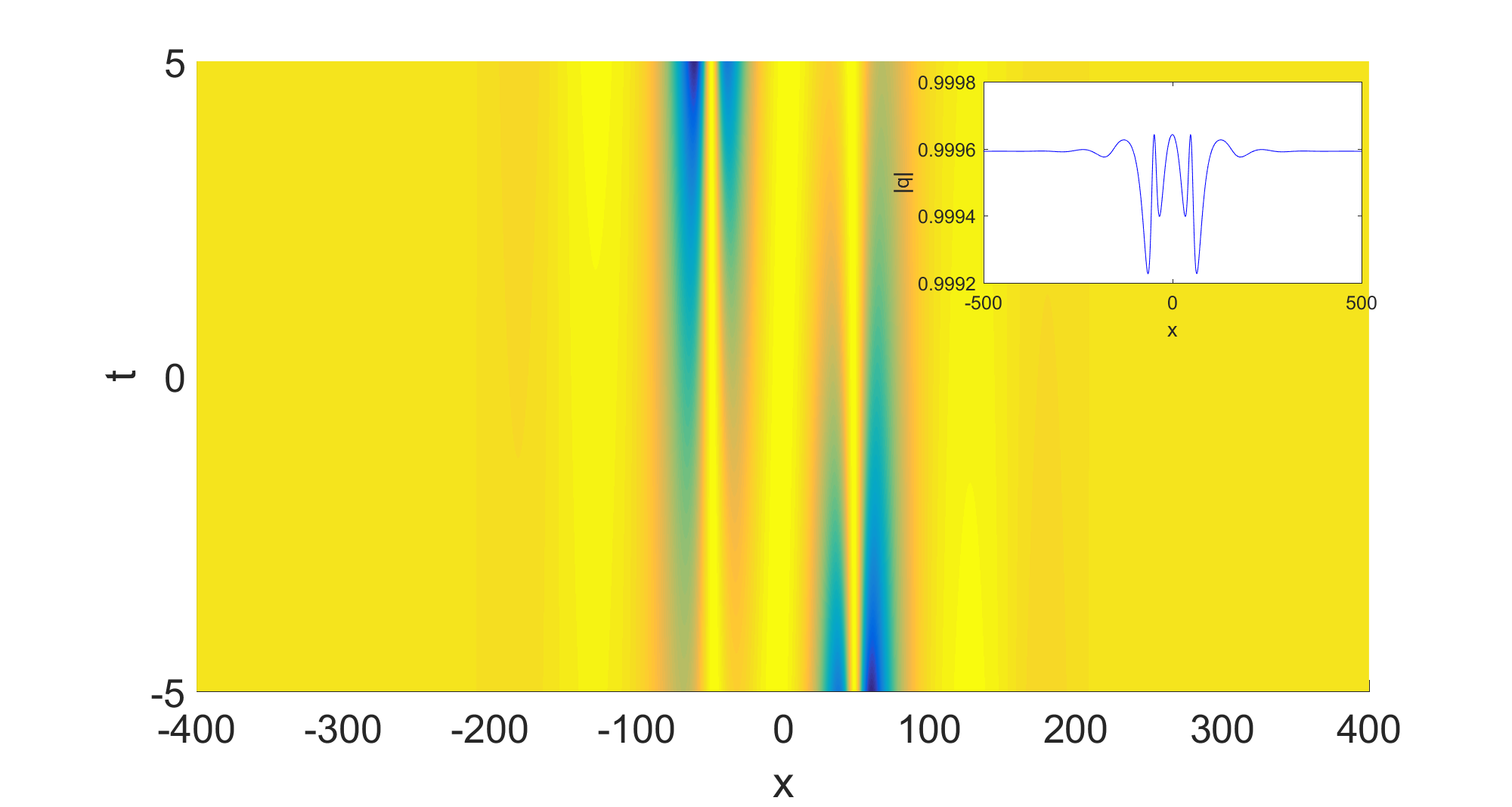}
        \caption{}
    \end{subfigure}

    \caption{Evolution of a near-unit-circle Type-I breather as the modulation angle $\alpha$ varies, with $R=1.01$ and $\gamma=\theta=0$. (a,b) $\alpha=\pi/4$; (c,d) $\alpha=\pi/11$; (e,f) $\alpha=\pi/100$. The left and right panels show $|p_{\mathrm I}|$ and $|q_{\mathrm I}|$, respectively.}
\end{figure}

Figure~5 shows how the waveform changes as $\alpha$ decreases at fixed $R=1.01$. The dominant spatial scale is inversely related to $|\sin\alpha|$, so the oscillations become progressively sparser. For $\alpha=\pi/4$, both components display dense periodic modulation. At $\alpha=\pi/11$, the spacing increases and the pattern becomes more localized. For $\alpha=\pi/100$, the solution is close to a rationally localized structure. Because $R\neq1$, this sequence does not reach the exact Peregrine limit, which requires $R\to1$ and $\alpha\to0$ simultaneously.

\subsection{The $(0,2)$-order configuration}

For $M=0$ and $N=2$, equation \eqref{qr7} gives
\begin{equation}\label{TypeI02}
\begin{aligned}
p_{\mathrm I}^{[0,2]}=e^{\mathrm{i}t}\left(1+\frac12H_2\right),\qquad
q_{\mathrm I}^{[0,2]}=\frac12e^{\mathrm{i}t}H_2.
\end{aligned}
\end{equation}
Compared with the $(0,1)$-order configuration, the $(0,2)$-order solution contains two spectral points and therefore supports a wider range of structures. Its morphology depends on the individual parameter pairs $(R_1,\alpha_1)$ and $(R_2,\alpha_2)$, and particularly on their relative symmetry or mismatch.
\begin{figure}[!tbh]
    \centering
    \begin{subfigure}[b]{0.32\textwidth}
        \centering
        \includegraphics[height=3.2cm]{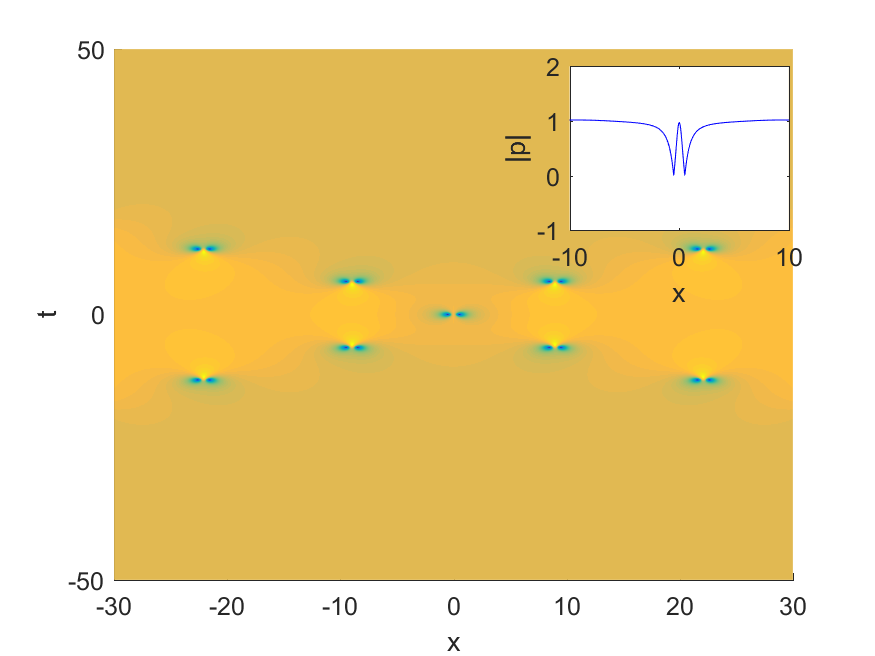}
        \caption{}
    \end{subfigure}
    \hfill
    \begin{subfigure}[b]{0.32\textwidth}
        \centering
        \includegraphics[height=3.2cm]{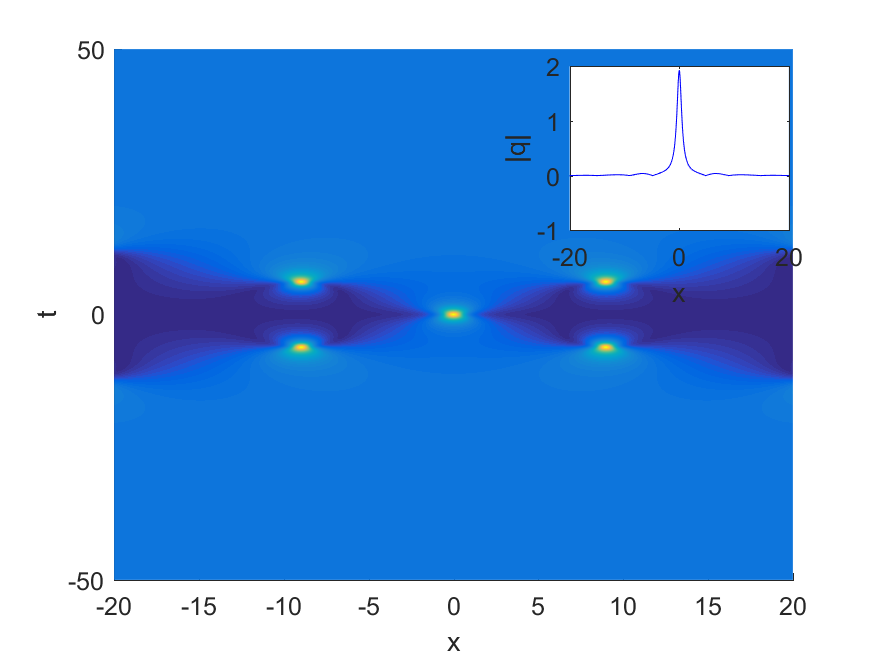}
        \caption{}
    \end{subfigure}
    \hfill
    \begin{subfigure}[b]{0.32\textwidth}
        \centering
        \includegraphics[height=3.2cm]{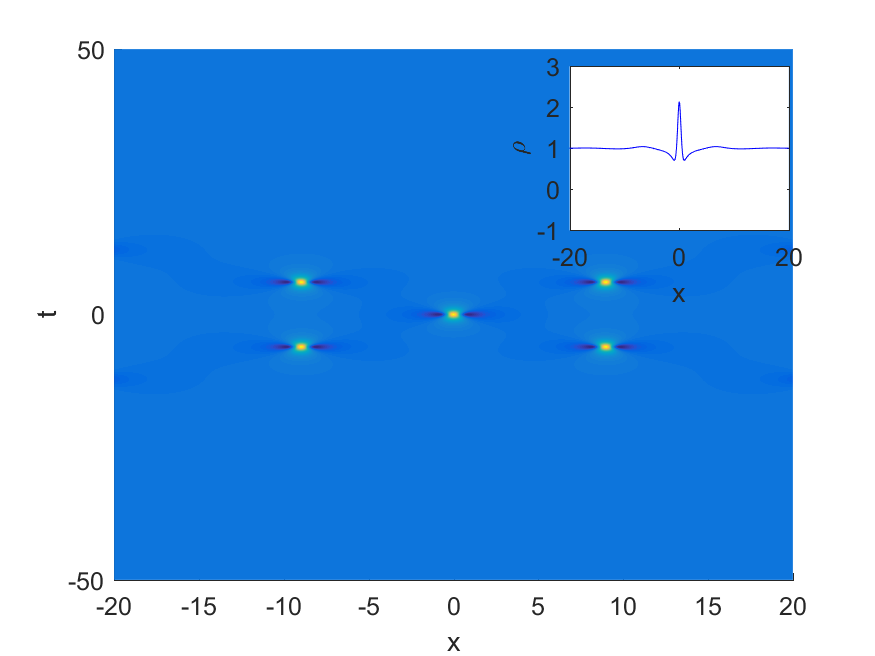}
        \caption{}
    \end{subfigure}

    \vspace{6pt}

    \begin{subfigure}[b]{0.32\textwidth}
        \centering
        \includegraphics[height=3.2cm]{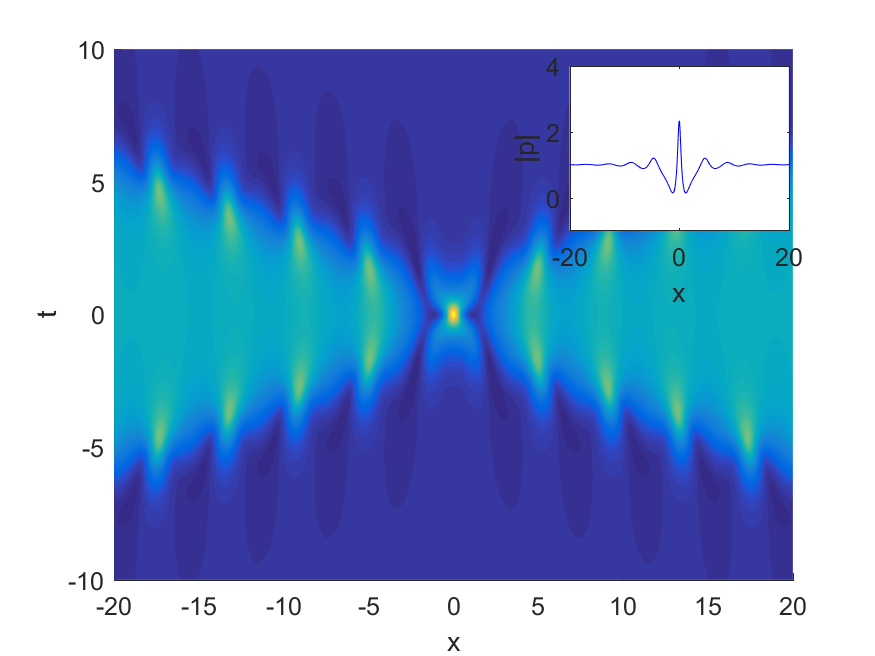}
        \caption{}
    \end{subfigure}
    \hfill
    \begin{subfigure}[b]{0.32\textwidth}
        \centering
        \includegraphics[height=3.2cm]{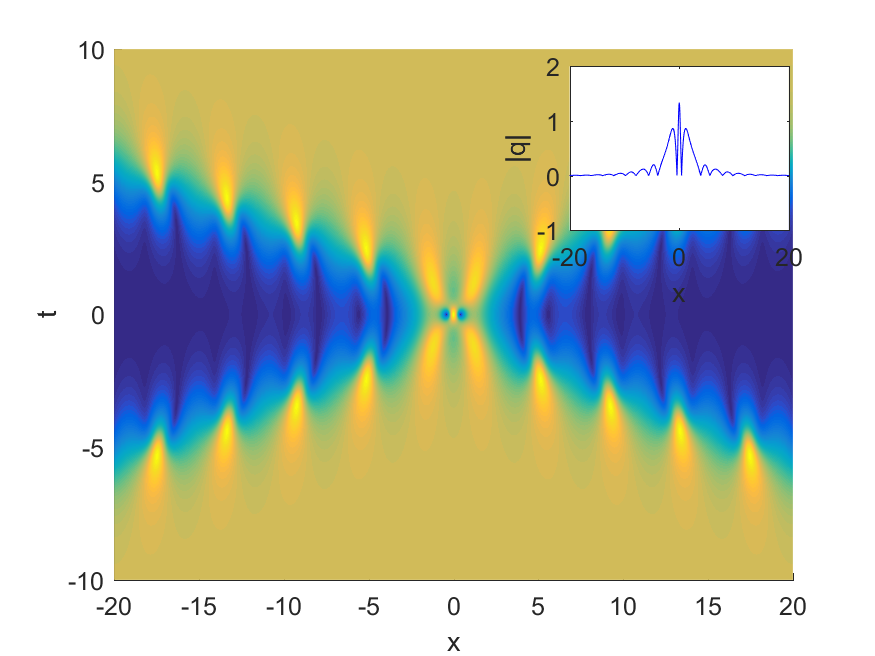}
        \caption{}
    \end{subfigure}
    \hfill
    \begin{subfigure}[b]{0.32\textwidth}
        \centering
        \includegraphics[height=3.2cm]{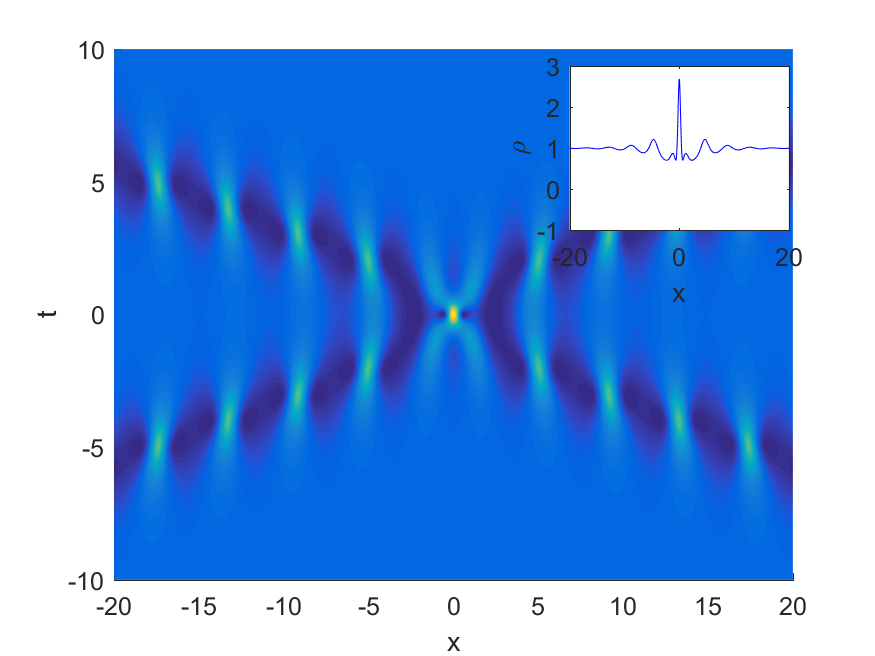}
        \caption{}
    \end{subfigure}

    \vspace{6pt}

    \begin{subfigure}[b]{0.32\textwidth}
        \centering
        \includegraphics[height=3.2cm]{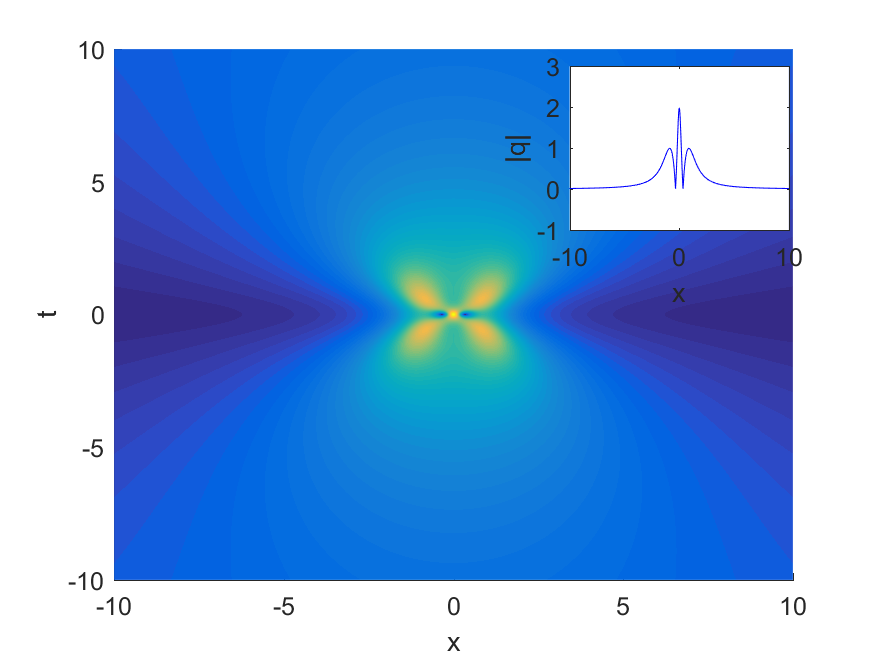}
        \caption{}
    \end{subfigure}
    \hfill
    \begin{subfigure}[b]{0.32\textwidth}
        \centering
        \includegraphics[height=3.2cm]{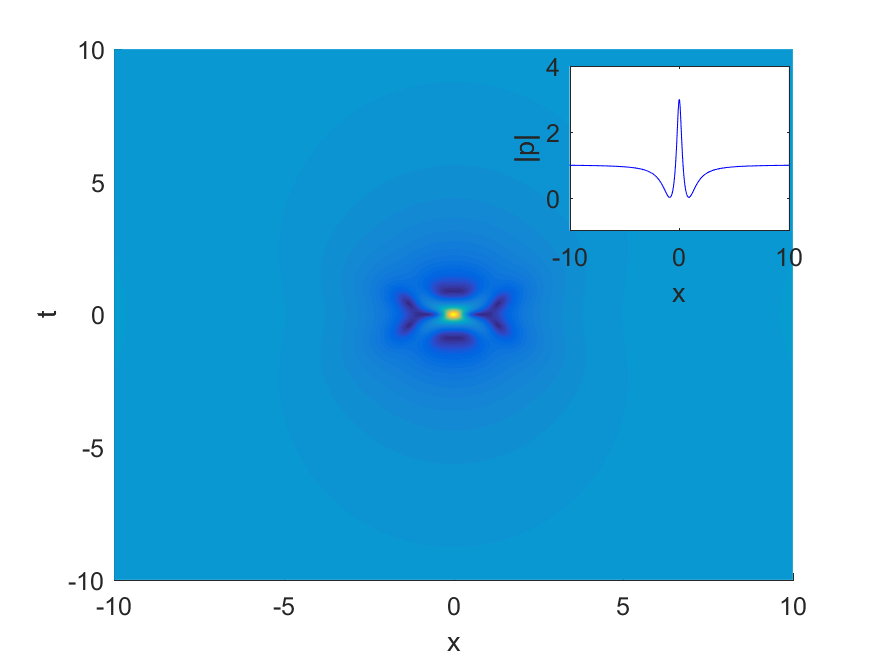}
        \caption{}
    \end{subfigure}
    \hfill
    \begin{subfigure}[b]{0.32\textwidth}
        \centering
        \includegraphics[height=3.2cm]{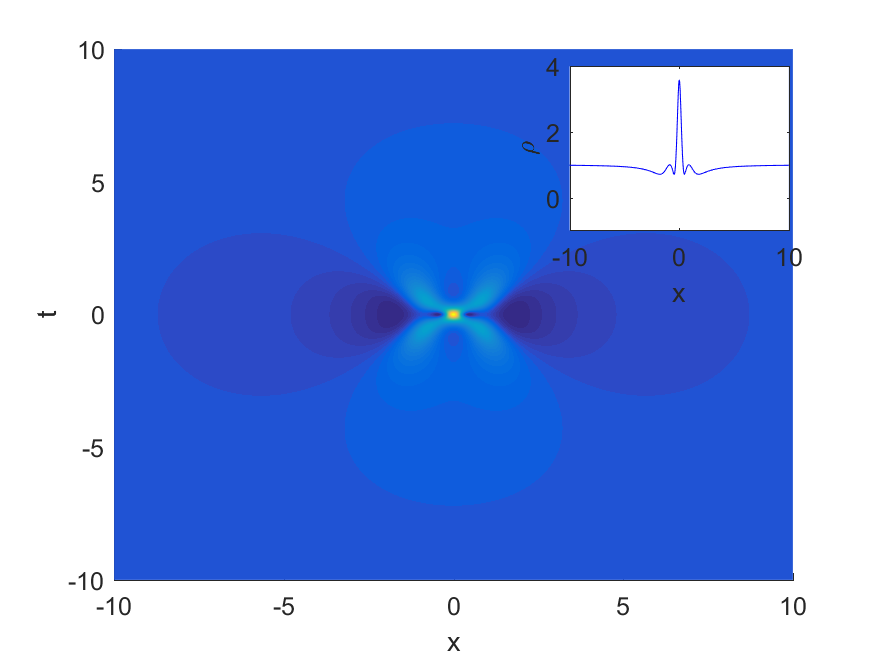}
        \caption{}
    \end{subfigure}
    \caption{Density plots of $(0,2)$-order Type-I configurations with $\theta_1=\theta_2=0$ and $\gamma_1=\gamma_2=0$. (a--c) $R_1=R_2=1.1$ and $\alpha_1=-\alpha_2=0.2$; (d--f) $R_1=R_2=1.2$ and $\alpha_1=-\alpha_2=4$; (g--i) $R_1=R_2=1.01$ and $\alpha_1=-\alpha_2=3$.}
\end{figure}


Figure~6 presents three representative $(0,2)$-order Type-I configurations generated under the mirror-symmetric spectral condition $R_1=R_2$ and $\alpha_1=-\alpha_2$. In Figures~6(a)--6(c), the spectral points lie close to the unit circle, which produces a symmetric quasi-Akhmediev-type multipeak pattern. The $p_I$ and $q_I$ components exhibit complementary peak--dip distributions, whereas $\rho_I$ preserves the overall composite structure. In Figures~6(d)--6(f), the larger spectral radius and angular parameters produce two obliquely propagating KM-like wave packets, whose superposition in the physical-component basis produces a symmetric X-shaped pattern. In Figures~6(g)--6(i), the spectral points approach the branch-point regime near $\xi=-1$, and the extended oscillatory pattern contracts into a near-rational localized structure. Since the two spectral points remain distinct, this configuration should be interpreted as a distinct-point two-fold solution close to the rational limit rather than as an exact second-order rational rogue wave of the standard higher-order NLS family \cite{OhtaYang2012}.

\begin{figure}[!tbh]
    \centering
    \begin{subfigure}[b]{0.32\textwidth}
        \centering
        \includegraphics[height=3.2cm]{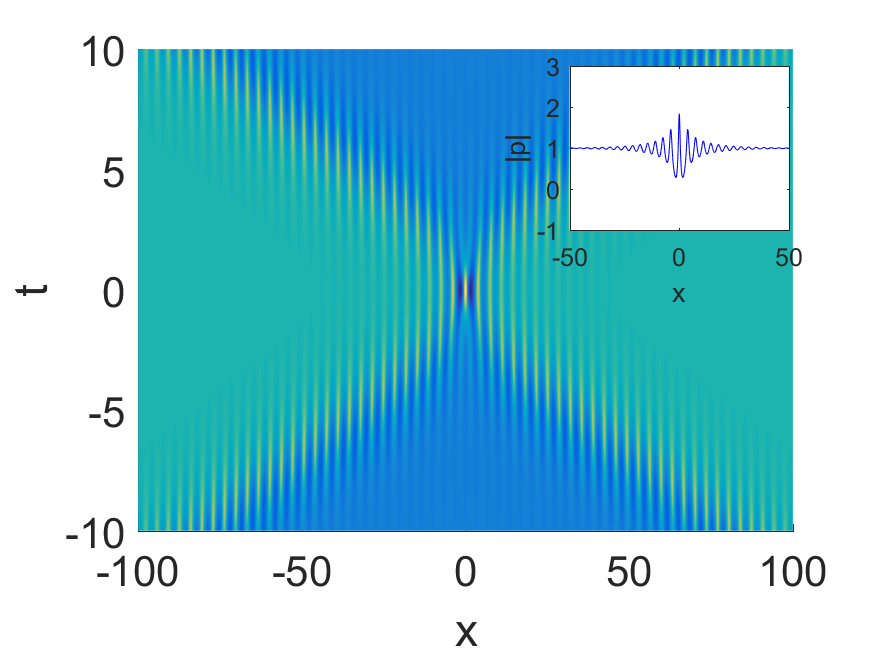}
        \caption{}
    \end{subfigure}
    \hfill
    \begin{subfigure}[b]{0.32\textwidth}
        \centering
        \includegraphics[height=3.2cm]{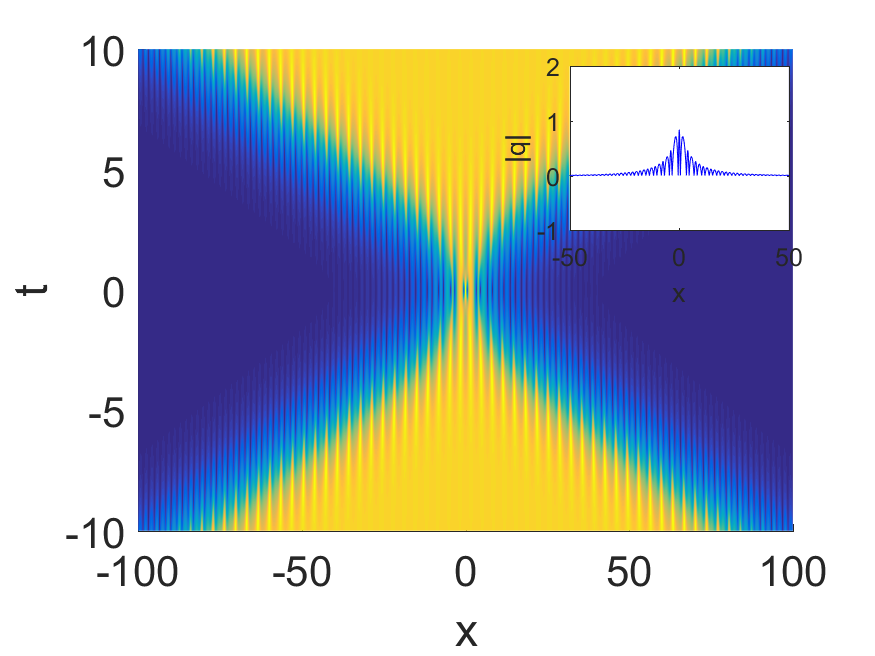}
        \caption{}
    \end{subfigure}
    \hfill
    \begin{subfigure}[b]{0.32\textwidth}
        \centering
        \includegraphics[height=3.2cm]{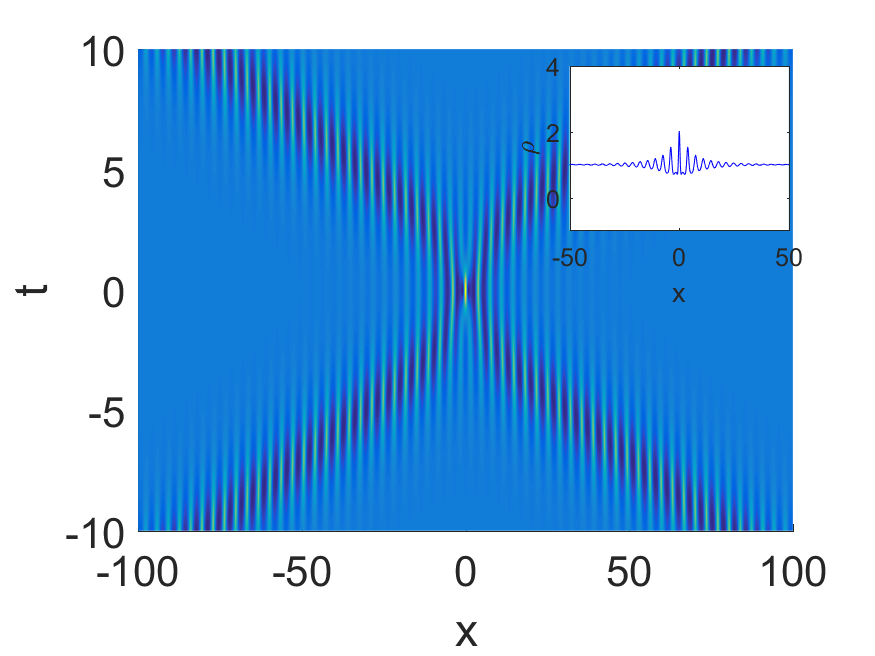}
        \caption{}
    \end{subfigure}

    \vspace{6pt}

    \begin{subfigure}[b]{0.32\textwidth}
        \centering
        \includegraphics[height=3.2cm]{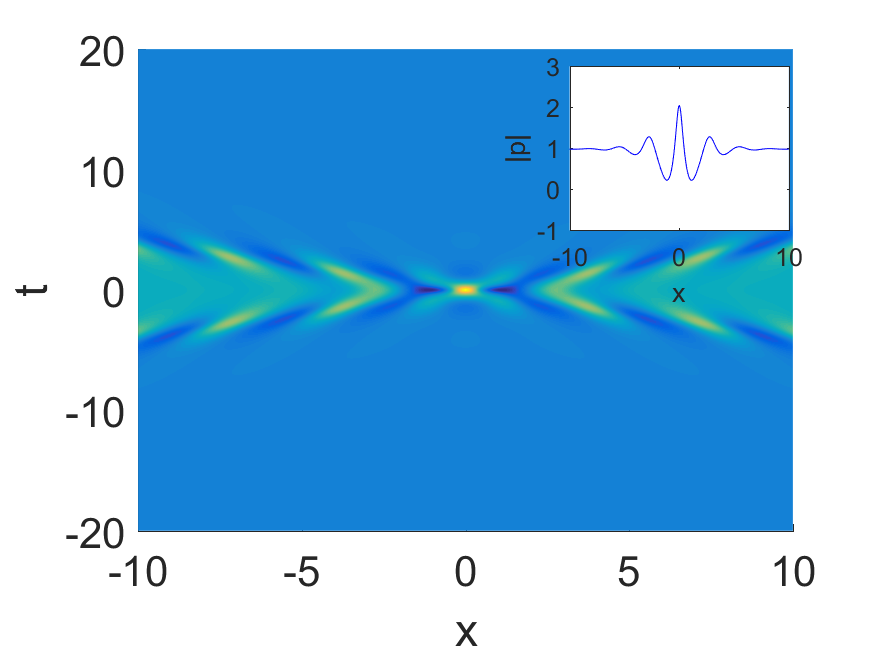}
        \caption{}
    \end{subfigure}
    \hfill
    \begin{subfigure}[b]{0.32\textwidth}
        \centering
        \includegraphics[height=3.2cm]{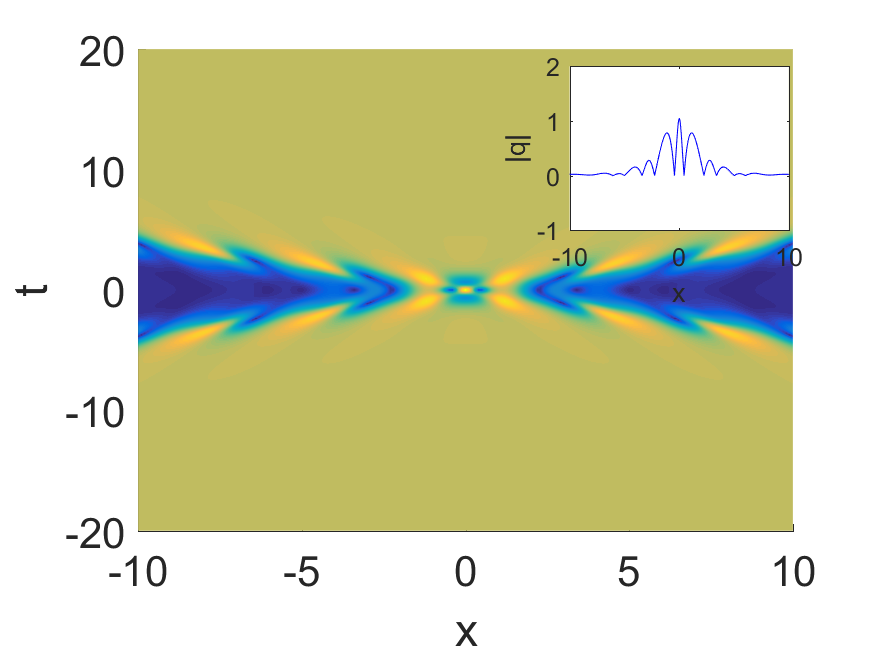}
        \caption{}
    \end{subfigure}
    \hfill
    \begin{subfigure}[b]{0.32\textwidth}
        \centering
        \includegraphics[height=3.2cm]{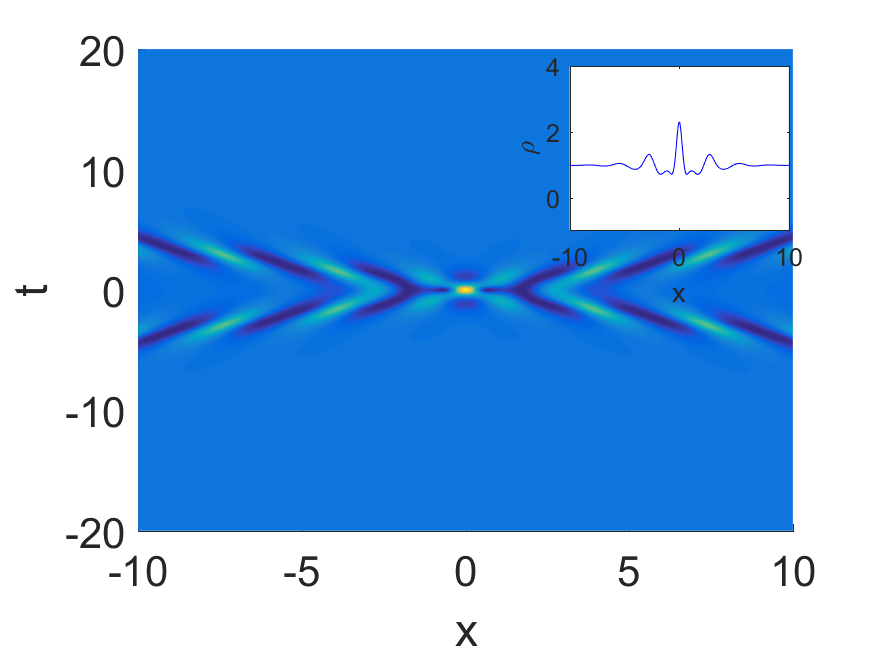}
        \caption{}
    \end{subfigure}
    \caption{Evolution from a symmetric breather pattern to a soliton-like localized pattern, with $\alpha_1=-\alpha_2=2$, $\theta_1=\theta_2=0$, and $\gamma_1=\gamma_2=0$. (a--c) $R_1=R_2=1.1$; (d--f) $R_1=R_2=2$.}
\end{figure}

When $R_1=R_2$ and $\alpha_1=-\alpha_2$, the $(0,2)$-order Type-I solution is symmetric in the $(x,t)$ plane. Figures~7(a)--7(c) show a symmetric breather pattern concentrated near the central overlap region. When the common radius is increased while the angular relation is preserved, the periodic modulation weakens and the localized ridges become narrower, as shown in Figures~7(d)--7(f).

\begin{figure}[!tbh]
    \begin{subfigure}[b]{0.32\textwidth}
        \centering
        \includegraphics[width=\textwidth, height=3.2cm]{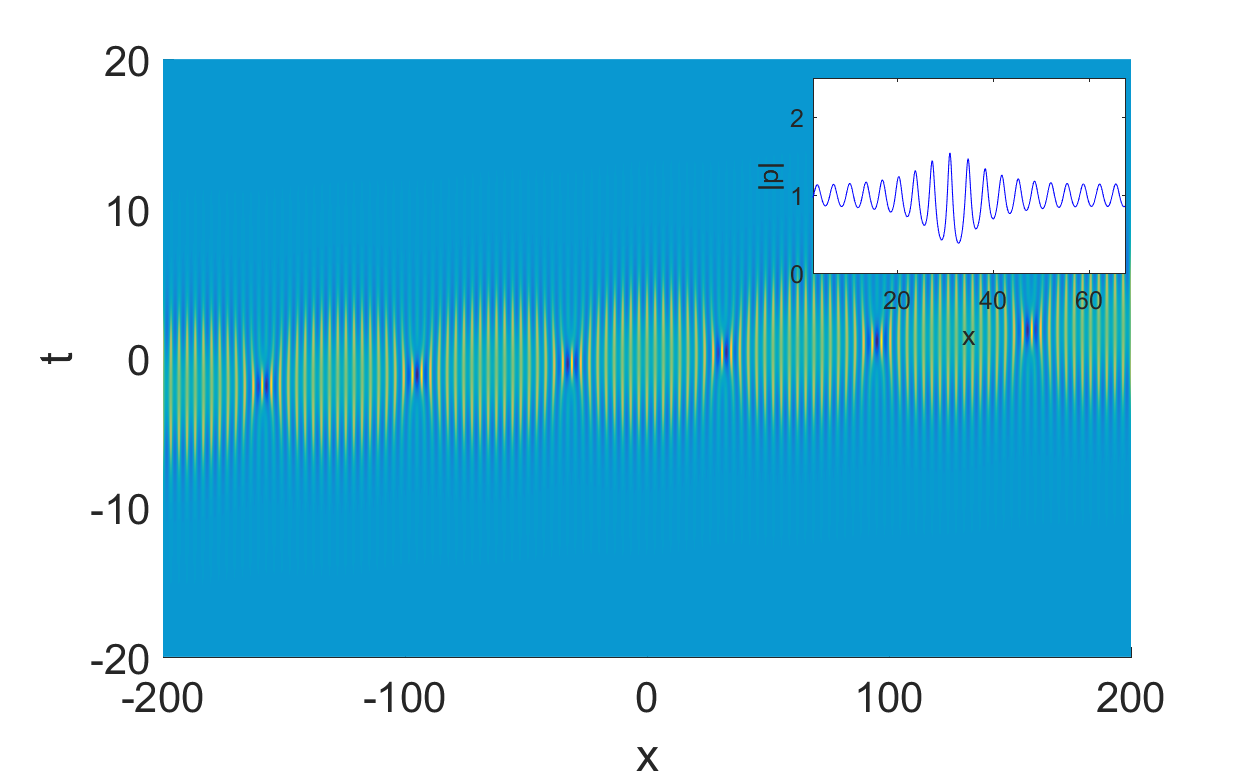}
        \caption{}
    \end{subfigure}
    \hfill
    \begin{subfigure}[b]{0.32\textwidth}
        \centering
        \includegraphics[width=\textwidth, height=3.2cm]{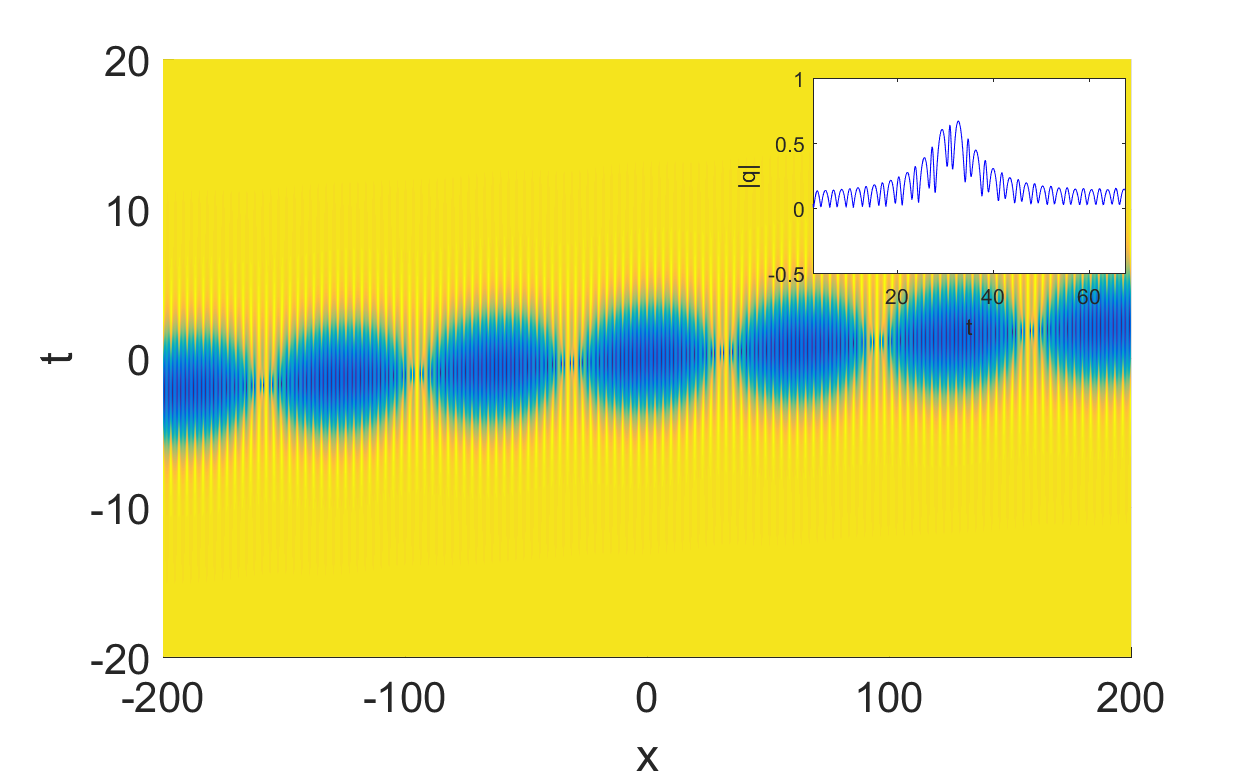}
        \caption{}
    \end{subfigure}
    \hfill
    \begin{subfigure}[b]{0.32\textwidth}
        \centering
        \includegraphics[width=\textwidth, height=3.2cm]{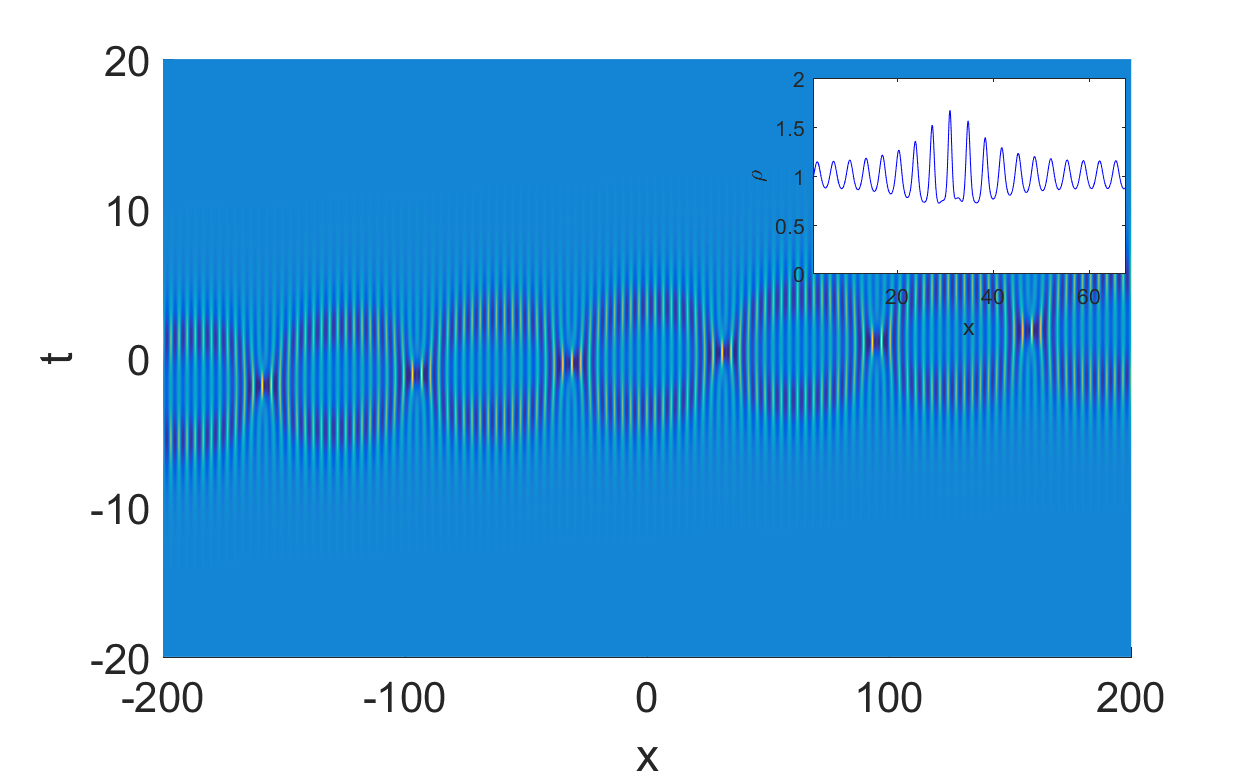}
        \caption{}
    \end{subfigure}

    \vspace{6pt}

    \begin{subfigure}[b]{0.32\textwidth}
        \centering
        \includegraphics[width=\textwidth, height=3.2cm]{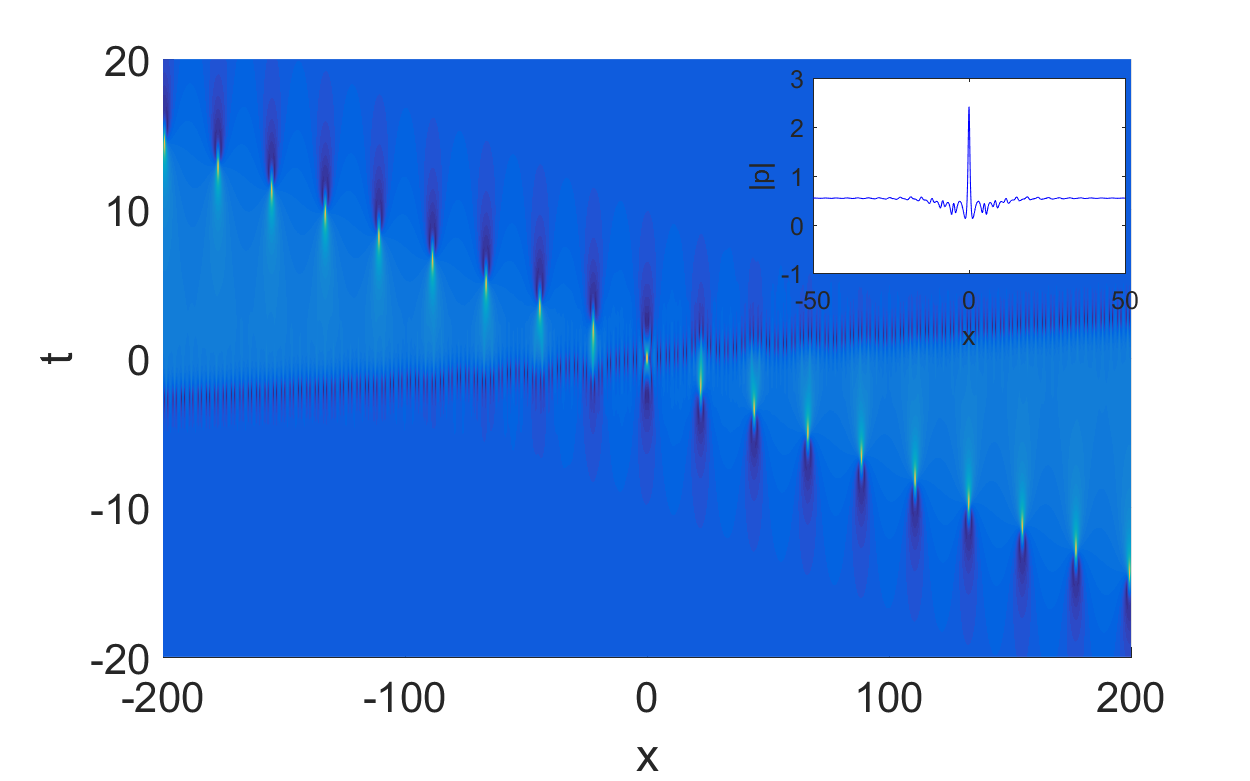}
        \caption{}
    \end{subfigure}
    \hfill
    \begin{subfigure}[b]{0.32\textwidth}
        \centering
        \includegraphics[width=\textwidth, height=3.2cm]{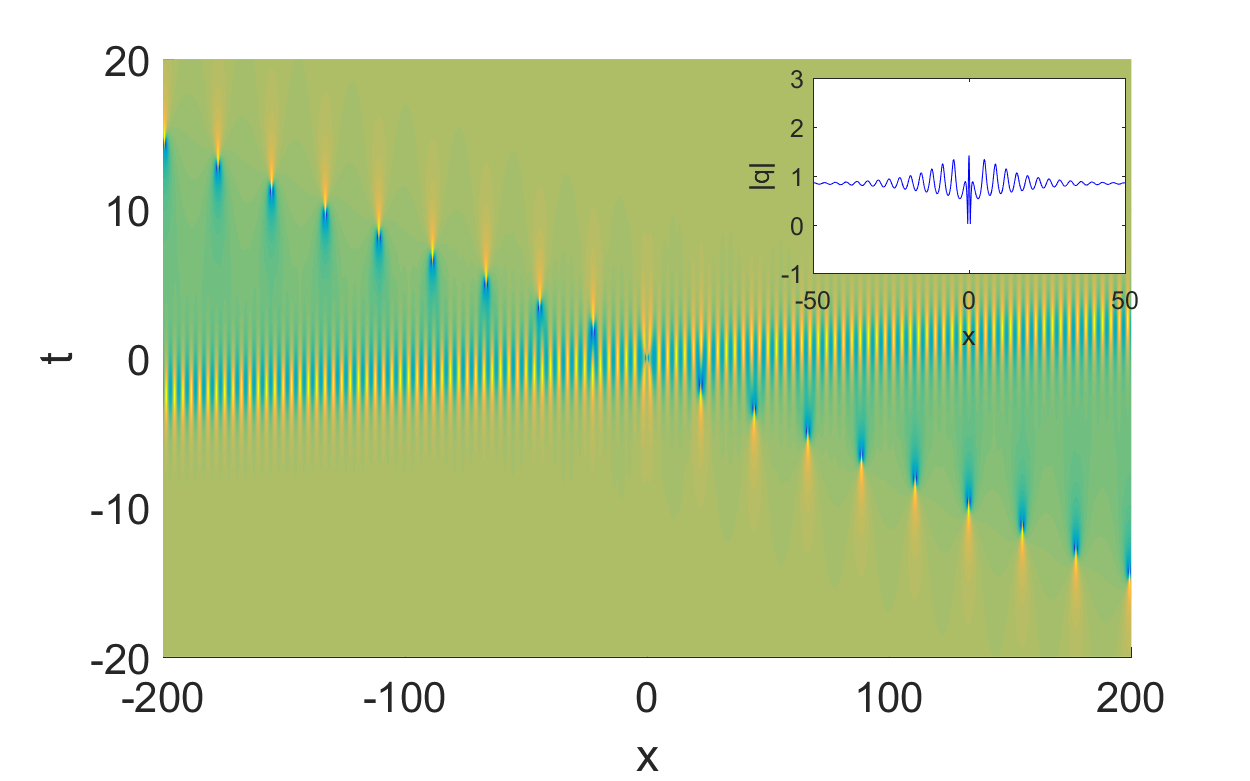}
        \caption{}
    \end{subfigure}
    \hfill
    \begin{subfigure}[b]{0.32\textwidth}
        \centering
        \includegraphics[width=\textwidth, height=3.2cm]{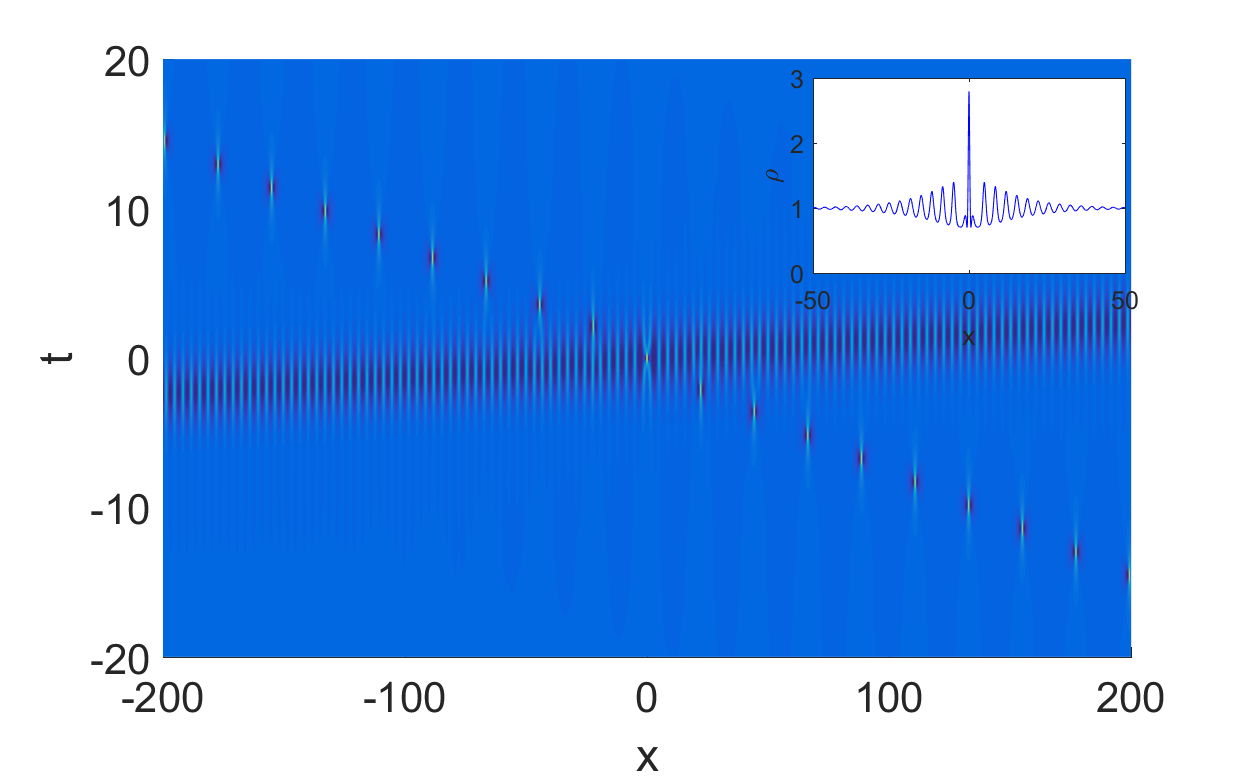}
        \caption{}
    \end{subfigure}
    \caption{Asymmetric two-breather configurations with $R_1=R_2=1.01$, $\theta_1=\theta_2=0$, and $\gamma_1=\gamma_2=0$. (a--c) $\alpha_1=5$ and $\alpha_2=-2$; (d--f) $\alpha_1=3$ and $\alpha_2=-2$.}
\end{figure}

When $R_1=R_2$ but $\alpha_1\neq-\alpha_2$, the mirror symmetry of the spectral pair is broken. Figure~8 shows that the localized ridges become tilted and their spacing changes. Reducing the angular mismatch modifies the ridge inclination and the relative peak positions. Thus, angular asymmetry alone can generate asymmetric and walk-off-like patterns even when the two spectral radii are equal.
\begin{figure}[!tbh]
    \centering
    \begin{subfigure}[b]{0.32\textwidth}
        \centering
        \includegraphics[height=3.2cm]{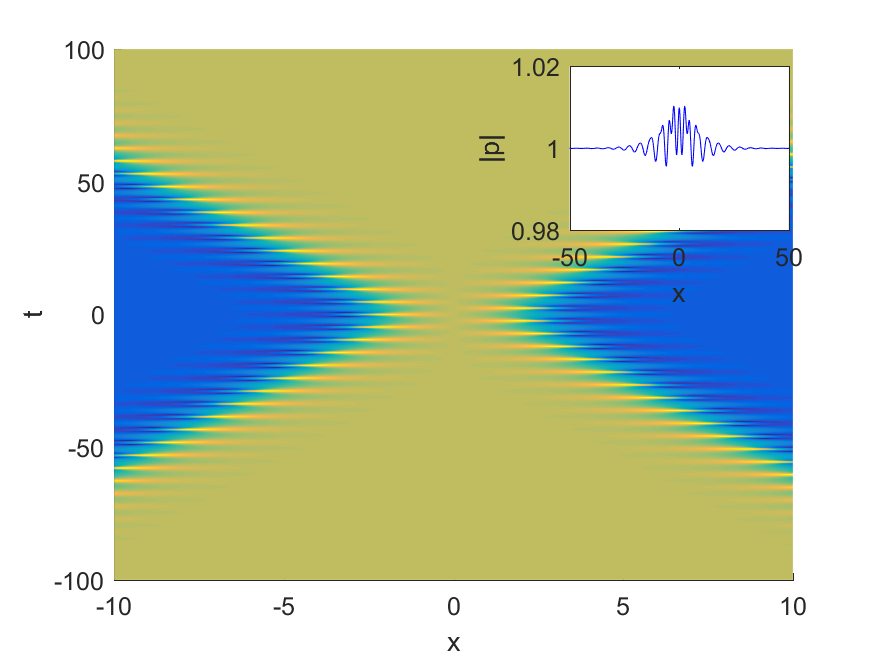}
        \caption{}
    \end{subfigure}
    \hfill
    \begin{subfigure}[b]{0.32\textwidth}
        \centering
        \includegraphics[height=3.2cm]{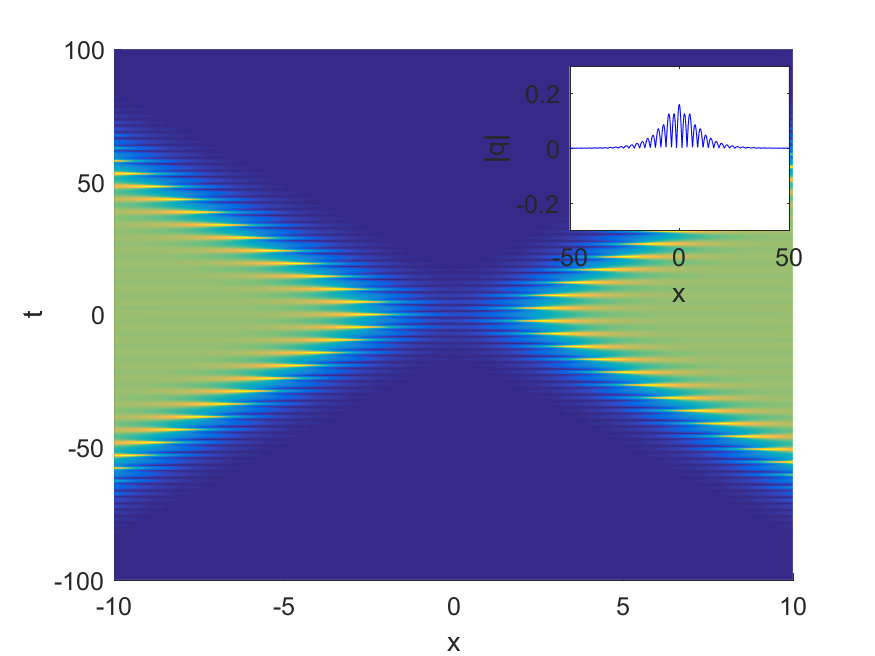}
        \caption{}
    \end{subfigure}
    \hfill
    \begin{subfigure}[b]{0.32\textwidth}
        \centering
        \includegraphics[height=3.2cm]{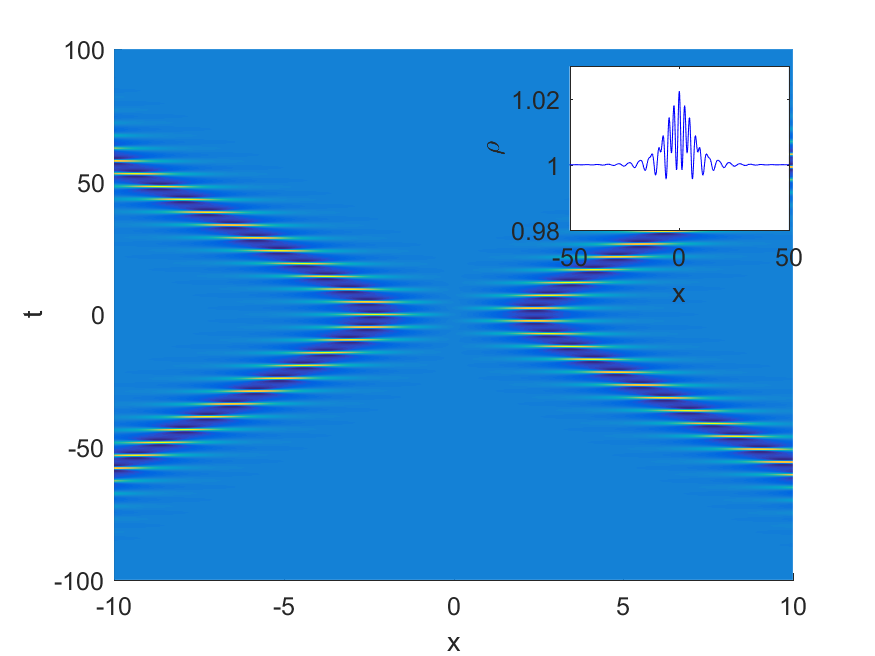}
        \caption{}
    \end{subfigure}

    \vspace{6pt}

    \begin{subfigure}[b]{0.32\textwidth}
        \centering
        \includegraphics[height=3.2cm]{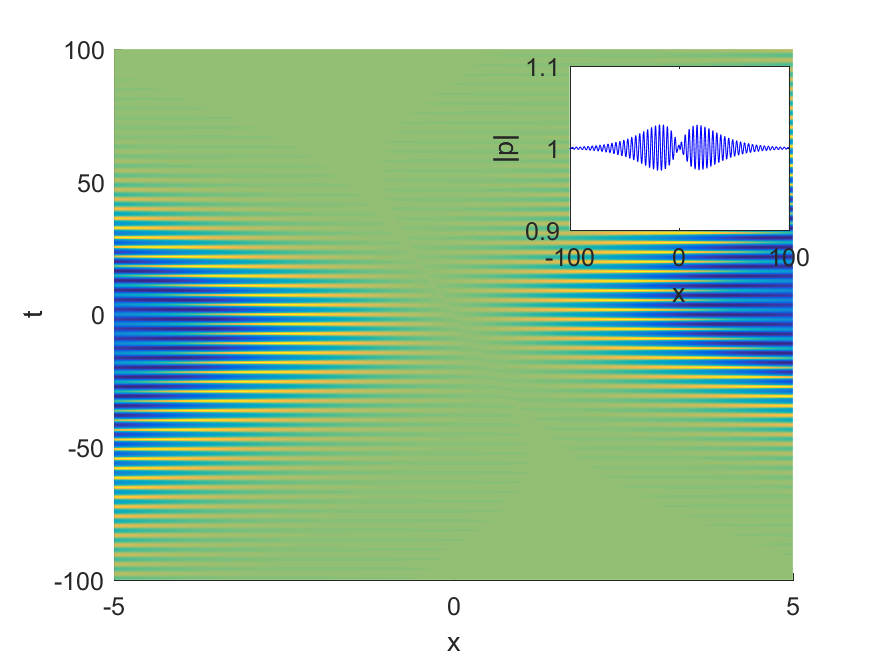}
        \caption{}
    \end{subfigure}
    \hfill
    \begin{subfigure}[b]{0.32\textwidth}
        \centering
        \includegraphics[height=3.2cm]{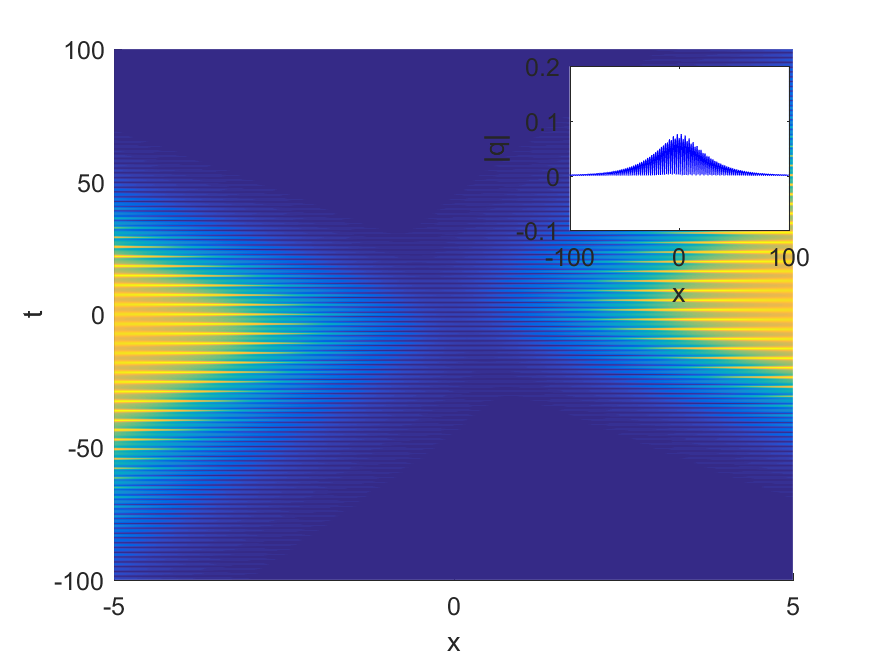}
        \caption{}
    \end{subfigure}
    \hfill
    \begin{subfigure}[b]{0.32\textwidth}
        \centering
        \includegraphics[height=3.2cm]{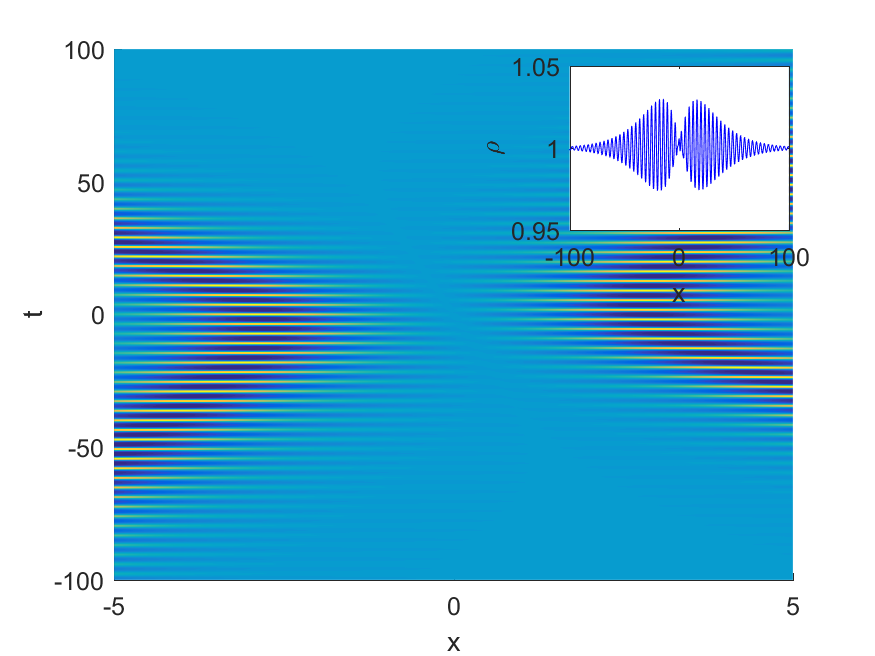}
        \caption{}
    \end{subfigure}

    \vspace{6pt}
    \begin{subfigure}[b]{0.32\textwidth}
        \centering
        \includegraphics[height=3.2cm]{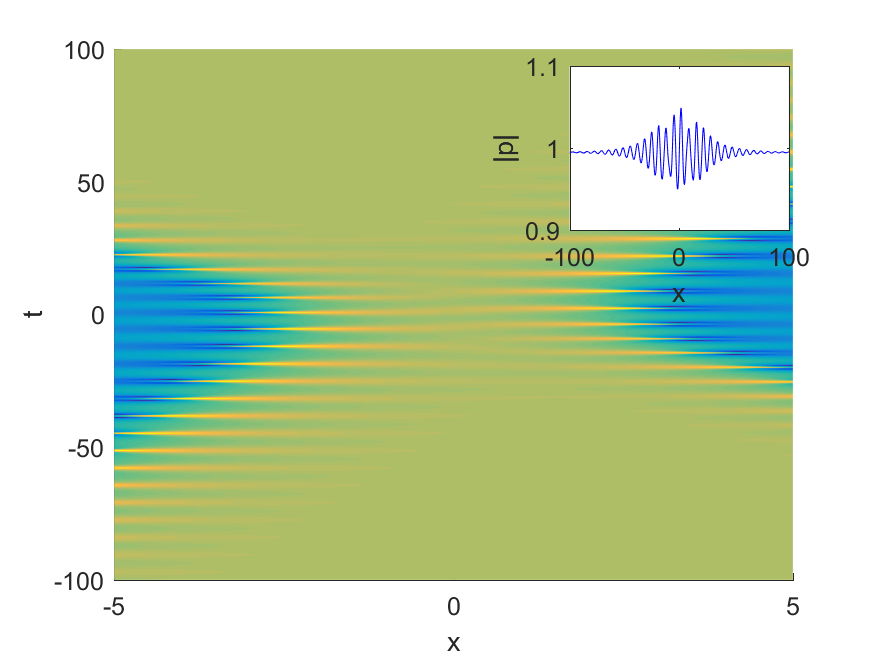}
        \caption{}
    \end{subfigure}
    \hfill
    \begin{subfigure}[b]{0.32\textwidth}
        \centering
        \includegraphics[height=3.2cm]{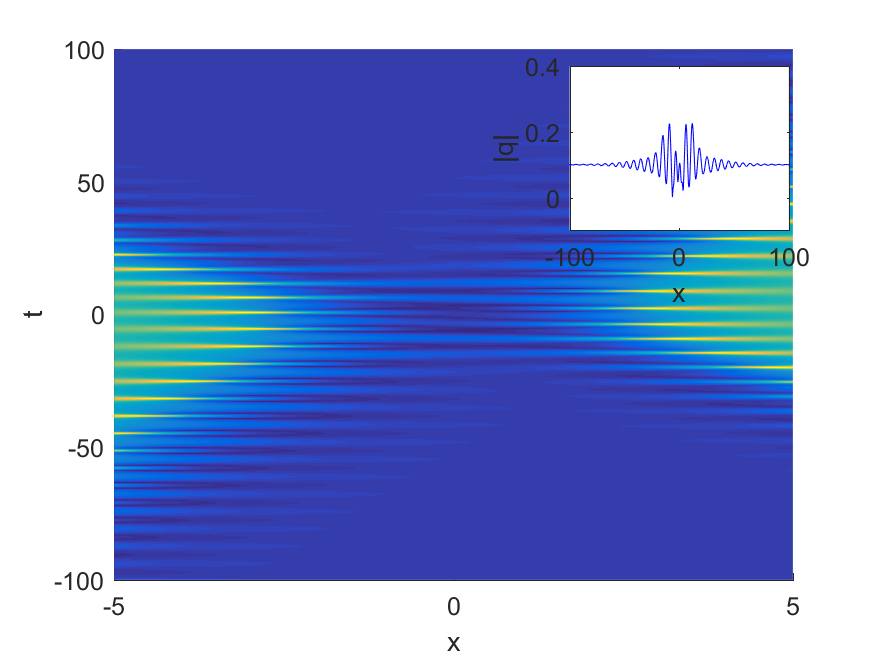}
        \caption{}
    \end{subfigure}
    \hfill
    \begin{subfigure}[b]{0.32\textwidth}
        \centering
        \includegraphics[height=3.2cm]{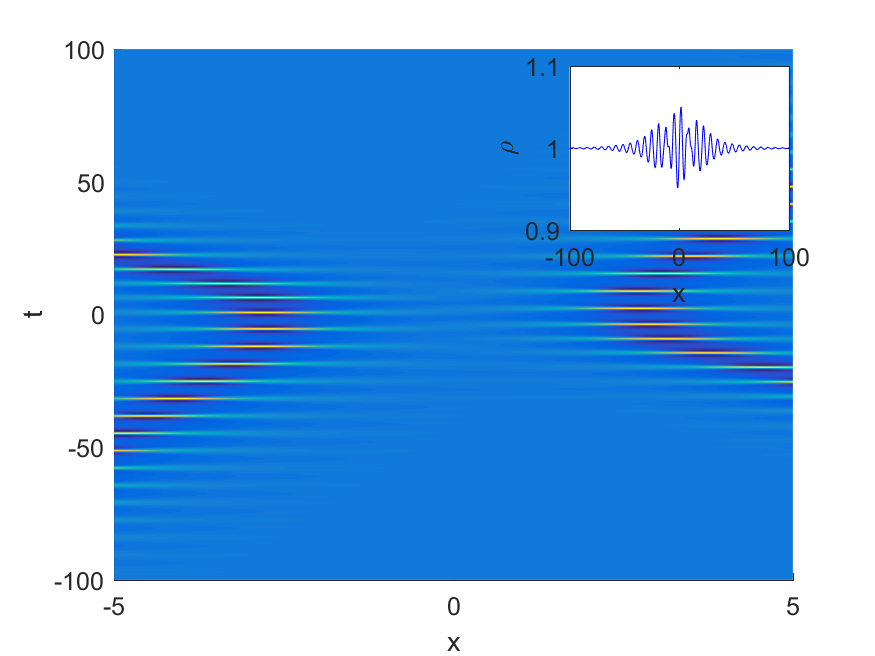}
        \caption{}
    \end{subfigure}
    \caption{Evolution of small perturbations on a finite background for the $(0,2)$-order Type-I configuration, with $\gamma_1=\gamma_2=0$. (a--c) $R_1=R_2=1.1$, $\alpha_1=-\alpha_2=0.7$, and $\theta_1=\theta_2=1.5$; (d--f) $R_1=1.1$, $R_2=1.05$, $\alpha_1=-\alpha_2=1.05$, and $\theta_1=\theta_2=1.6$; (g--i) $R_1=1.08$, $R_2=1.03$, $\alpha_1=0.6$, $\alpha_2=-0.5$, and $\theta_1=\theta_2=1.5$.}
\end{figure}

Figure~9 shows the evolution of small localized perturbations on a finite background. Under the symmetric conditions $R_1=R_2$ and $\alpha_1=-\alpha_2$, the perturbation develops into two branches that separate approximately symmetrically, as shown in Figures~9(a)--9(c). This behavior is reminiscent of the superregular-breather scenario, in which a small localized perturbation generates a pair of quasi-Akhmediev constituents propagating in opposite directions \cite{ZakharovGelash2013,GelashZakharov2014,KiblerEtAl2015}. Here, however, this connection is used only as a structural comparison, because the present fields arise from the normal-mode reconstruction of the CNLS-p system. When the radii differ but the angular relation is retained, the two branches have different localization scales and the splitting becomes unbalanced (Figures~9(d)--9(f)). When both the radii and angular parameters are mismatched, the peak distribution becomes more asymmetric and the propagation paths become more strongly tilted (Figures~9(g)--9(i)).

These examples show that the $(0,2)$-order Type-I configuration supports symmetric and asymmetric two-breather patterns, near-periodic structures, soliton-like localized ridges, near-rational patterns, and small-perturbation splitting. Their morphology is governed by the relative positions of the two distinct spectral points.

\subsection{The $(1,2)$-order configuration}

For $M=1$ and $N=2$, equation \eqref{qr7} gives
\begin{equation}\label{TypeI12}
\begin{aligned}
p_{\mathrm I}^{[1,2]}=\frac12e^{\mathrm{i}t}(2+H_2+H_1),\qquad
q_{\mathrm I}^{[1,2]}=\frac12e^{\mathrm{i}t}(H_2-H_1).
\end{aligned}
\end{equation}
Here, $H_2$ is associated with the spectral parameters $\lambda_1$ and $\lambda_2$, whereas $H_1$ is associated with $\lambda_3$. When the two spectral points of the two-fold block are selected on the same localized branch and the two-fold correction vanishes in the far field, the asymptotic limits are
\begin{equation*}
H_2\longrightarrow0,
\qquad
H_1\longrightarrow\frac{\lambda_3^*}{\lambda_3}-1,
\qquad x\to\pm\infty.
\end{equation*}
Therefore,
\begin{equation*}
\begin{aligned}
p_{\mathrm I}^{[1,2]}
\longrightarrow
\frac{1}{2}e^{\mathrm{i}t}
\left(1+\frac{\lambda_3^*}{\lambda_3}\right),\qquad
q_{\mathrm I}^{[1,2]}
\longrightarrow
\frac{1}{2}e^{\mathrm{i}t}
\left(1-\frac{\lambda_3^*}{\lambda_3}\right).
\end{aligned}
\label{eq:typeI12-background}
\end{equation*}
The corresponding background intensities satisfy
\begin{equation*}
\left|p_{\mathrm I,\mathrm{bg}}^{[1,2]}\right|^2
+
\left|q_{\mathrm I,\mathrm{bg}}^{[1,2]}\right|^2
=1,
\end{equation*}
and hence $\rho_{\mathrm I}^{[1,2]}\to1$. For mixed-branch spectral choices, such as $R_1>1$ and $R_2<1$, the far-field value of the two-fold correction must be evaluated separately. In such cases, the individual component backgrounds are not determined solely by $\lambda_3$.

In the KM regime $\alpha_3=0$, the $p_{\mathrm I}$ background vanishes, whereas $q_{\mathrm I}\to e^{\mathrm{i}t}$. For general off-axis spectral parameters, both physical components may retain nonzero backgrounds. Because the first-order and two-fold building blocks contain different numbers of spectral points, their localized deviations from the background remain sensitive to spectral mismatch. We next present three representative parameter regimes.

\begin{figure}[!tbh]
    \centering
    \begin{subfigure}[b]{0.32\textwidth}
        \centering
        \includegraphics[width=\textwidth, height=3.2cm]{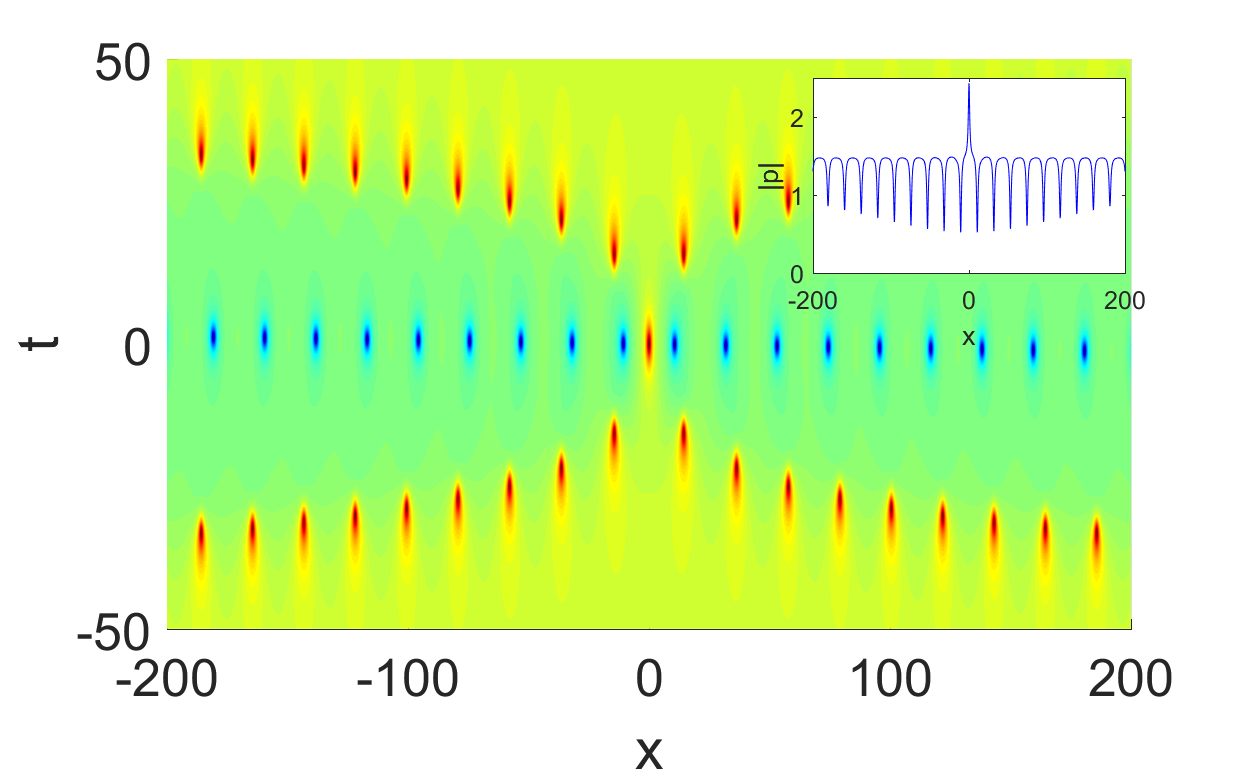}
        \caption{}
    \end{subfigure}
    \hfill
    \begin{subfigure}[b]{0.32\textwidth}
        \centering
        \includegraphics[width=\textwidth, height=3.2cm]{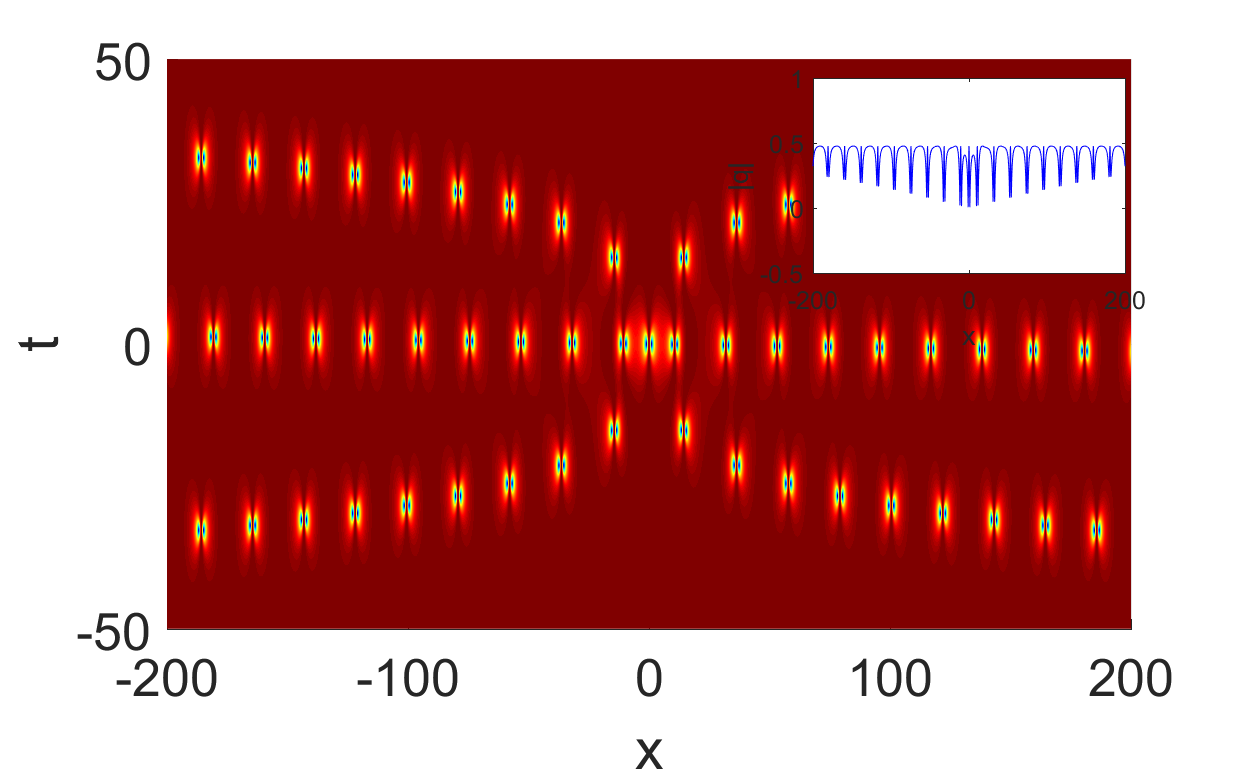}
        \caption{}
    \end{subfigure}
    \hfill
    \begin{subfigure}[b]{0.32\textwidth}
        \centering
        \includegraphics[width=\textwidth, height=3.2cm]{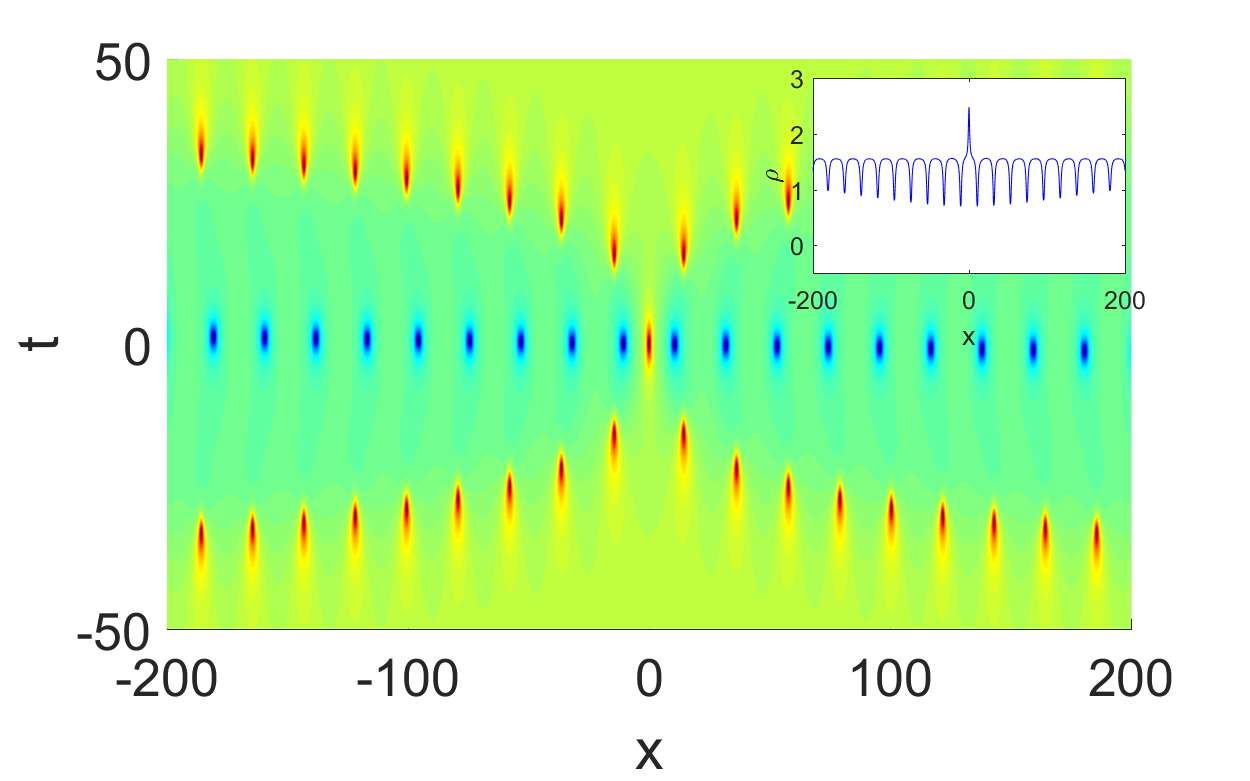}
        \caption{}
    \end{subfigure}

    \vspace{6pt}

    \begin{subfigure}[b]{0.32\textwidth}
        \centering
        \includegraphics[width=\textwidth, height=3.2cm]{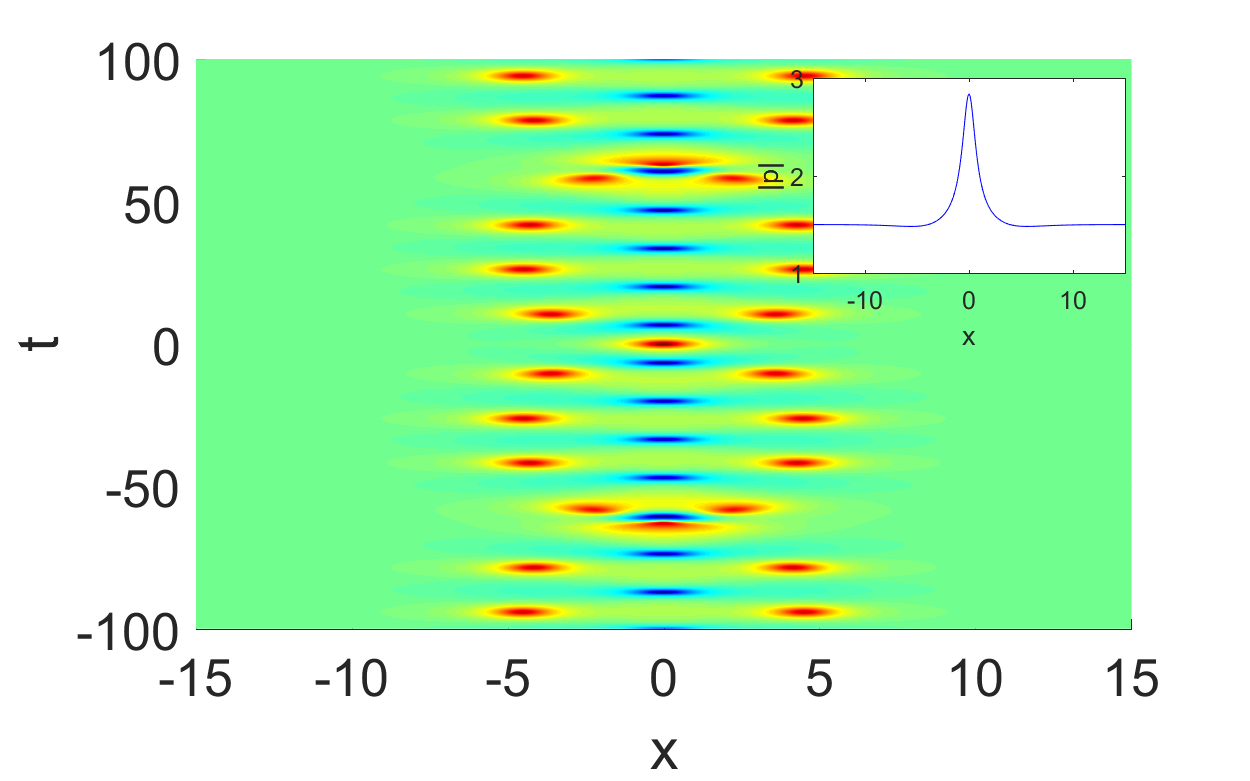}
        \caption{}
    \end{subfigure}
    \hfill
    \begin{subfigure}[b]{0.32\textwidth}
        \centering
        \includegraphics[width=\textwidth, height=3.2cm]{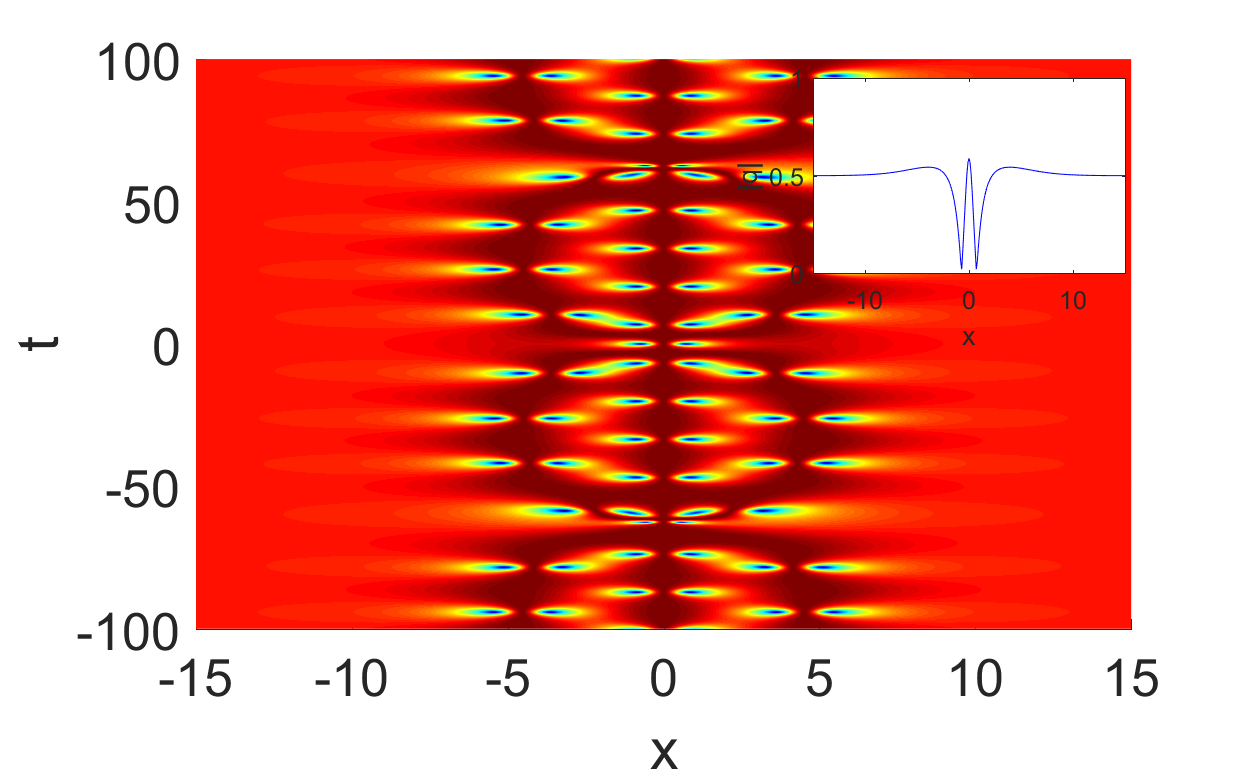}
        \caption{}
    \end{subfigure}
    \hfill
    \begin{subfigure}[b]{0.32\textwidth}
        \centering
        \includegraphics[width=\textwidth, height=3.2cm]{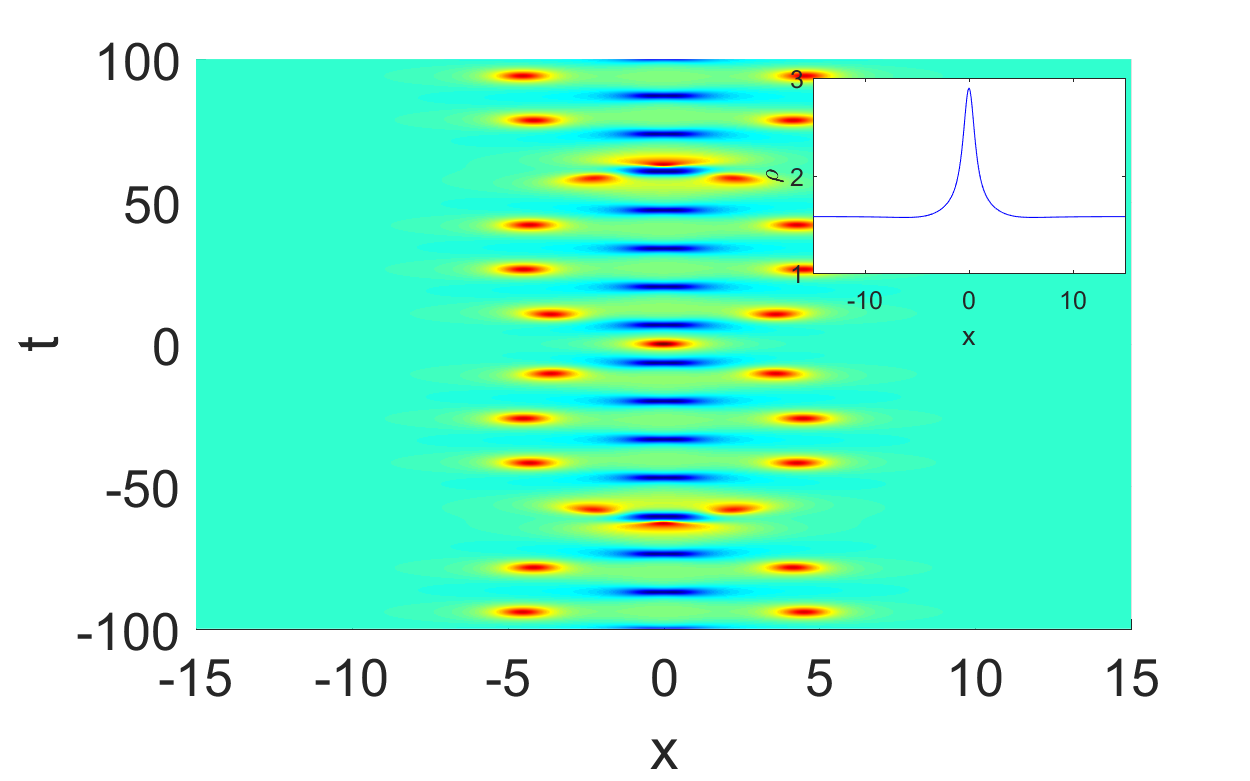}
        \caption{}
    \end{subfigure}

    \vspace{6pt}

    \begin{subfigure}[b]{0.32\textwidth}
        \centering
        \includegraphics[width=\textwidth, height=3.2cm]{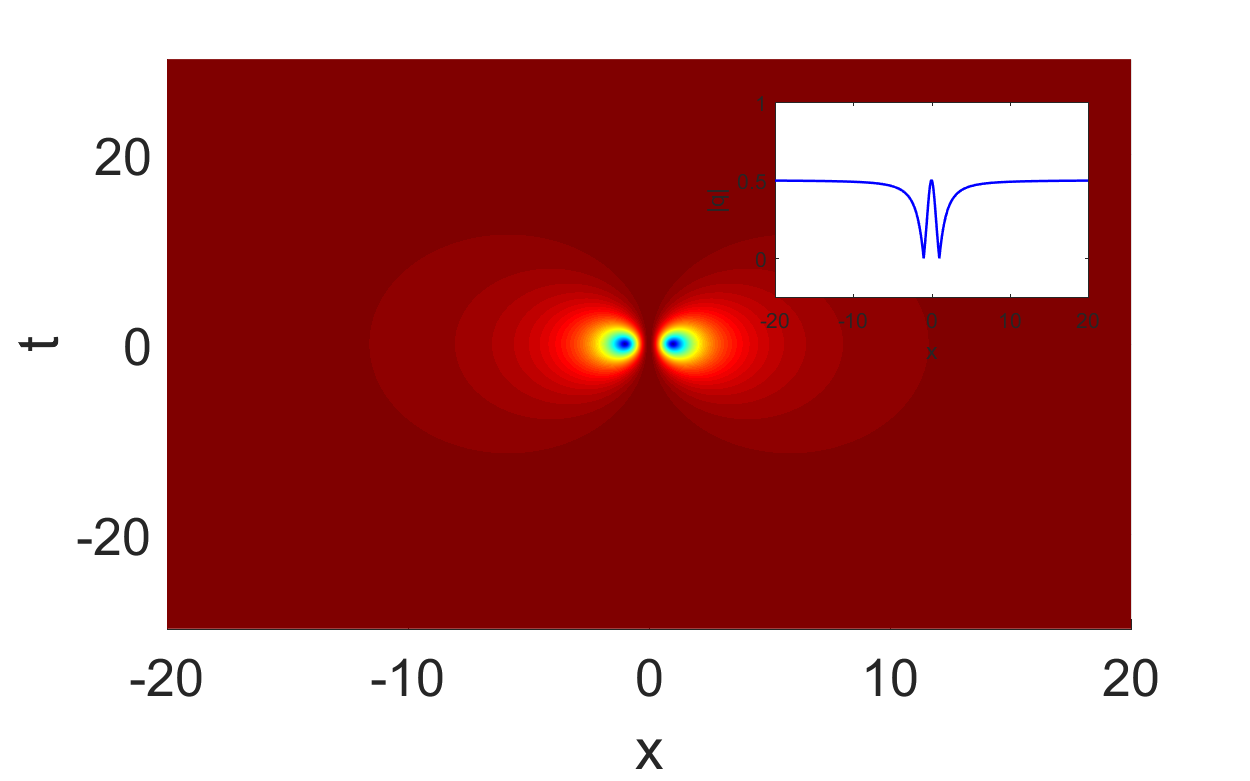}
        \caption{}
    \end{subfigure}
    \hfill
    \begin{subfigure}[b]{0.32\textwidth}
        \centering
        \includegraphics[width=\textwidth, height=3.2cm]{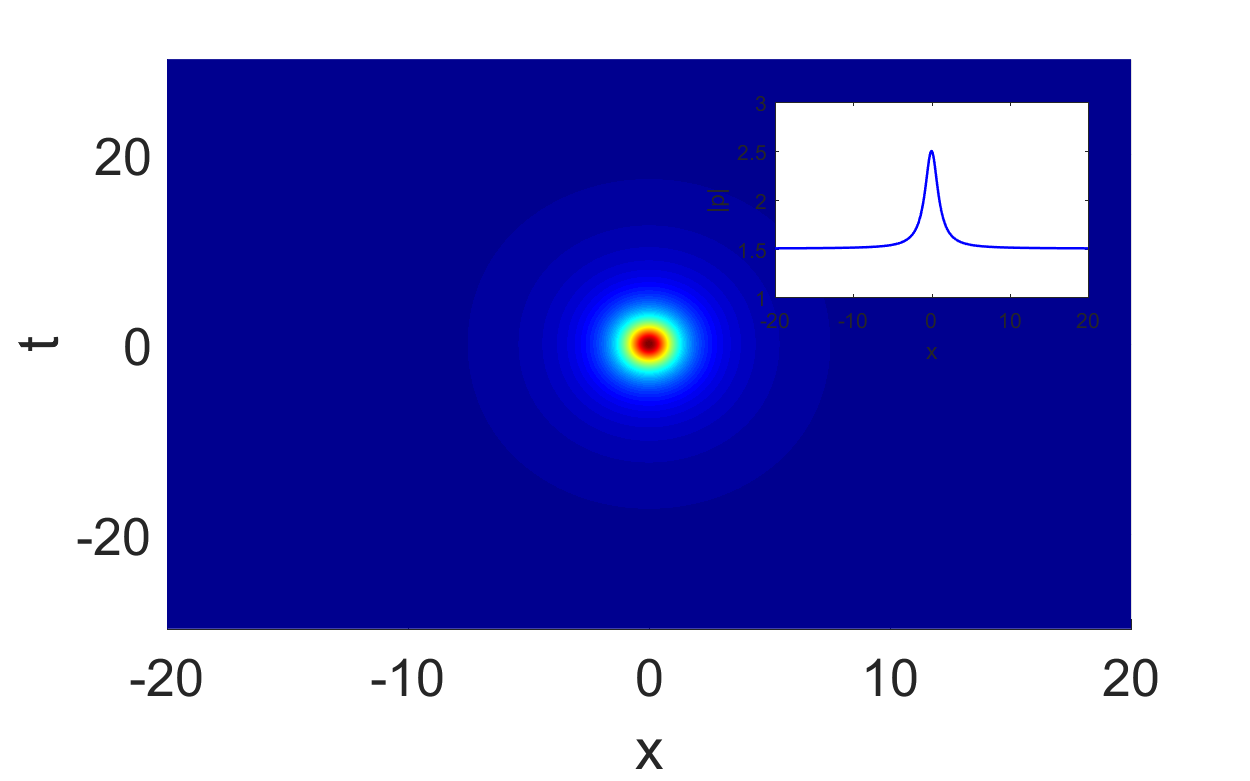}
        \caption{}
    \end{subfigure}
    \hfill
    \begin{subfigure}[b]{0.32\textwidth}
        \centering
        \includegraphics[width=\textwidth, height=3.2cm]{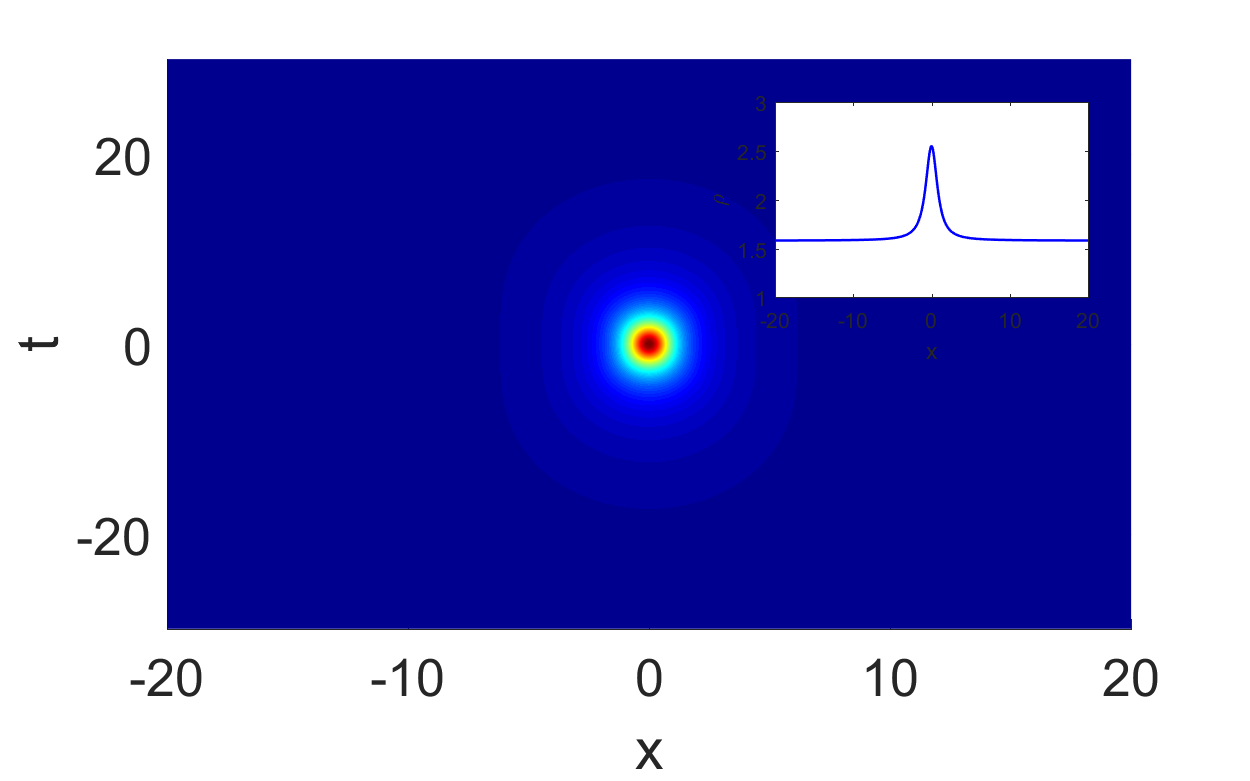}
        \caption{}
    \end{subfigure}
    \caption{Representative $(1,2)$-order Type-I configurations with $\gamma_1=\gamma_2=\gamma_3=0$ and $\theta_1=\theta_2=\theta_3=0$. (a--c) near-Akhmediev configuration with $R_1=R_2=R_3=1.001$, $\alpha_1=0.3$, $\alpha_2=-0.3$, and $\alpha_3=0.3$; (d--f) KM configuration with $R_1=2$, $R_2=1.8$, $R_3=2$, and $\alpha_1=\alpha_2=\alpha_3=0$; (g--i) near-rational configuration with $R_1=R_2=R_3=1.001$, $\alpha_1=0.001$, $\alpha_2=-0.001$, and $\alpha_3=0.001$.}
\end{figure}

\begin{figure}[!tbh]
    \centering
    \begin{subfigure}[b]{0.32\textwidth}
        \centering
        \includegraphics[width=\textwidth, height=3.2cm]{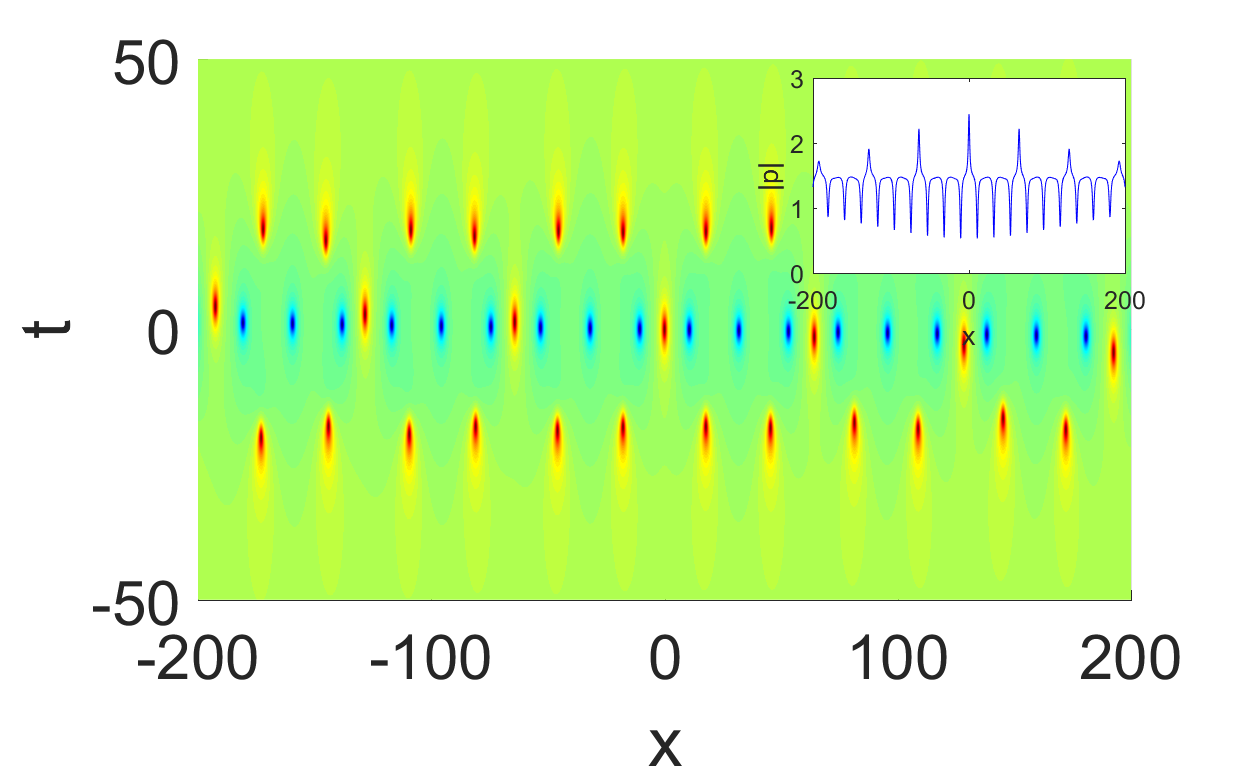}
        \caption{}
    \end{subfigure}
    \hfill
    \begin{subfigure}[b]{0.32\textwidth}
        \centering
        \includegraphics[width=\textwidth, height=3.2cm]{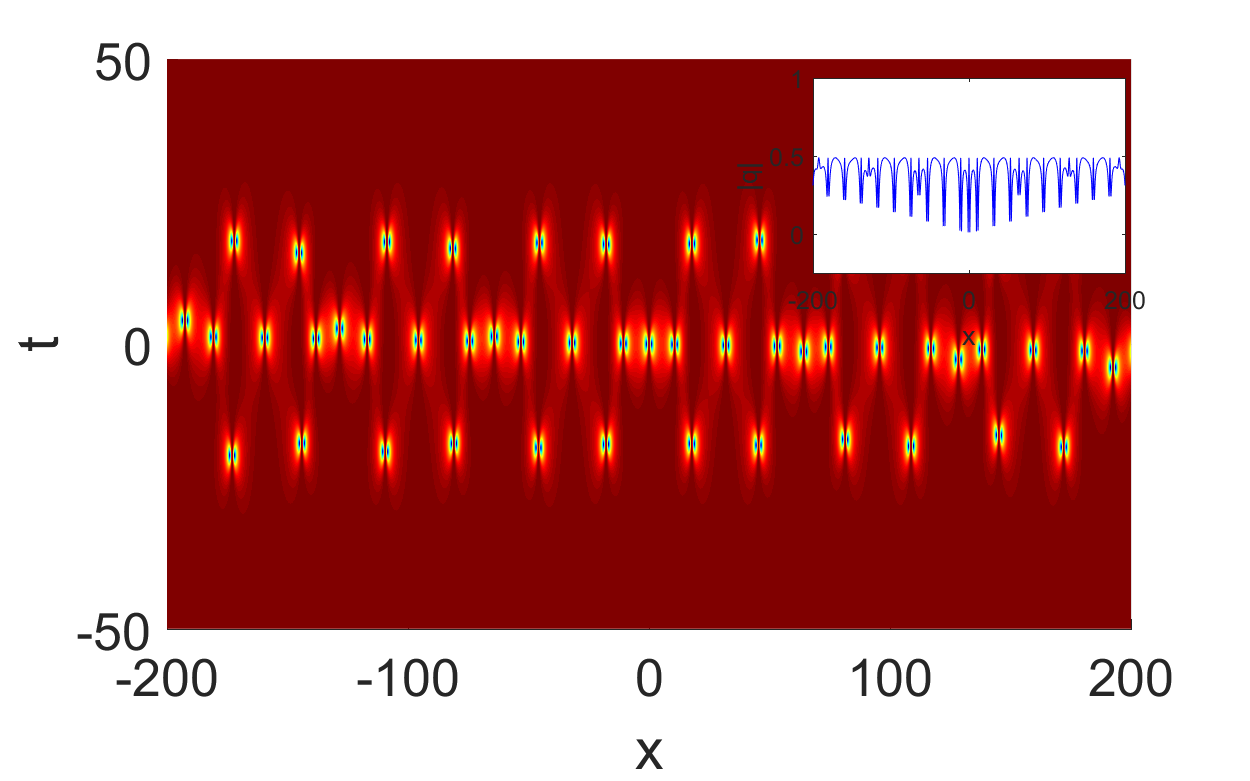}
        \caption{}
    \end{subfigure}
    \hfill
    \begin{subfigure}[b]{0.32\textwidth}
        \centering
        \includegraphics[width=\textwidth, height=3.2cm]{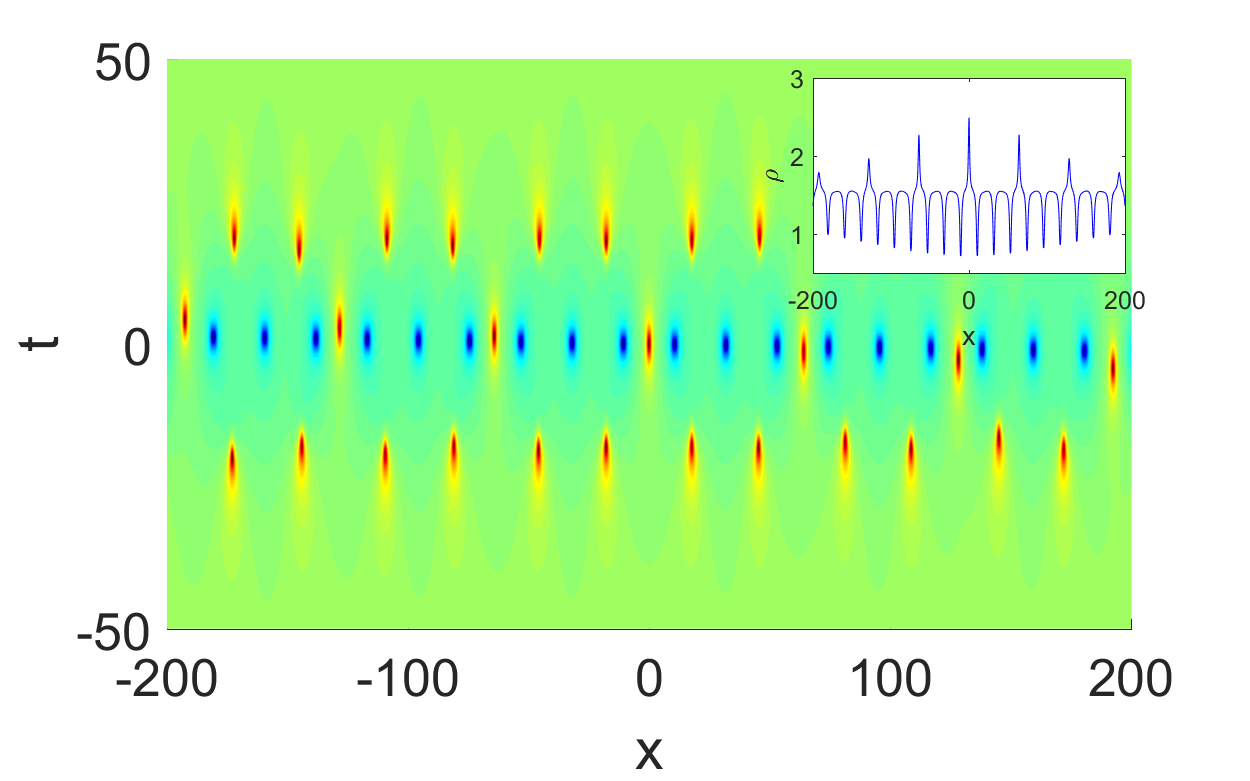}
        \caption{}
    \end{subfigure}

    \vspace{6pt}

    \begin{subfigure}[b]{0.32\textwidth}
        \centering
        \includegraphics[width=\textwidth, height=3.2cm]{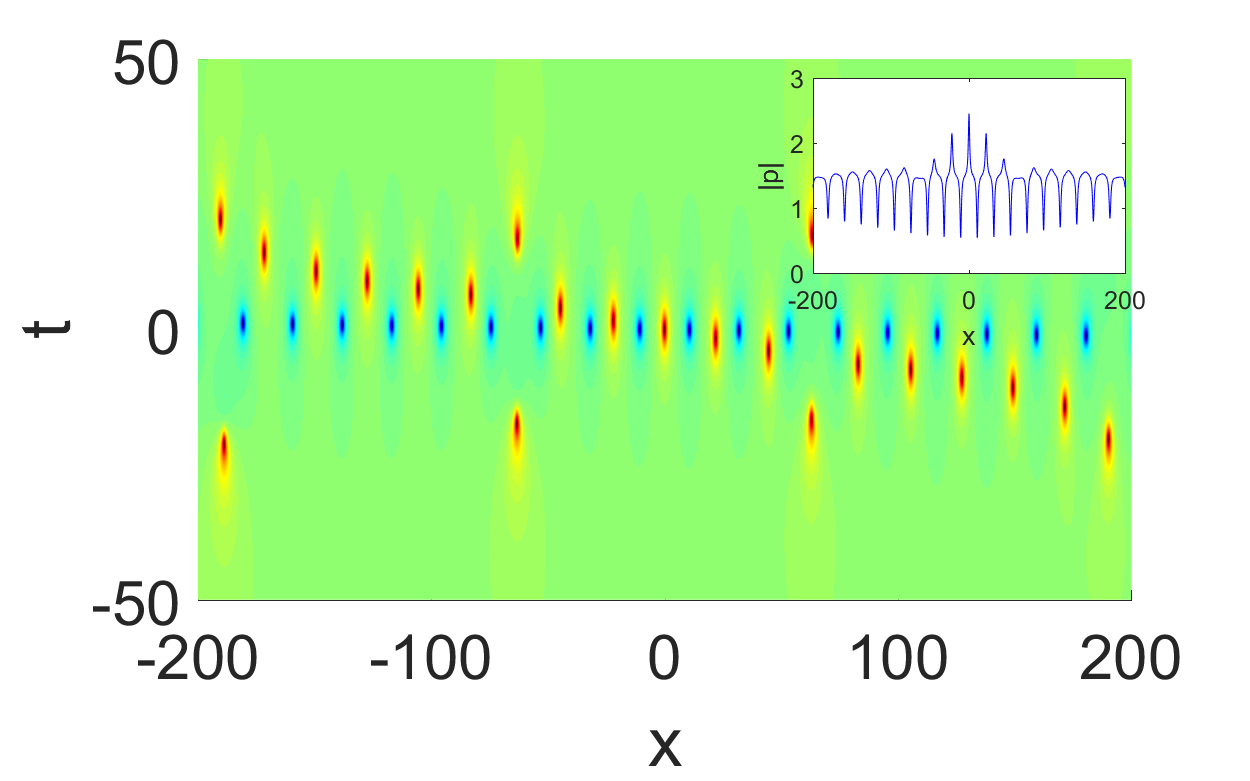}
        \caption{}
    \end{subfigure}
    \hfill
    \begin{subfigure}[b]{0.32\textwidth}
        \centering
        \includegraphics[width=\textwidth, height=3.2cm]{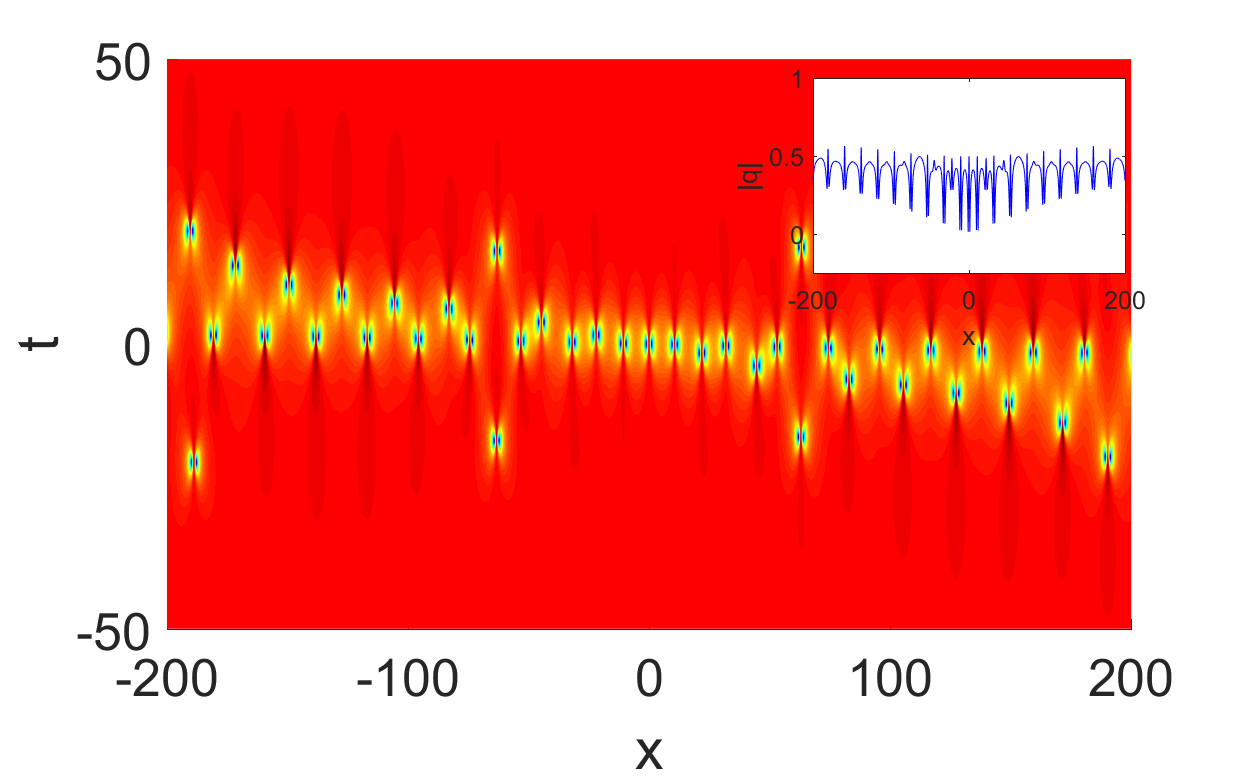}
        \caption{}
    \end{subfigure}
    \hfill
    \begin{subfigure}[b]{0.32\textwidth}
        \centering
        \includegraphics[width=\textwidth, height=3.2cm]{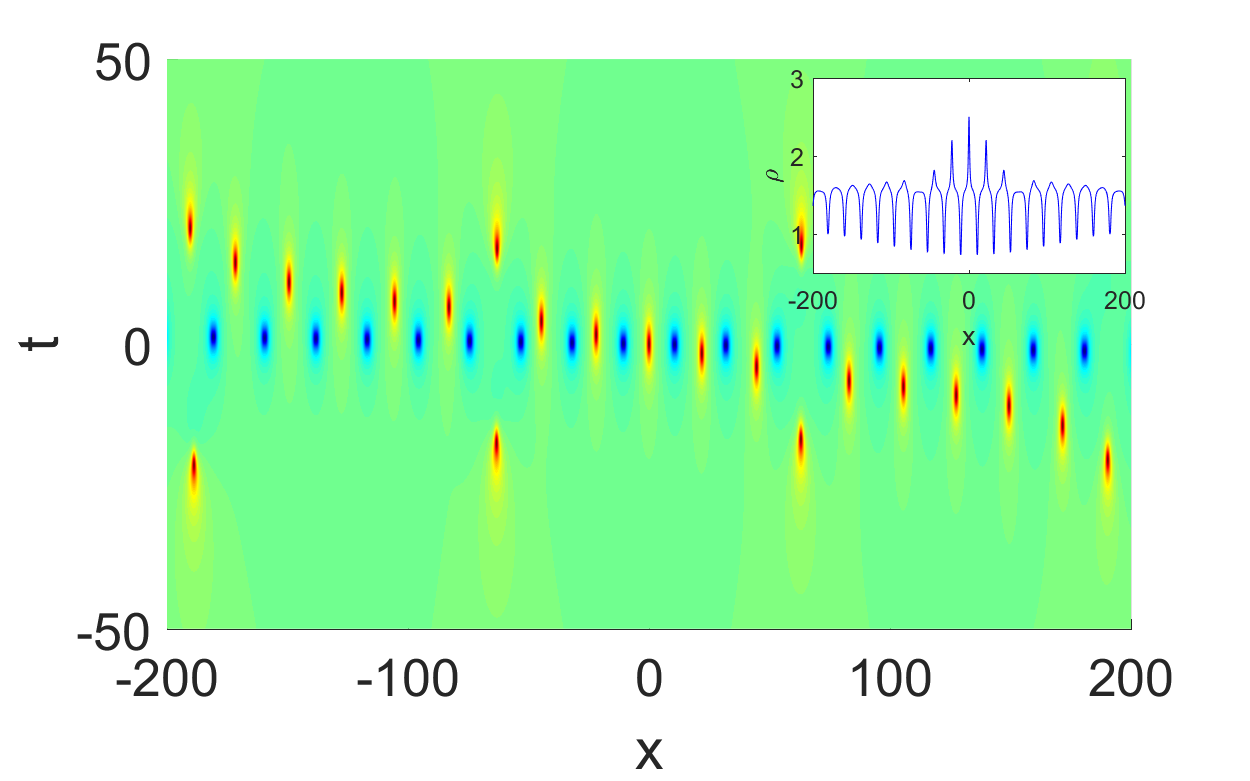}
        \caption{}
    \end{subfigure}

    \caption{Spatial beating in the near-Akhmediev $(1,2)$-order Type-I configuration, with $\gamma_1=\gamma_2=\gamma_3=0$ and $\theta_1=\theta_2=\theta_3=0$. (a--c) $R_1=R_2=R_3=1.001$, $\alpha_1=\alpha_3=0.3$, and $\alpha_2=-0.2$; (d--f) $R_1=R_3=1.001$, $R_2=1.01$, $\alpha_1=\alpha_2=0.3$, and $\alpha_3=-0.05$.}
\end{figure}

\begin{figure}[!tbh]
    \centering
    \begin{subfigure}[b]{0.32\textwidth}
        \centering
        \includegraphics[width=\textwidth, height=3.2cm]{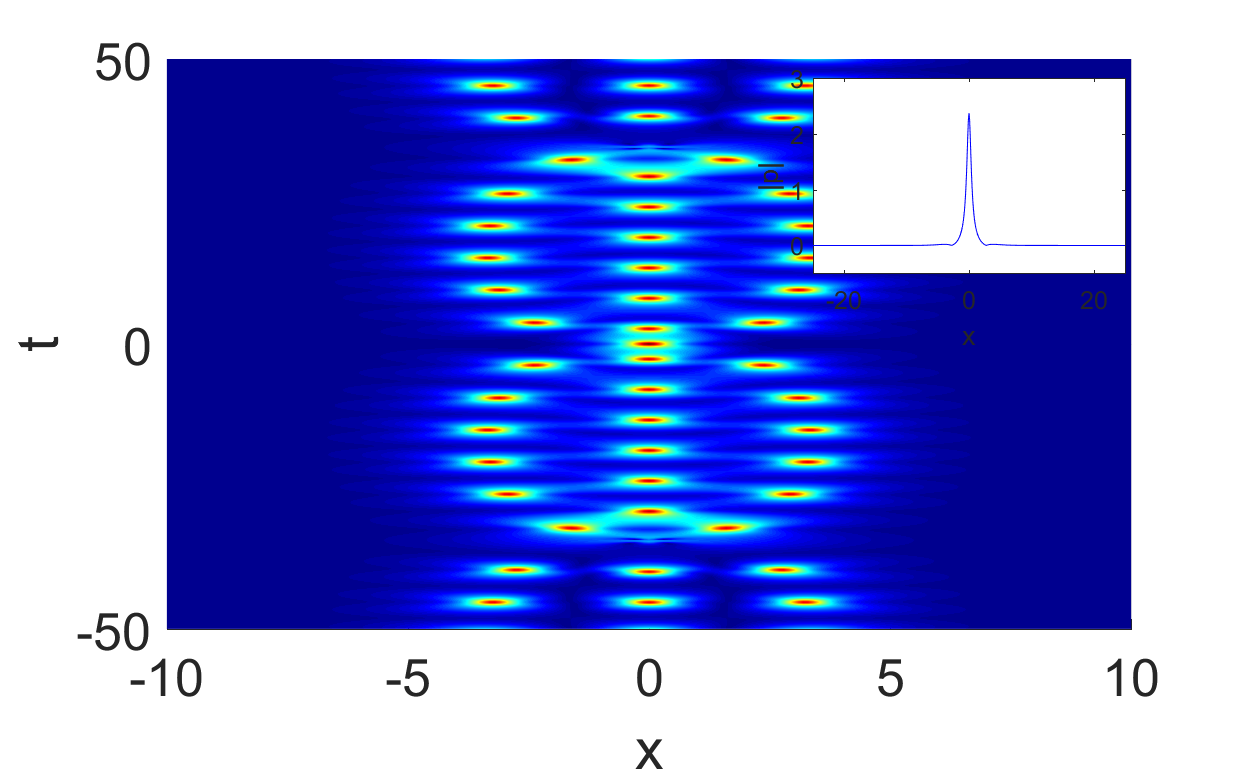}
        \caption{}
    \end{subfigure}
    \hfill
    \begin{subfigure}[b]{0.32\textwidth}
        \centering
        \includegraphics[width=\textwidth, height=3.2cm]{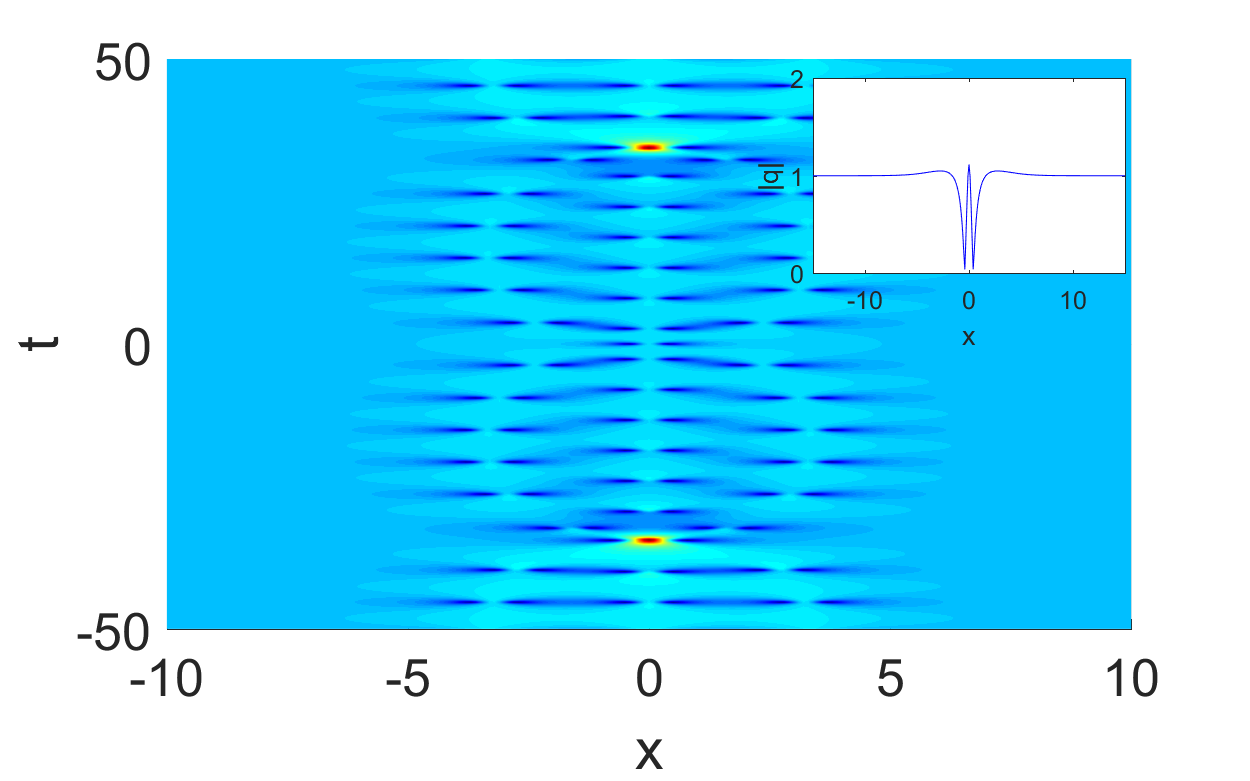}
        \caption{}
    \end{subfigure}
    \hfill
    \begin{subfigure}[b]{0.32\textwidth}
        \centering
        \includegraphics[width=\textwidth, height=3.2cm]{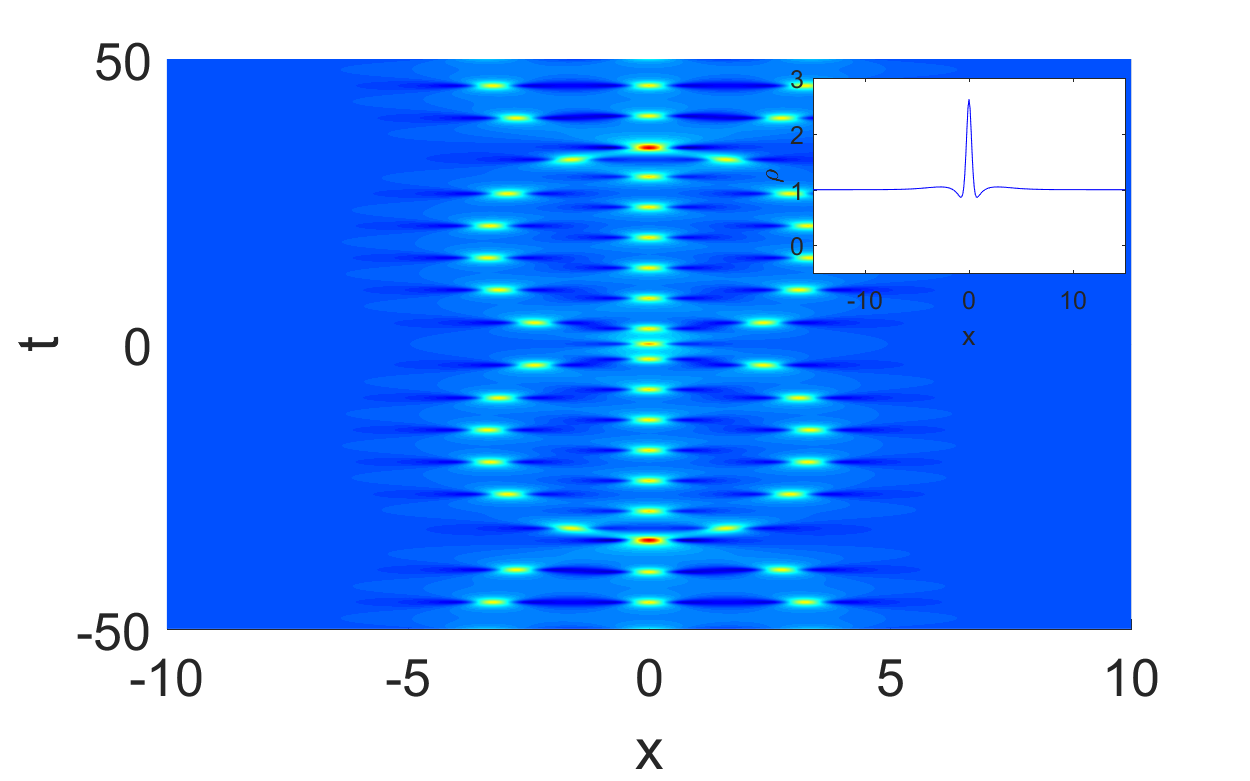}
        \caption{}
    \end{subfigure}

    \vspace{6pt}

    \begin{subfigure}[b]{0.32\textwidth}
        \centering
        \includegraphics[width=\textwidth, height=3.2cm]{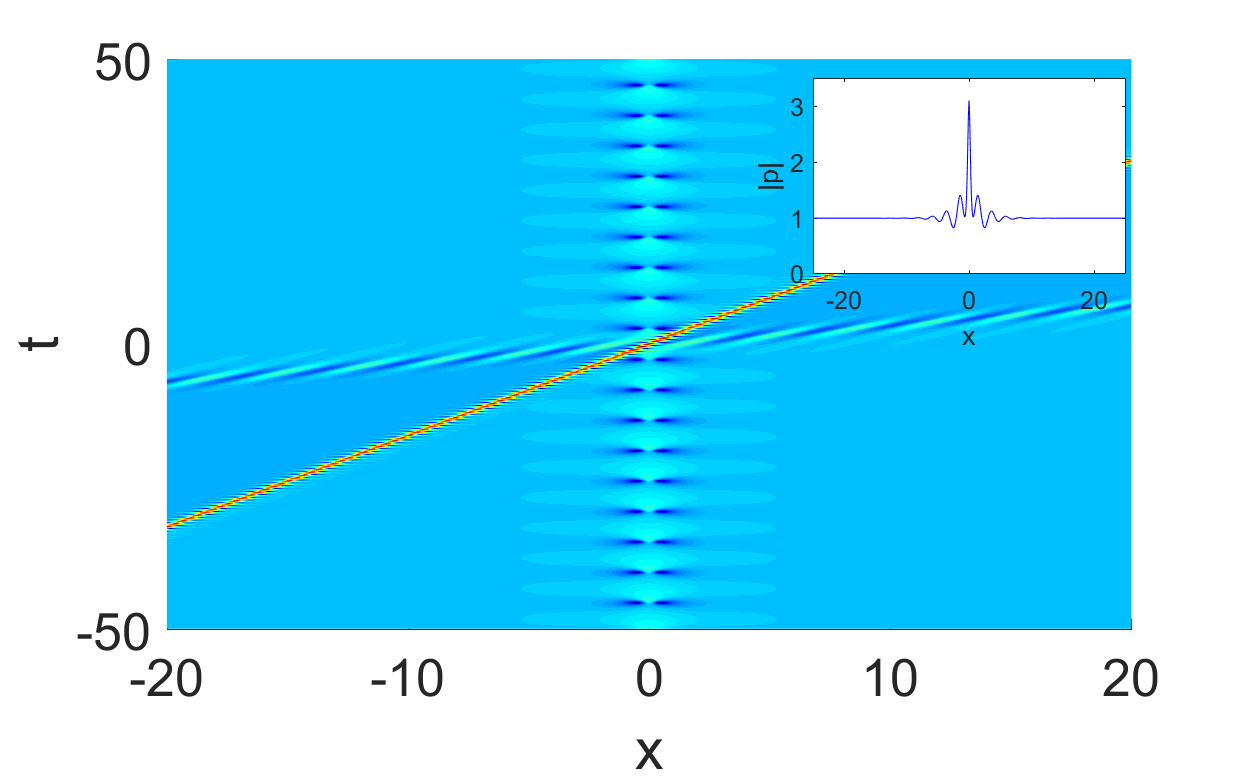}
        \caption{}
    \end{subfigure}
    \hfill
    \begin{subfigure}[b]{0.32\textwidth}
        \centering
        \includegraphics[width=\textwidth, height=3.2cm]{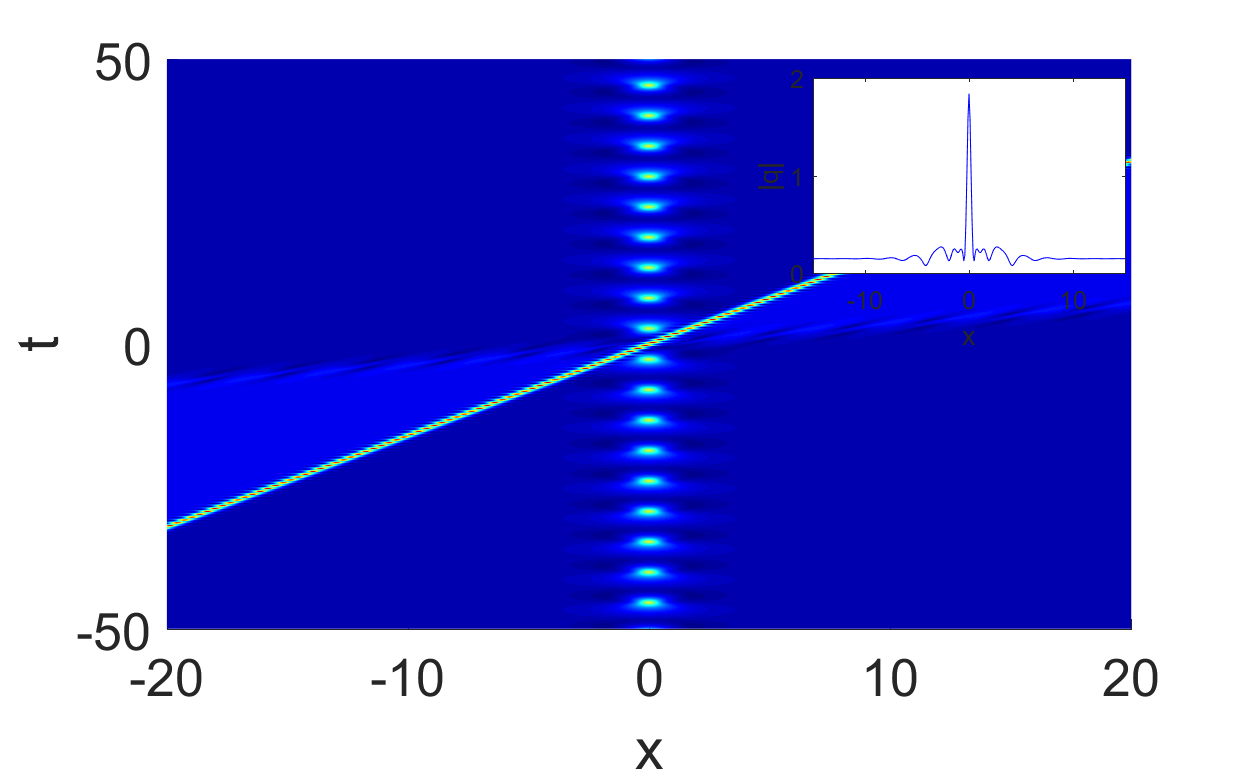}
        \caption{}
    \end{subfigure}
    \hfill
    \begin{subfigure}[b]{0.32\textwidth}
        \centering
        \includegraphics[width=\textwidth, height=3.2cm]{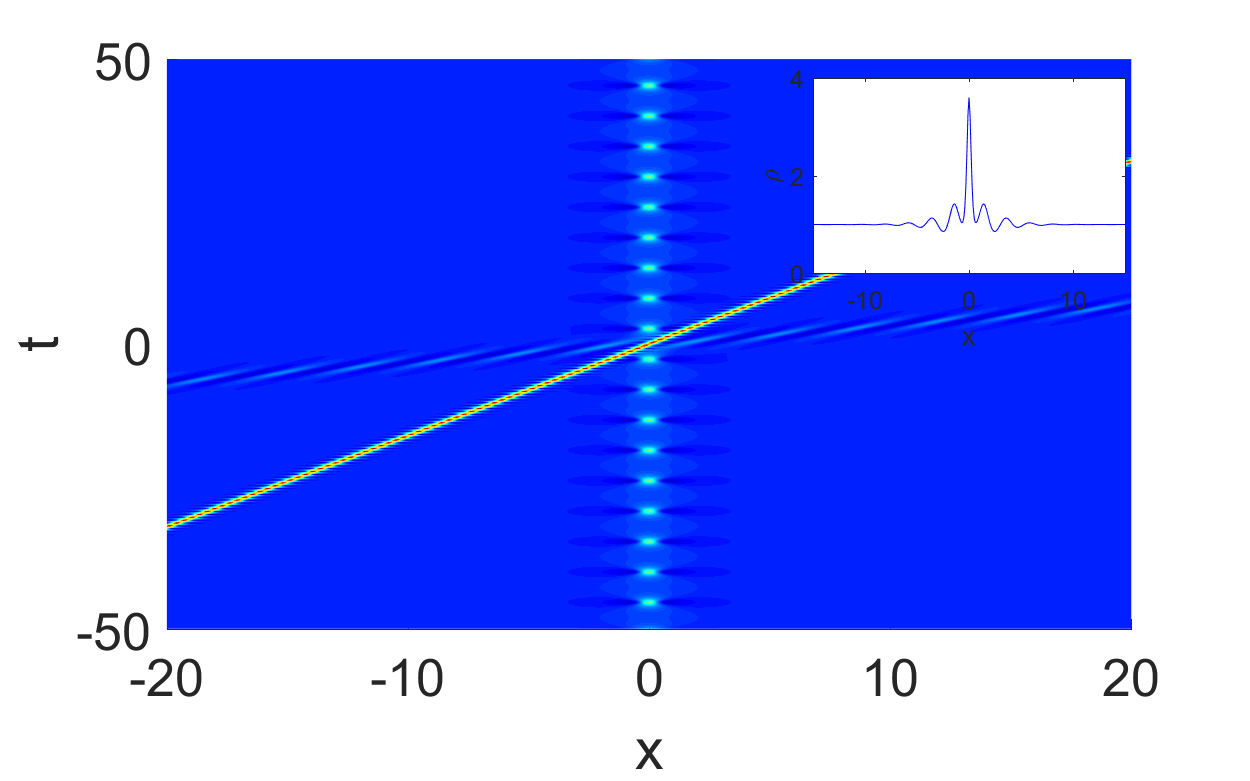}
        \caption{}
    \end{subfigure}

    \caption{Representative temporal-compression and walk-off-like patterns in the $(1,2)$-order Type-I configuration, with $\gamma_1=\gamma_2=\gamma_3=0$ and $\theta_1=\theta_2=\theta_3=0$. (a--c) $R_1=1.68$, $R_2=1.58$, $R_3=1.65$, and $\alpha_1=\alpha_2=\alpha_3=0$; (d--f) $R_1=2.5 $, $R_2=0.2$, $R_3=1.65$, $\alpha_1=1.3$, $\alpha_2=-0.12$, and $\alpha_3=0$.}
\end{figure}

Figures~10(a)--10(c) show a near-Akhmediev configuration with spatially extended, nearly periodic modulation. The $p_{\mathrm I}$ and $q_{\mathrm I}$ components display different interference patterns, while $\rho_{\mathrm I}$ retains the overall periodic structure. Figures~10(d)--10(f) show a KM configuration that is localized in $x$ and periodic in $t$. The two physical components exhibit complementary temporal peak--dip structures, while $\rho_{\mathrm I}$ displays the corresponding periodic compression. Figures~10(g)--10(i) show a near-rational configuration with a sharply localized central structure. Since the spectral points remain distinct, this case should be interpreted as a distinct-point near-rational configuration rather than an exact higher-order rational rogue wave.

In the near-Akhmediev region, small radial or angular mismatches modify the characteristic spatial scales of the first-order and two-fold building blocks, thereby producing spatial beating in the reconstructed physical components.

In Figures~11(a)--11(c), the asymmetric angular distribution of the spectral points produces weak bunching of the interference fringes and a slowly varying modulation of the otherwise near-periodic pattern. In Figures~11(d)--11(f), a small radial mismatch is introduced between the two spectral points of the two-fold block through $R_1\neq R_2$. Together with the different angular parameter of the first-order block, this mismatch generates a more visible slow envelope superimposed on the rapid spatial oscillations. As a result, alternating dense and sparse fringe regions appear in both physical components, with a more pronounced tilted redistribution in $q_{\mathrm I}$. These examples show that spatial beating may arise from the combined effects of radial and angular spectral mismatch while all spectral points remain close to the unit circle.

In the KM region, the same order mismatch appears mainly as temporal compression. Cross-region parameter choices additionally generate tilted, walk-off-like structures.

Figures~12(a)--12(c) show a representative KM configuration. The two physical components form temporally periodic localized structures, while $\rho_{\mathrm I}$ displays the corresponding periodic compression. Figures~12(d)--12(f) show a mixed-regime configuration generated by off-axis spectral parameters. A narrow temporally oscillatory band is intersected by obliquely propagating localized ridges, producing an asymmetric interference pattern with walk-off-like features. The $p_{\mathrm I}$ component retains a background close to unit amplitude, whereas the $q_{\mathrm I}$ component has a comparatively small but nonzero background. The combined amplitude $\rho_{\mathrm I}$ preserves the tilted trajectories and satisfies $\rho_{\mathrm I}\to1$ in the far field.

The $(1,2)$-order Type-I configuration therefore displays spatial beating near the Akhmediev regime, temporal compression in the KM regime, and asymmetric tilted structures for cross-region spectral parameters. These examples complement the $(0,1)$- and $(0,2)$-order configurations and illustrate how order and spectral mismatch affect the reconstructed physical components.

\section{Type-II breather solutions}

The Type-II configuration introduces a relative spatiotemporal displacement between two symmetric-sector building blocks.

\begin{thm}\label{thm:typeII}
Let $g_1(x,t)$ and $g_2(x,t)$ be symmetric-sector solutions of the CNLS-p system, so that each pair $(p,q)=(g_i,g_i)$ solves \eqref{qr1}. For arbitrary real constants $x_1,x_2,t_1,t_2$, define
\begin{equation}\label{typeII-theorem}
\begin{aligned}
p(x,t)&=g_1(x-x_1,t-t_1)+g_2(x-x_2,t-t_2),\\
q(x,t)&=g_1(x-x_1,t-t_1)-g_2(x-x_2,t-t_2).
\end{aligned}
\end{equation}
Then $(p,q)$ solves the CNLS-p system \eqref{qr1}.
\end{thm}

\begin{proof}
Set $u_+=2g_1(x-x_1,t-t_1)$ and $u_-=2g_2(x-x_2,t-t_2)$. Translation invariance implies that both $u_+$ and $u_-$ satisfy the scalar NLS equation \eqref{decoupledNLS}. Equation \eqref{typeII-theorem} then follows from the inverse normal-mode transformation \eqref{inverse-normalmodes}, so $(p,q)$ solves \eqref{qr1}.
\end{proof}

Using the building blocks $g_N=p[N]=q[N]$ and $g_M=p[M]=q[M]$, the general Type-II configuration is
\begin{equation}\label{TypeII-general}
\begin{aligned}
p_{\mathrm{II}}^{[M,N]}&=g_N(x-x_1,t-t_1)+g_M(x-x_2,t-t_2),\\
q_{\mathrm{II}}^{[M,N]}&=g_N(x-x_1,t-t_1)-g_M(x-x_2,t-t_2).
\end{aligned}
\end{equation}
We define the corresponding combined amplitude by
\begin{equation*}\label{rhoII}
\rho_{\mathrm{II}}=\left(|p_{\mathrm{II}}|^2+|q_{\mathrm{II}}|^2\right)^{1/2}.
\end{equation*}
Only the relative displacement
\begin{equation*}
\Delta x=x_N-x_M,\qquad
\Delta t=t_N-t_M
\end{equation*}
affects the internal pattern. A common displacement merely translates the complete solution. In addition to shifting the envelope, a temporal displacement introduces a relative carrier-phase shift and can therefore modify the background cancellation.

\subsection{The $(1,1)$-order configuration}

The first-order Type-II configuration is
\begin{equation}\label{TypeII11}
\begin{aligned}
p_{\mathrm{II}}^{[1,1]}&=g_1^{(1)}(x-x_1,t-t_1)+g_1^{(2)}(x-x_2,t-t_2),\\
q_{\mathrm{II}}^{[1,1]}&=g_1^{(1)}(x-x_1,t-t_1)-g_1^{(2)}(x-x_2,t-t_2),
\end{aligned}
\end{equation}
where $g_1^{(1)}$ and $g_1^{(2)}$ are associated with $\lambda_1$ and $\lambda_2$, respectively.

For zero relative displacement (Figures~13(a)--13(c)), the principal amplification regions of the two building blocks overlap near the origin. Constructive superposition produces a pronounced maximum in $|p_{\mathrm{II}}|$, whereas partial cancellation produces a central depression in
$|q_{\mathrm{II}}|$. The combined amplitude $\rho_{\mathrm{II}}$ displays the superposed intensity distributions of the two constituents.

For nonzero relative displacement (Figures~13(d)--13(f)), the KM-localized region and the principal amplification region of the Akhmediev-type constituent are displaced relative to each other. Their overlap is reduced, producing separated peak groups in $|p_{\mathrm{II}}|$ and weaker cancellation in $|q_{\mathrm{II}}|$. Thus, the observed changes result from relative translation and normal-mode interference rather than the collision of two directly interacting localized waves.

\begin{figure}[htbp]
    \centering
    \begin{subfigure}[b]{0.32\textwidth}
        \centering
        \includegraphics[height=3.2cm]{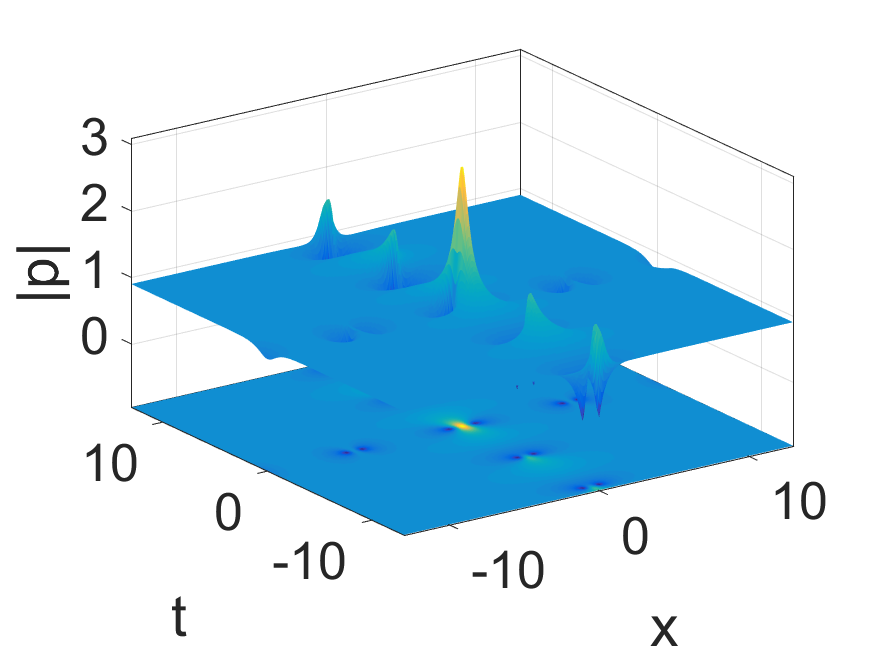}
        \caption{}
    \end{subfigure}
    \hfill
    \begin{subfigure}[b]{0.32\textwidth}
        \centering
        \includegraphics[height=3.2cm]{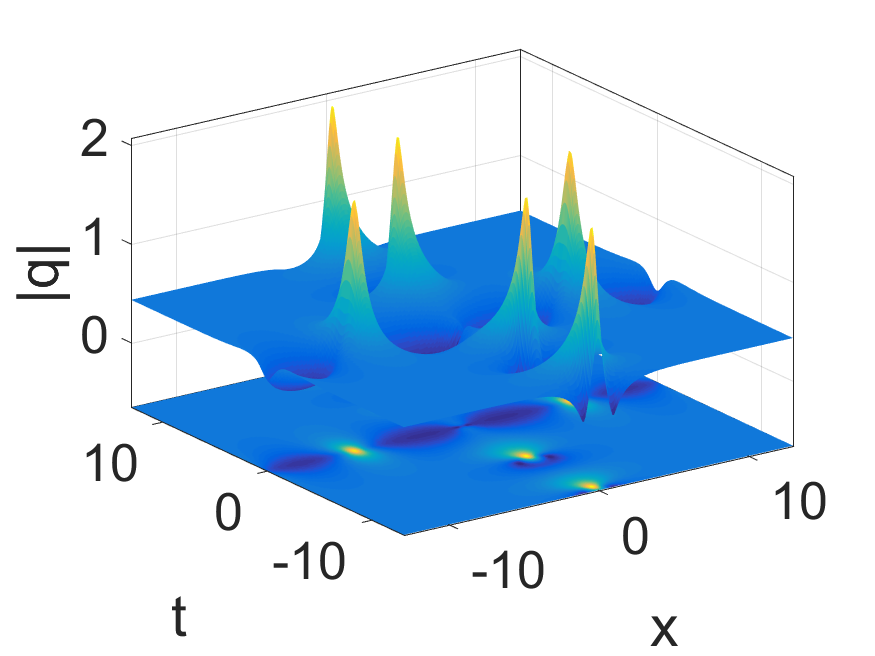}
        \caption{}
    \end{subfigure}
    \hfill
    \begin{subfigure}[b]{0.32\textwidth}
        \centering
        \includegraphics[height=3.2cm]{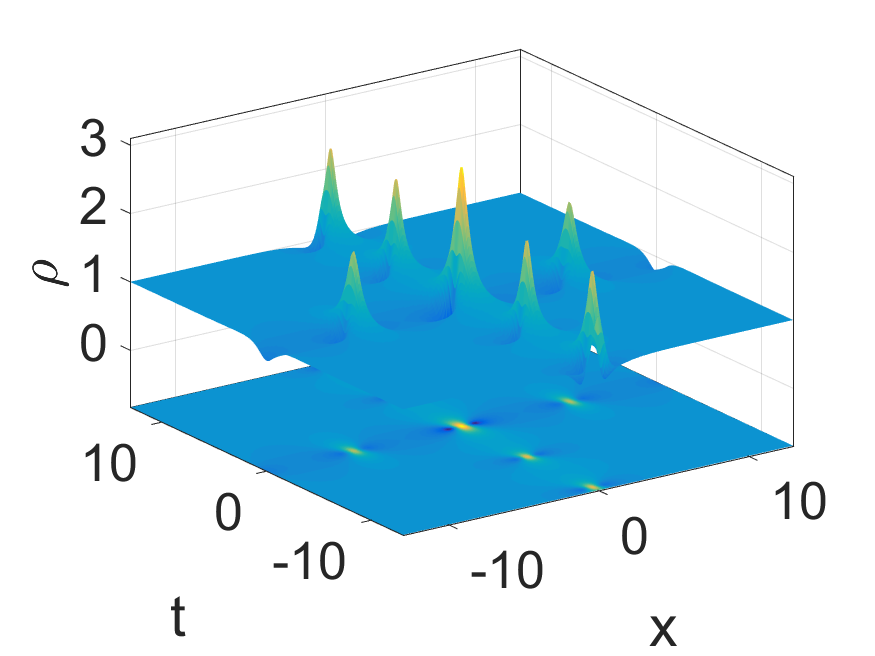}
        \caption{}
    \end{subfigure}

    \vspace{6pt}

    \begin{subfigure}[b]{0.32\textwidth}
        \centering
        \includegraphics[height=3.2cm]{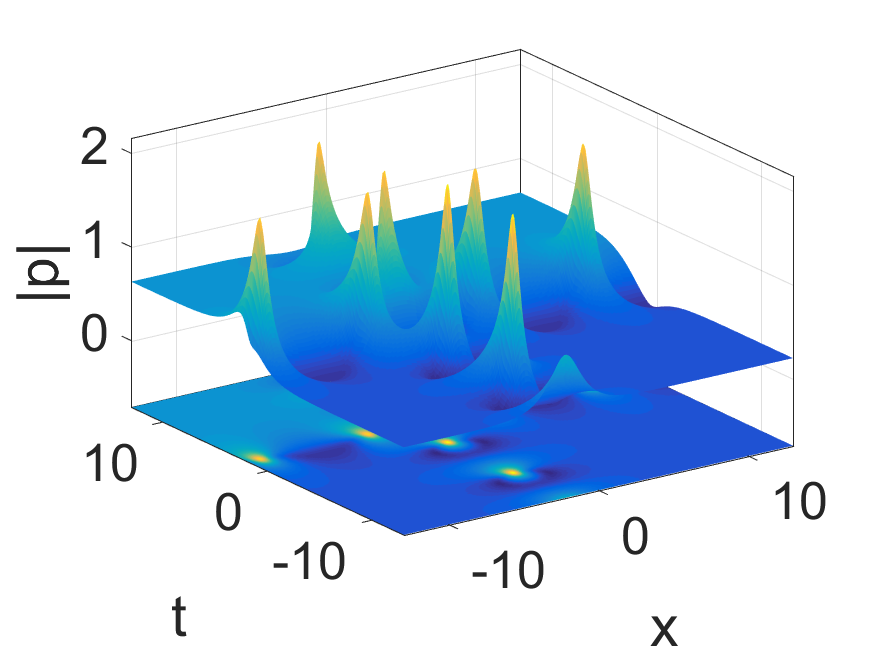}
        \caption{}
    \end{subfigure}
    \hfill
    \begin{subfigure}[b]{0.32\textwidth}
        \centering
        \includegraphics[height=3.2cm]{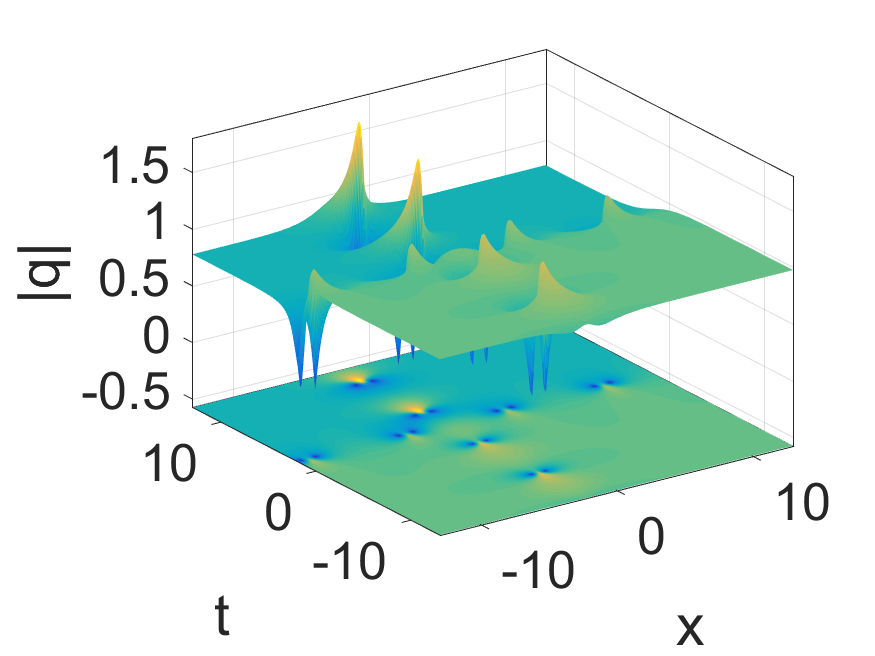}
        \caption{}
    \end{subfigure}
    \hfill
    \begin{subfigure}[b]{0.32\textwidth}
        \centering
        \includegraphics[height=3.2cm]{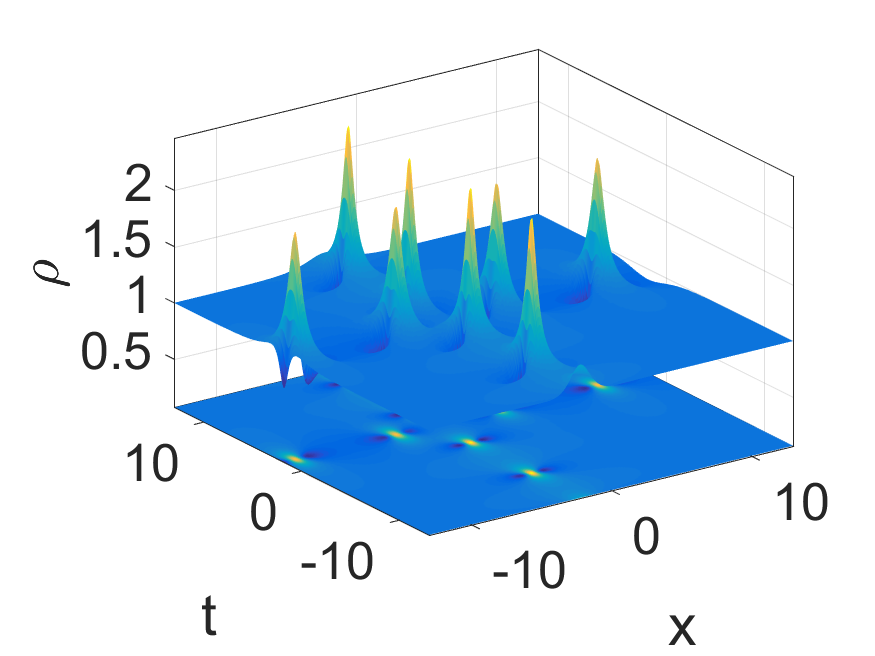}
        \caption{}
    \end{subfigure}

    \caption{$(1,1)$-order Type-II configurations with $\lambda_1=1.10\mathrm{i}$ and $\lambda_2=0.90\mathrm{i}$. (a--c) $(x_1,t_1,x_2,t_2)=(0,0,0,0)$; (d--f) $(x_1,t_1,x_2,t_2)=(-2.2,-1.8,2.2,1.8)$.}
\end{figure}

\subsection{The $(2,2)$-order configuration}

The second-order Type-II configuration is
\begin{equation}\label{TypeII22}
\begin{aligned}
p_{\mathrm{II}}^{[2,2]}&=g_2^{(1)}(x-x_1,t-t_1)+g_2^{(2)}(x-x_2,t-t_2),\\
q_{\mathrm{II}}^{[2,2]}&=g_2^{(1)}(x-x_1,t-t_1)-g_2^{(2)}(x-x_2,t-t_2),
\end{aligned}
\end{equation}
where $g_2^{(1)}$ is associated with $(\lambda_1,\lambda_2)$ and $g_2^{(2)}$ is associated with $(\lambda_3,\lambda_4)$.

\begin{figure}[!tbh]
    \centering
    \begin{subfigure}[b]{0.32\textwidth}
        \centering
        \includegraphics[height=3.2cm]{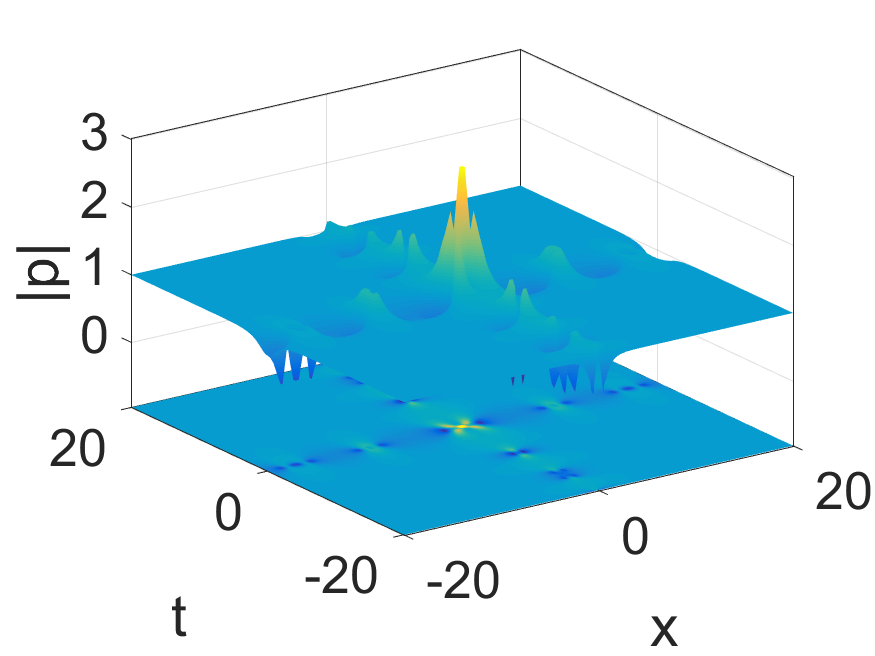}
        \caption{}
    \end{subfigure}
    \hfill
    \begin{subfigure}[b]{0.32\textwidth}
        \centering
        \includegraphics[height=3.2cm]{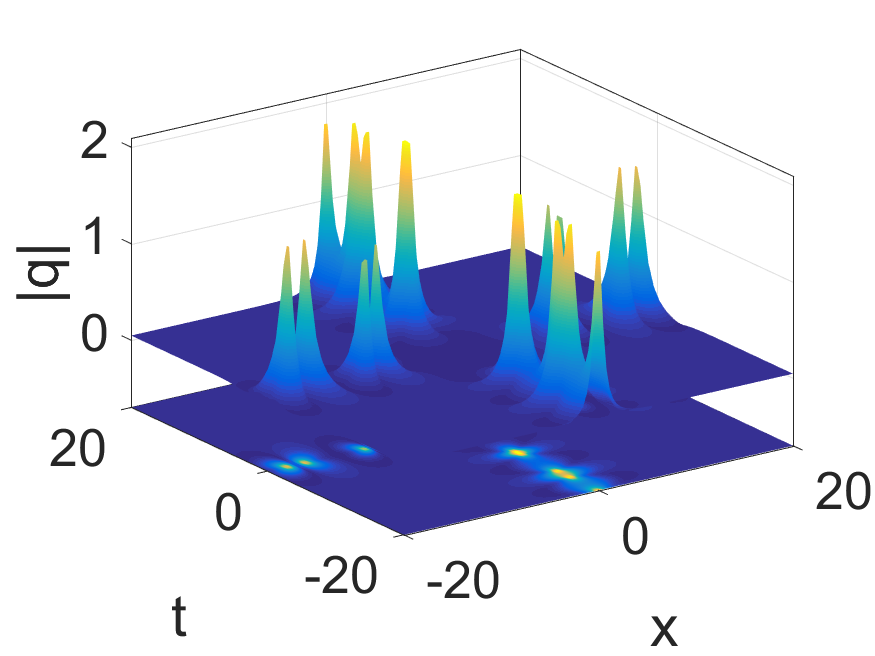}
        \caption{}
    \end{subfigure}
    \hfill
    \begin{subfigure}[b]{0.32\textwidth}
        \centering
        \includegraphics[height=3.2cm]{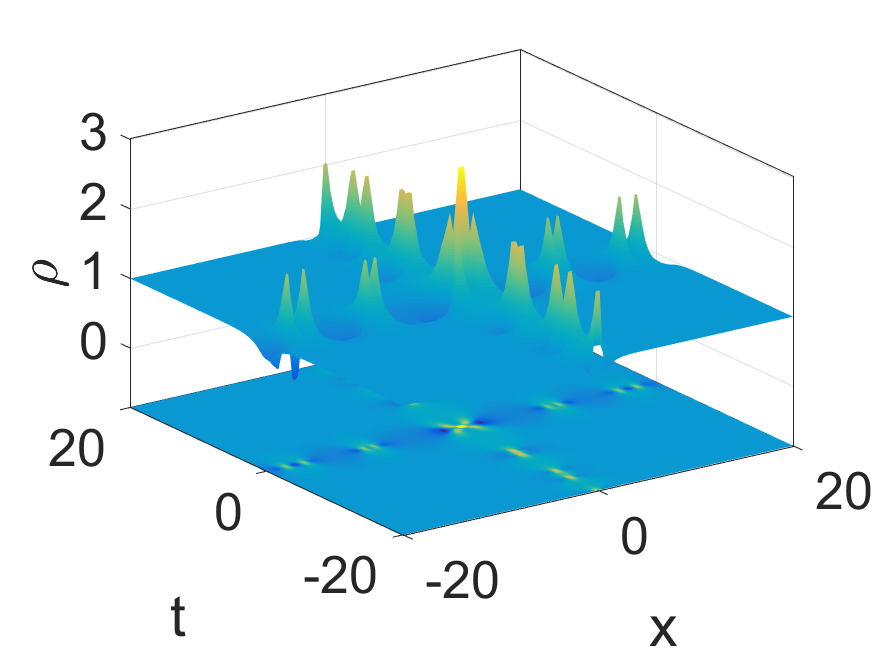}
        \caption{}
    \end{subfigure}

    \vspace{6pt}

    \begin{subfigure}[b]{0.32\textwidth}
        \centering
        \includegraphics[height=3.2cm]{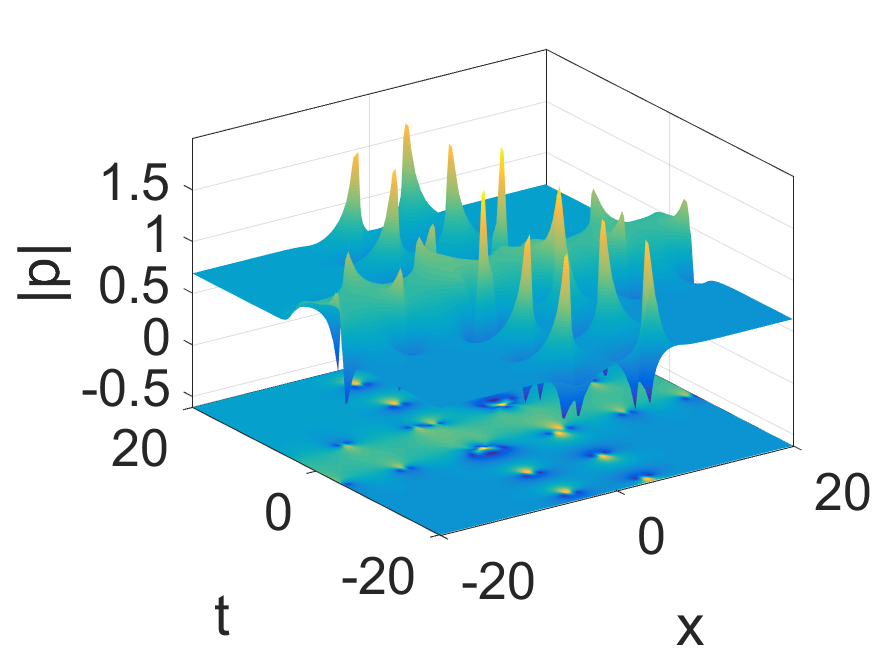}
        \caption{}
    \end{subfigure}
    \hfill
    \begin{subfigure}[b]{0.32\textwidth}
        \centering
        \includegraphics[height=3.2cm]{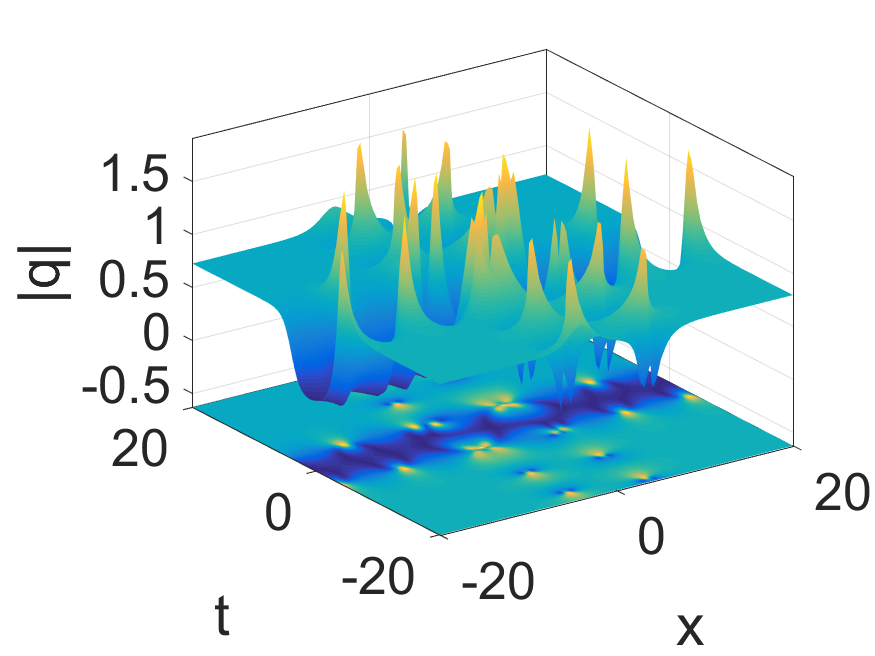}
        \caption{}
    \end{subfigure}
    \hfill
    \begin{subfigure}[b]{0.32\textwidth}
        \centering
        \includegraphics[height=3.2cm]{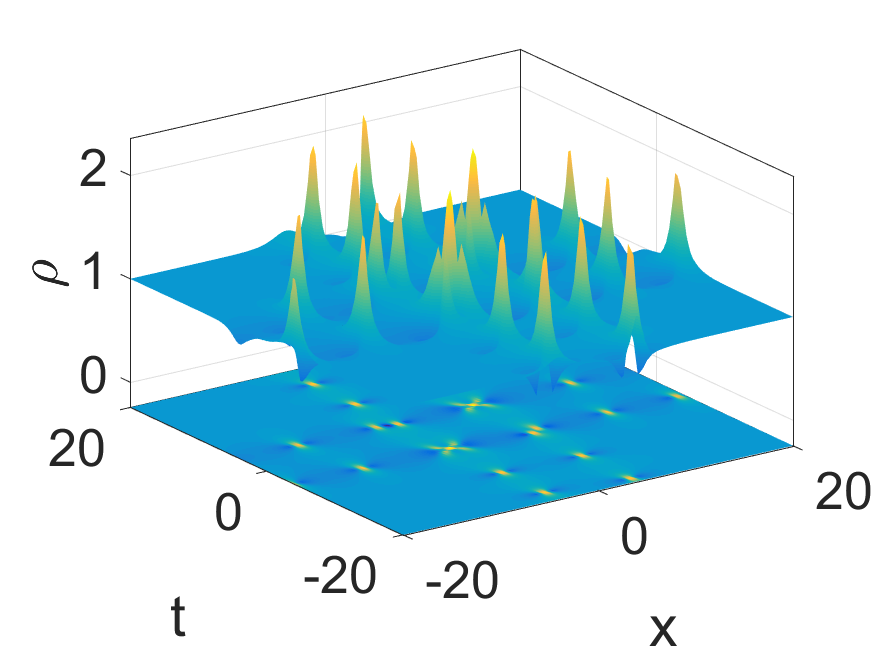}
        \caption{}
    \end{subfigure}

    \caption{$(2,2)$-order Type-II configurations with $\lambda_1=1.10\mathrm{i}$, $\lambda_2=0.90\mathrm{i}$, $\lambda_3=1.08\mathrm{i}$, and $\lambda_4=0.92\mathrm{i}$. (a--c) $(x_1,t_1,x_2,t_2)=(0,0,0,0)$; (d--f) $(x_1,t_1,x_2,t_2)=(-4,-4,4,4)$.}
\end{figure}

Figure~14 shows the configurations for $\lambda_1=1.10\mathrm{i}$, $\lambda_2=0.90\mathrm{i}$, $\lambda_3=1.08\mathrm{i}$, and $\lambda_4=0.92\mathrm{i}$. Each two-fold building block therefore combines a KM-type spectral point with an Akhmediev-type spectral point.

With zero displacement (Figures~14(a)--14(c)), the two mixed-regime two-fold patterns overlap near the origin. Their internal multipeak structures interfere constructively in the sum component and undergo stronger cancellation in the difference component. The combined amplitude retains the intensity contributions from both multipeak patterns.

For nonzero displacement (Figures~14(d)--14(f)), the two multipeak patterns are translated in opposite directions. The sum component displays two more clearly distinguishable peak groups, whereas cancellation in the difference component is reduced. Compared with the $(1,1)$-order case, the reconstructed pattern is more intricate because each building block already contains two distinct spectral scales.

\subsection{The $(1,2)$-order configuration}

The mixed-order Type-II configuration is
\begin{equation}\label{TypeII12}
\begin{aligned}
p_{\mathrm{II}}^{[1,2]}&=g_2(x-x_1,t-t_1)+g_1(x-x_2,t-t_2),\\
q_{\mathrm{II}}^{[1,2]}&=g_2(x-x_1,t-t_1)-g_1(x-x_2,t-t_2),
\end{aligned}
\end{equation}
where $g_1$ is associated with $\lambda_1$, whereas $g_2$ is associated with $(\lambda_2,\lambda_3)$.

\begin{figure}[!tbh]
    \centering
    \begin{subfigure}[b]{0.32\textwidth}
        \centering
        \includegraphics[height=3.2cm]{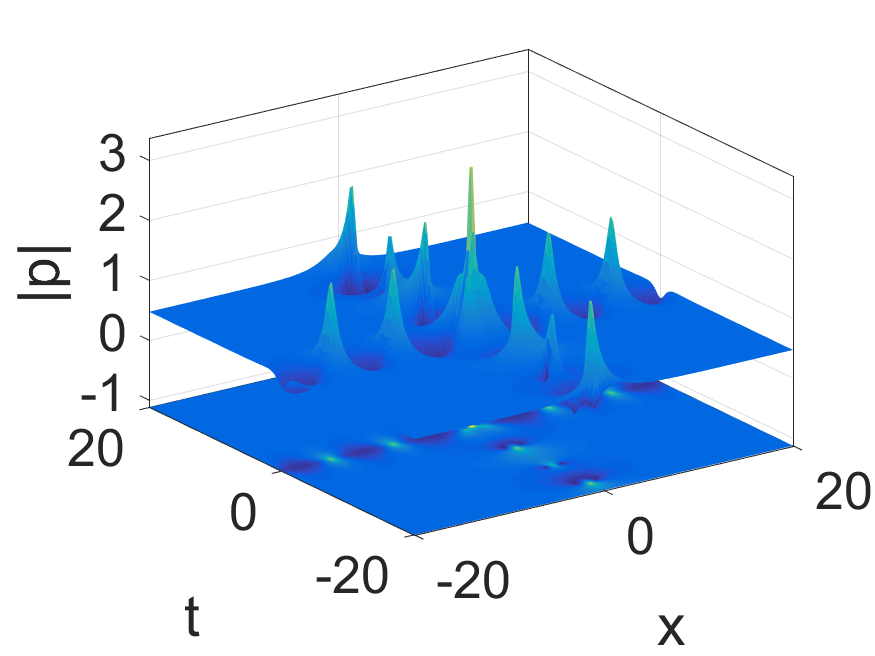}
        \caption{}
    \end{subfigure}
    \hfill
    \begin{subfigure}[b]{0.32\textwidth}
        \centering
        \includegraphics[height=3.2cm]{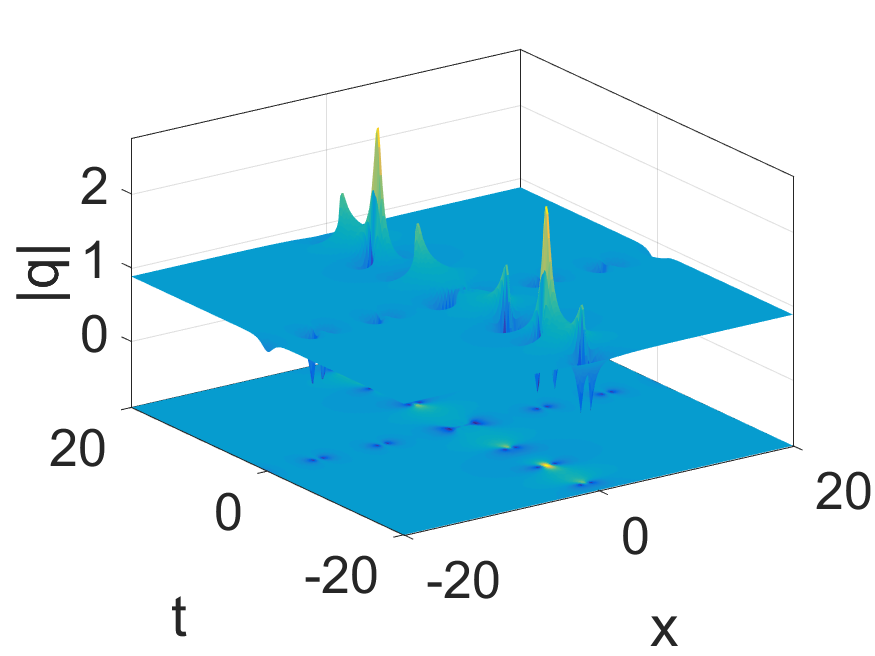}
        \caption{}
    \end{subfigure}
    \hfill
    \begin{subfigure}[b]{0.32\textwidth}
        \centering
        \includegraphics[height=3.2cm]{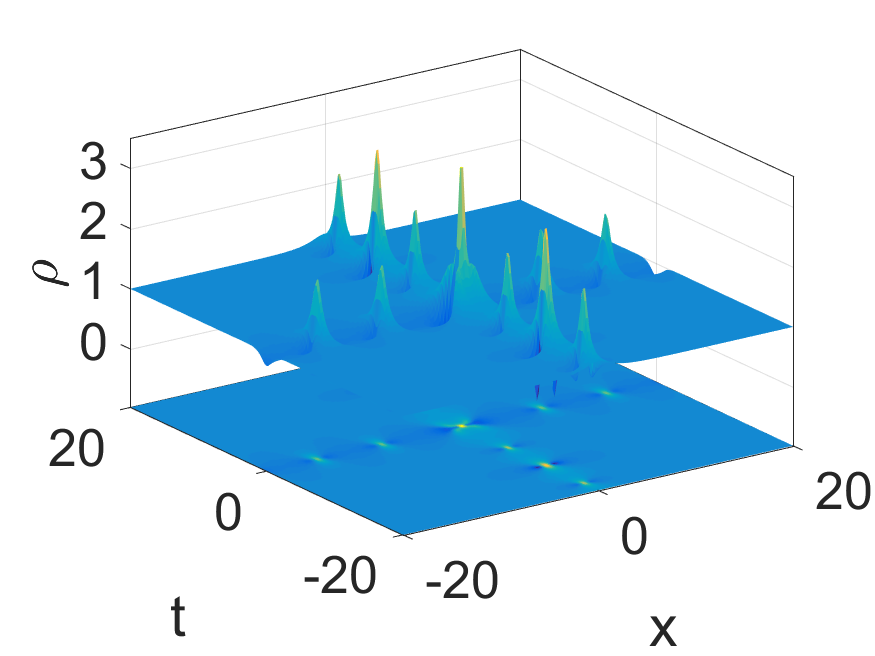}
        \caption{}
    \end{subfigure}

    \vspace{6pt}

    \begin{subfigure}[b]{0.32\textwidth}
        \centering
        \includegraphics[height=3.2cm]{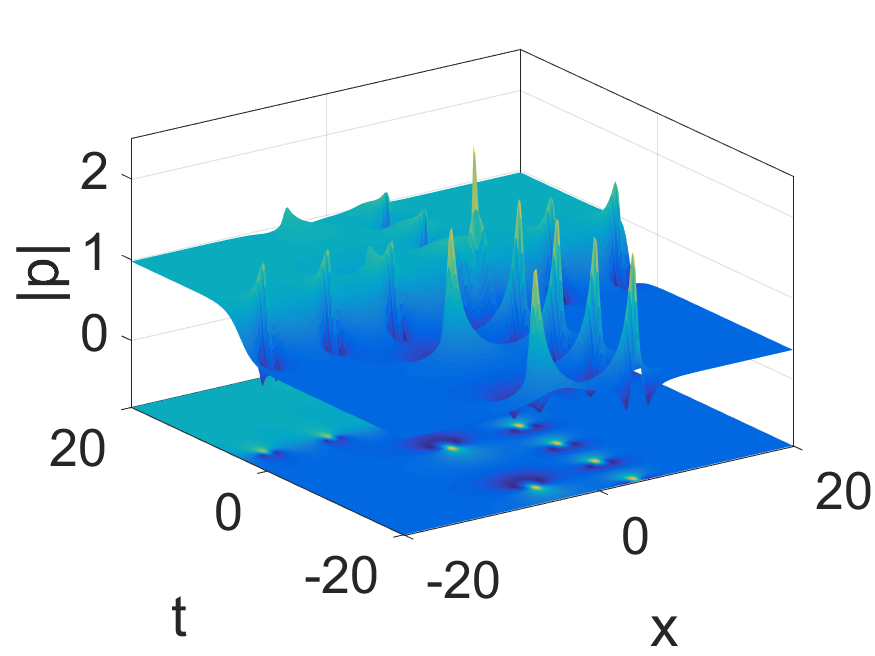}
        \caption{}
    \end{subfigure}
    \hfill
    \begin{subfigure}[b]{0.32\textwidth}
        \centering
        \includegraphics[height=3.2cm]{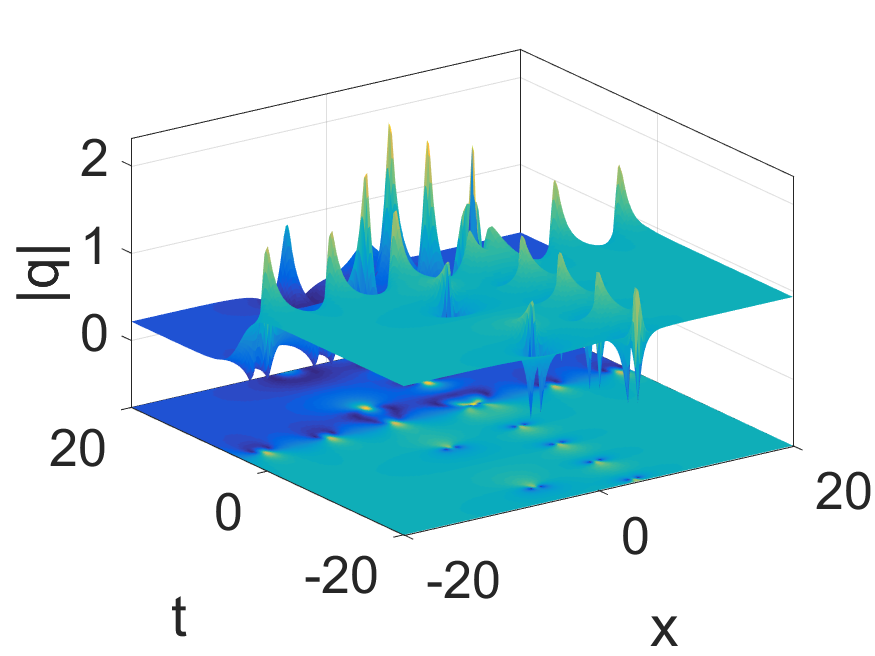}
        \caption{}
    \end{subfigure}
    \hfill
    \begin{subfigure}[b]{0.32\textwidth}
        \centering
        \includegraphics[height=3.2cm]{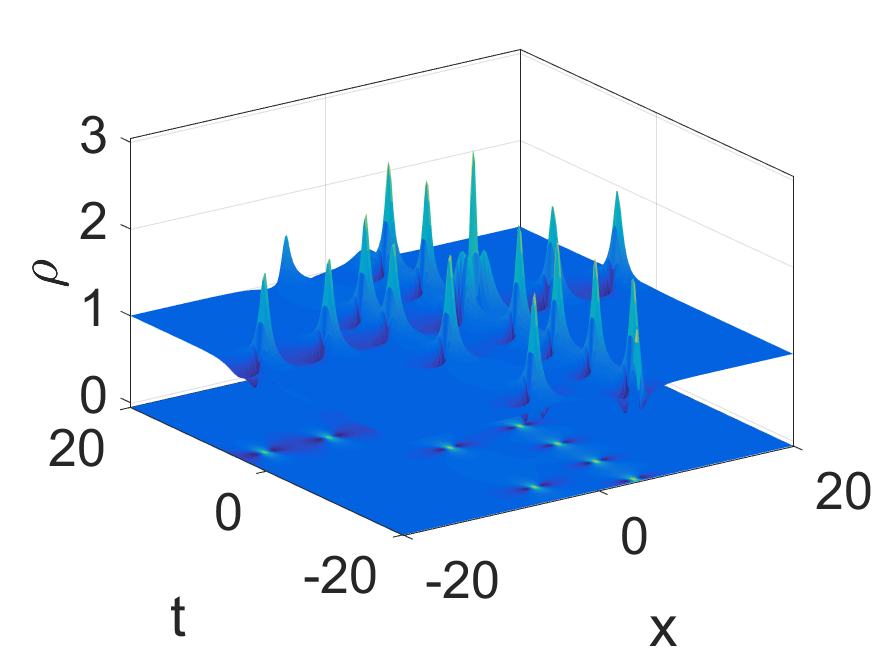}
        \caption{}
    \end{subfigure}

    \caption{$(1,2)$-order Type-II configurations with $\lambda_1=1.03\mathrm{i}$, $\lambda_2=1.12\mathrm{i}$, and $\lambda_3=0.88\mathrm{i}$. (a--c) $(x_1,t_1,x_2,t_2)=(0,0,0,0)$; (d--f) $(x_1,t_1,x_2,t_2)=(-4,-4,4,4)$.}
\end{figure}

Figure~15 shows the configurations for $\lambda_1=1.03\mathrm{i}$, $\lambda_2=1.12\mathrm{i}$, and $\lambda_3=0.88\mathrm{i}$. Thus, the first-order block is KM-type, whereas the two-fold block contains both KM-type and Akhmediev-type spectral contributions.

With zero displacement (Figures~15(a)--15(c)), the first-order and two-fold patterns overlap. Because their Darboux orders and characteristic scales differ, the sum and difference components exhibit asymmetric peak distributions. The sum component contains a dominant central peak with weaker side structures, whereas the difference component shows nonuniform cancellation.

For nonzero displacement (Figures~15(d)--15(f)), the first-order KM pattern and the mixed-regime two-fold pattern are translated relative to each other. Their reduced overlap weakens the cancellation in the difference component and separates the principal peak groups. This configuration illustrates how Darboux-order mismatch, spectral-regime mismatch, and relative translation jointly determine the reconstructed physical fields.

\section{Componentwise Fourier spectral analysis}

Exact spatial Fourier spectra of scalar NLS breathers, including the KM, Akhmediev, and Peregrine families, have been derived previously \cite{Karjanto2023}. Exact physical spectra of nondegenerate vector KM solitons have also been studied for the Manakov system \cite{CheChenLiu2022}. The purpose of the present section is therefore not to rederive the scalar KM spectrum as an isolated result, but to determine how this spectrum is distributed between the two physical components generated by the Type-I normal-mode reconstruction of the CNLS-p system.

For the first-order Type-I KM configuration, the background-subtracted localized parts of the two components have the same continuous Fourier spectrum. However, the full $q_I$ component retains a nonzero plane-wave background and therefore contains an additional
zero-wavenumber Dirac-$\delta$ contribution
\begin{equation*}
\widehat{q}_I(k,t)
=
-2\pi e^{\mathrm{i}t}\delta(k)
+
\widehat{p}_I(k,t).
\end{equation*}
The Akhmediev configuration is periodic in space and is naturally described by a Fourier series, whereas the localized parts of the KM and Peregrine configurations admit ordinary continuous Fourier transforms.
As a typical example, we only consider the KM configuration, other configurations can also derived in the similar way.

From Section~4, the Type-I KM solution is
\begin{equation}\label{qr8}
\begin{aligned}
p_{\mathrm I}(x,t)
&=b\,e^{\mathrm{i}t}
\frac{b\cos\tau+\mathrm{i}a\sin\tau}
{a\cosh X-\cos\tau},\\
q_{\mathrm I}(x,t)
&=-a\,e^{\mathrm{i}t}
\frac{\cosh X-a\cos\tau-\mathrm{i}b\sin\tau}
{a\cosh X-\cos\tau},
\end{aligned}
\end{equation}
where
\begin{equation*}\label{KM-parameters}
a=\cosh z,\qquad
b=\sinh z,\qquad
X=2bx,\qquad
\tau=t\sinh(2z).
\end{equation*}
As $|x|\to\infty$,
\begin{equation*}\label{KM-asymptotics}
p_{\mathrm I}(x,t)\to0,
\qquad
q_{\mathrm I}(x,t)\to-e^{\mathrm{i}t}.
\end{equation*}
We therefore define the localized part of the second component by
\begin{equation*}\label{q-localized}
q_{\mathrm I}^{\mathrm{loc}}(x,t)
=q_{\mathrm I}(x,t)+e^{\mathrm{i}t}.
\end{equation*}
For the Type-I KM configuration,
\begin{equation}\label{localized-identity}
q_{\mathrm I}^{\mathrm{loc}}(x,t)=p_{\mathrm I}(x,t).
\end{equation}
Hence, only one continuous Fourier transform must be evaluated.

We use the convention
\begin{equation}\label{Fourier-convention}
\widehat f(k,t)=\int_{-\infty}^{\infty}f(x,t)e^{-\mathrm{i}kx}\,dx.
\end{equation}
Equation \eqref{qr8} gives
\begin{equation}\label{qr9}
\begin{aligned}
\widehat p_{\mathrm I}(k,t)
={}&b e^{\mathrm{i}t}(b\cos\tau+\mathrm{i}a\sin\tau)\\
&\times
\int_{-\infty}^{\infty}
\frac{e^{-\mathrm{i}kx}}
{a\cosh(2bx)-\cos\tau}\,dx.
\end{aligned}
\end{equation}
Set
\begin{equation*}\label{spectral-substitution}
y=2bx,\qquad
\omega=\frac{k}{2b},\qquad
\varphi(t)=\arccos\left(\frac{\cos\tau}{a}\right)\in(0,\pi).
\end{equation*}
Since $dx=dy/(2b)$ and $a\cosh y-\cos\tau=a(\cosh y-\cos\varphi)$, equation \eqref{qr9} becomes
\begin{equation*}\label{qr9b}
\widehat p_{\mathrm I}(k,t)
=\frac{e^{\mathrm{i}t}}{2a}
(b\cos\tau+\mathrm{i}a\sin\tau)
\int_{-\infty}^{\infty}
\frac{e^{-\mathrm{i}\omega y}}
{\cosh y-\cos\varphi}\,dy.
\end{equation*}

The integrand has simple poles at
\begin{equation*}
y_n^{\pm}=\pm\mathrm{i}\varphi+2\pi\mathrm{i}n,
\qquad n\in\mathbb{Z}.
\end{equation*}
A rectangular contour and the residue theorem give
\begin{equation*}\label{basic-Fourier-integral}
\int_{-\infty}^{\infty}
\frac{e^{-\mathrm{i}\omega y}}
{\cosh y-\cos\varphi}\,dy
=
\frac{2\pi\sinh[\omega(\pi-\varphi)]}
{\sin\varphi\,\sinh(\pi\omega)}.
\end{equation*}
Therefore,
\begin{equation}\label{qr10}
\widehat p_{\mathrm I}(k,t)
=
\frac{\pi e^{\mathrm{i}t}}{a}
(b\cos\tau+\mathrm{i}a\sin\tau)
\frac{\sinh[\omega(\pi-\varphi)]}
{\sin\varphi\,\sinh(\pi\omega)}.
\end{equation}

Up to the normalization and parameterization used here, Eq.~(6.12) is consistent with the scalar KM spectrum reported in Ref.~\cite{Karjanto2023}. The additional result for the CNLS-p system is the componentwise identity in Eq.~(6.6), together with the zero-wavenumber background contribution in Eq.~(6.14).
At $k=0$, the quotient in \eqref{qr10} is understood through the continuous limit
\begin{equation*}
\lim_{\omega\to0}
\frac{\sinh[\omega(\pi-\varphi)]}
{\sinh(\pi\omega)}
=\frac{\pi-\varphi}{\pi}.
\end{equation*}
Thus,
\begin{equation*}\label{p-spectrum-zero}
\widehat p_{\mathrm I}(0,t)
=
\frac{e^{\mathrm{i}t}}{a}
(b\cos\tau+\mathrm{i}a\sin\tau)
\frac{\pi-\varphi}{\sin\varphi}.
\end{equation*}
Since $\mathcal{F}[1]=2\pi\delta(k)$ under the convention \eqref{Fourier-convention}, the complete spectra, understood in the sense of distributions, are
\begin{equation}\label{qr11}
\begin{aligned}
\widehat p_{\mathrm I}(k,t)
&=
\frac{\pi e^{\mathrm{i}t}}{a}
(b\cos\tau+\mathrm{i}a\sin\tau)
\frac{\sinh[\omega(\pi-\varphi)]}
{\sin\varphi\,\sinh(\pi\omega)},\\
\widehat q_{\mathrm I}(k,t)
&=-2\pi e^{\mathrm{i}t}\delta(k)
+\widehat p_{\mathrm I}(k,t).
\end{aligned}
\end{equation}

\begin{figure}[!tbh]
    \centering
    \begin{subfigure}[b]{0.48\textwidth}
        \centering
        \includegraphics[width=\textwidth, height=4.2cm]{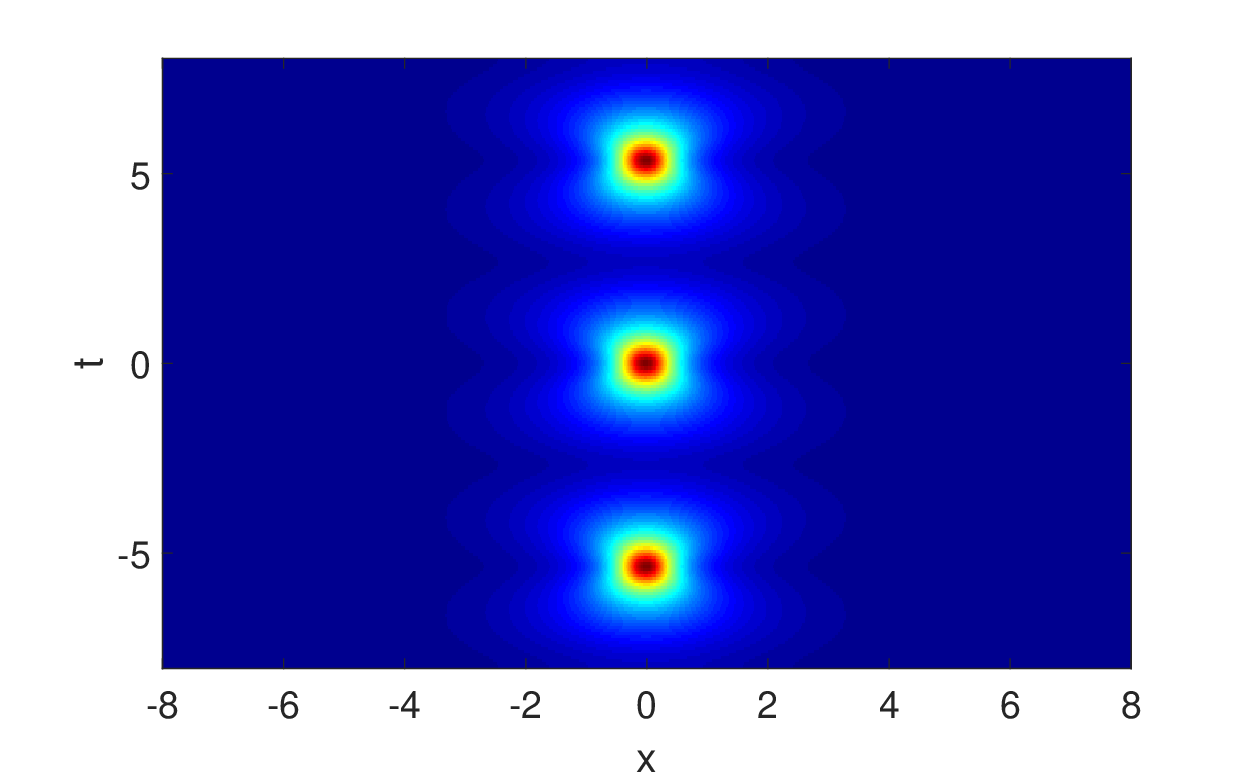}
        \caption{}
    \end{subfigure}
    \hfill
    \begin{subfigure}[b]{0.48\textwidth}
        \centering
        \includegraphics[width=\textwidth, height=4.2cm]{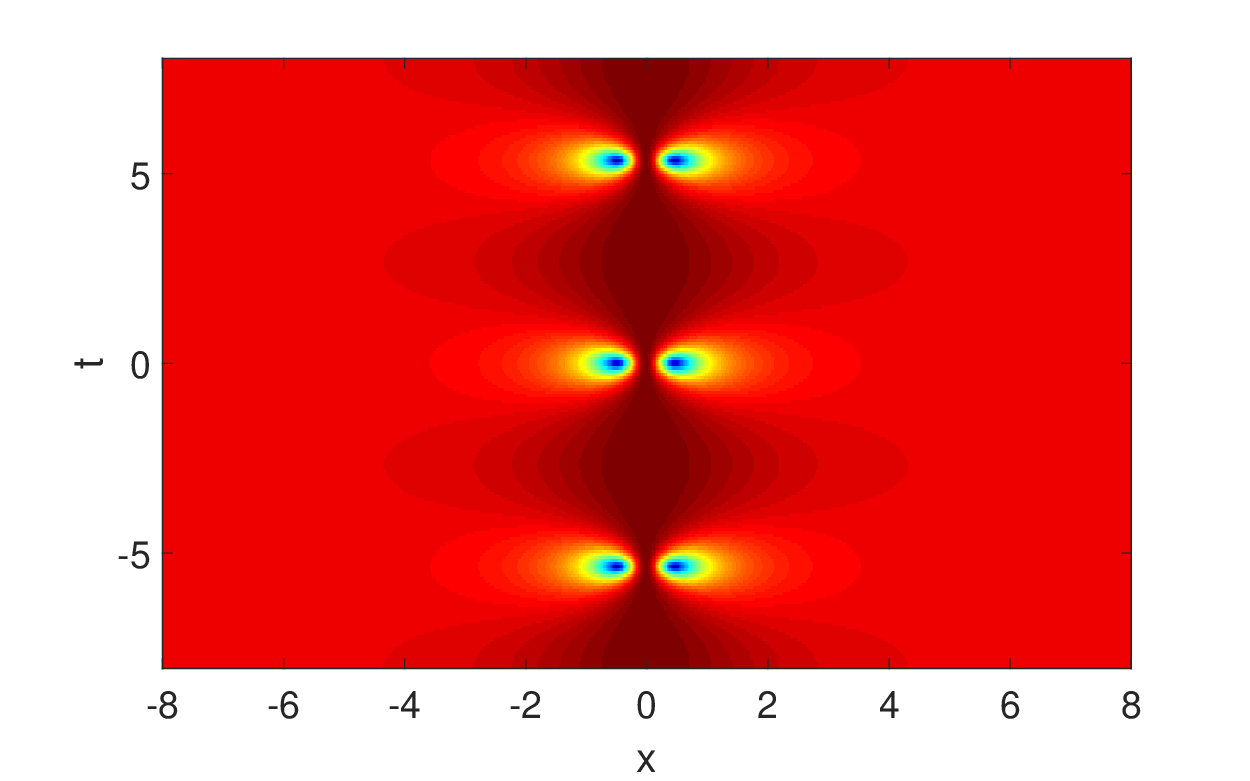}
        \caption{}
    \end{subfigure}
    \caption{Spatiotemporal evolution of the Type-I KM configuration for $z=0.5$. (a) $|p_{\mathrm I}|$, which is exponentially localized in $x$; (b) $|q_{\mathrm I}|$, which approaches the background value $1$ as $|x|\to\infty$.}
\end{figure}

Figure~16 shows the spatiotemporal profiles. The $p_{\mathrm I}$ component is periodic in $t$ and decays exponentially as $|x|\to\infty$, whereas $q_{\mathrm I}$ approaches the background value $|q_{\mathrm I}|=1$. At the compression instants $\tau=2\pi n$, $n\in\mathbb{Z}$, the $q_{\mathrm I}$ component has two exact zeros at
\begin{equation*}
x=\pm\frac{z}{2\sinh z}.
\end{equation*}
Between these instants, the depths and positions of the two dips vary periodically. This complementary component structure follows directly from the identity \eqref{localized-identity}.

The full $p_{\mathrm I}$ component has a continuous Fourier spectrum because it vanishes as $|x|\to\infty$. The background-subtracted component $q_{\mathrm I}^{\mathrm{loc}}$ has the same continuous spectrum. The complete $q_{\mathrm I}$ spectrum additionally contains the zero-wavenumber contribution $-2\pi e^{\mathrm{i}t}\delta(k)$, which originates from the retained plane-wave background.

\begin{figure}[!tbh]
    \centering
    \begin{subfigure}[b]{0.48\textwidth}
        \centering
        \includegraphics[width=\textwidth, height=4.2cm]{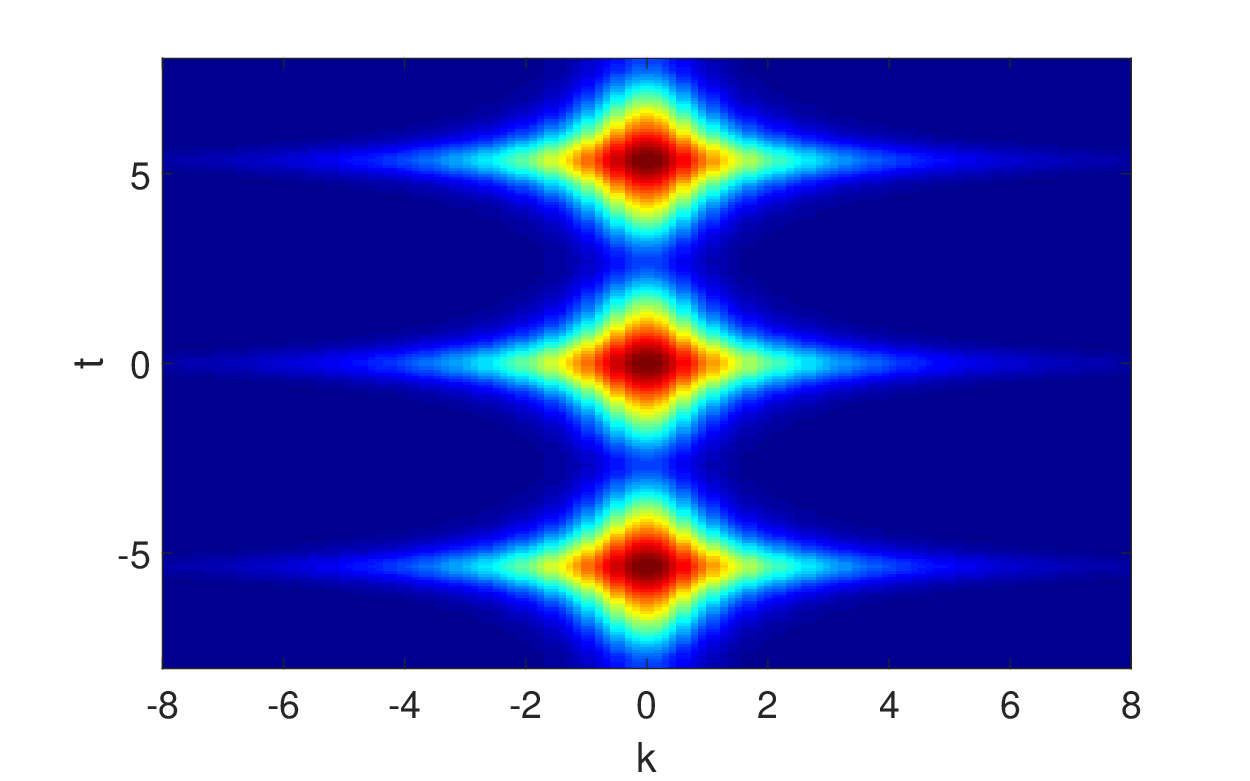}
        \caption{}
    \end{subfigure}
    \hfill
    \begin{subfigure}[b]{0.48\textwidth}
        \centering
        \includegraphics[width=\textwidth, height=4.2cm]{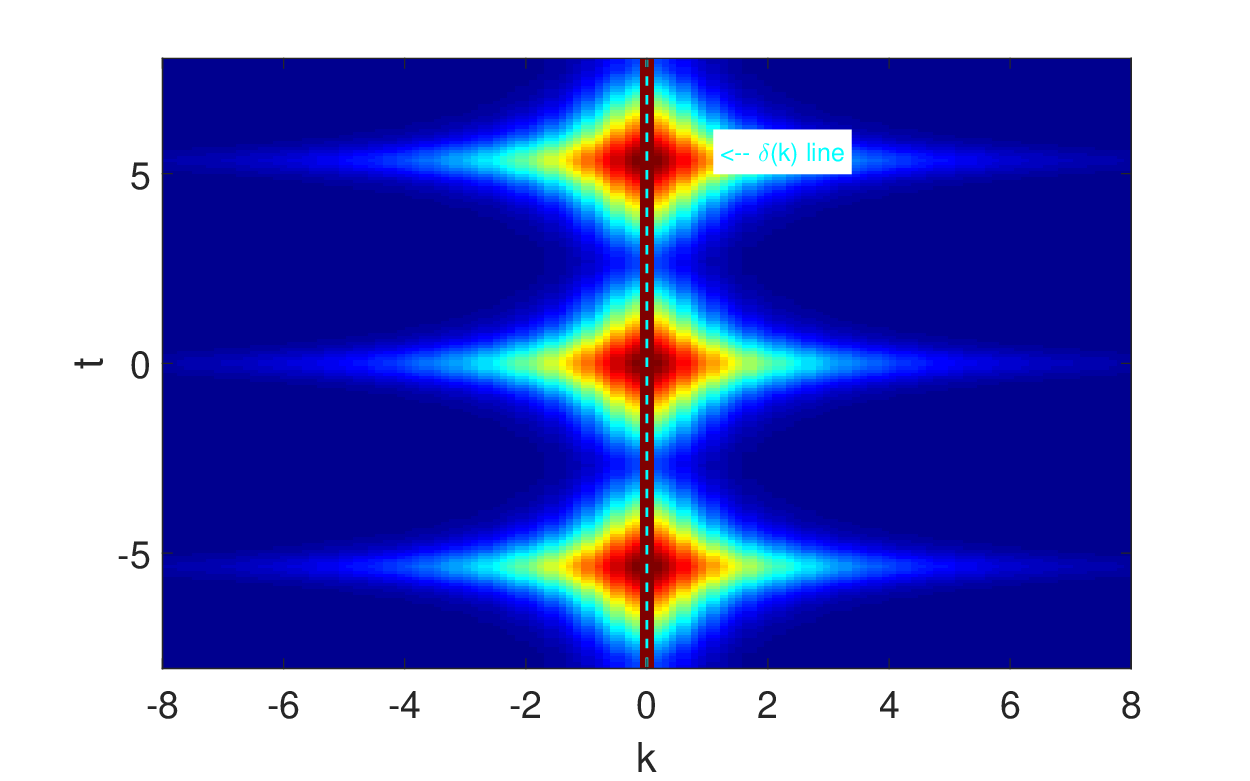}
        \caption{}
    \end{subfigure}
    \caption{Two-dimensional spectral evolution. (a) $|\widehat p_{\mathrm I}(k,t)|$, which is symmetric about $k=0$ and periodically broadens and narrows in time; (b) a finite-window numerical representation of the full $q_{\mathrm I}$ spectrum. The strong zero-wavenumber bin in (b) represents the retained constant background. Its finite height depends on the computational interval and the Fourier-transform normalization.}
\end{figure}

Figure~17 shows the two-dimensional spectral evolution. The continuous spectrum $|\widehat p_{\mathrm I}(k,t)|$ is symmetric about $k=0$ and periodically broadens and narrows in time. The localized part of $q_{\mathrm I}$ has the same continuous spectrum. In Figure~17(b), the Dirac contribution is represented by the zero-wavenumber bin of a finite-window numerical transform rather than by an ordinary finite-amplitude function.

To verify the continuous-spectrum formula \eqref{qr10}, we compare the analytical result for $p_{\mathrm I}(x,0)$ with a numerical Fourier transform computed on a finite spatial interval. Figure~18 shows close agreement over $|k|\leq5$.

\begin{figure}[!tbh]
    \centering
    \begin{subfigure}[b]{0.55\textwidth}
        \centering
        \includegraphics[width=\textwidth, height=4.2cm]{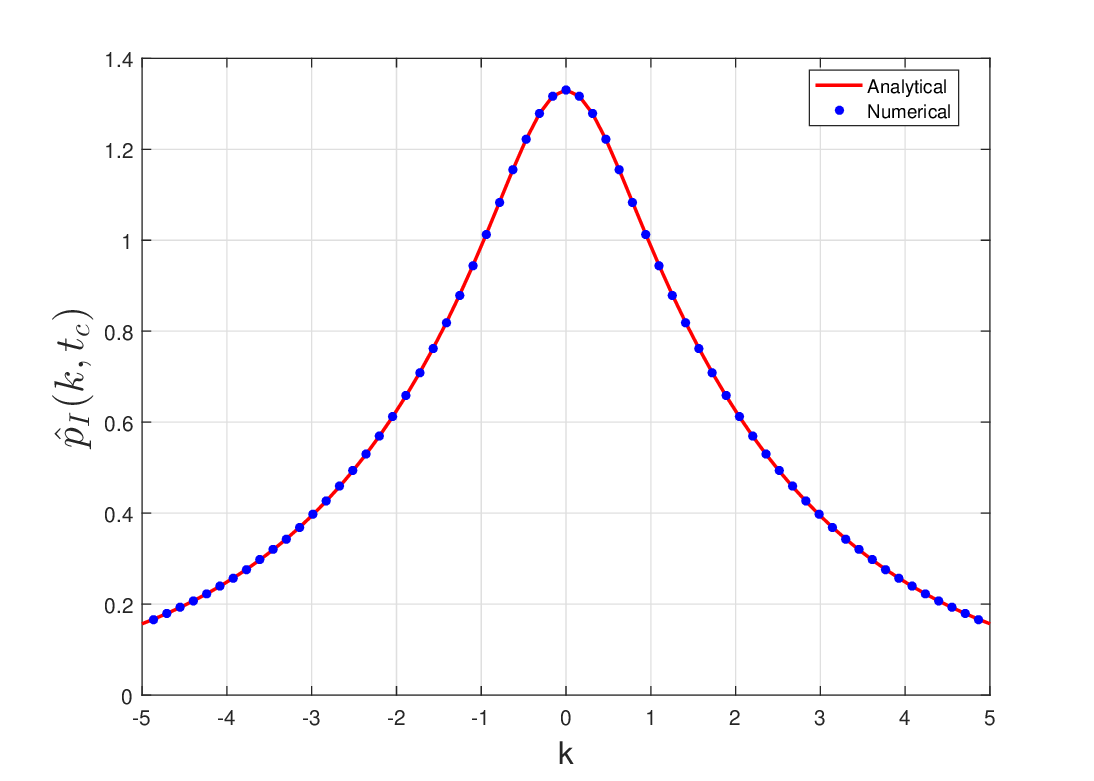}
        \caption{}
    \end{subfigure}
    \caption{Verification of the analytical continuous spectrum $|\widehat p_{\mathrm I}(k,0)|$. The analytical expression in \eqref{qr10} (red solid curve) is compared with the numerical Fourier transform of $p_{\mathrm I}(x,0)$ (blue dots). The two results agree closely over $|k|\leq5$.}
\end{figure}

\section{Numerical verification}
Similar to the standard numerical frameworks for NLS equations \cite{TA-1984,WeidemanHerbst1986},
we verify the Type-I KM solution by direct numerical integration using the split-step Fourier method (SSFM). The exact analytical fields at $t=0$ are used as the initial condition. The same numerical framework can also be applied to the Type-II configurations.
\begin{figure}[!tbh]
    \centering
    \begin{subfigure}[b]{0.48\textwidth}
        \centering
        \includegraphics[width=\textwidth, height=4.2cm]{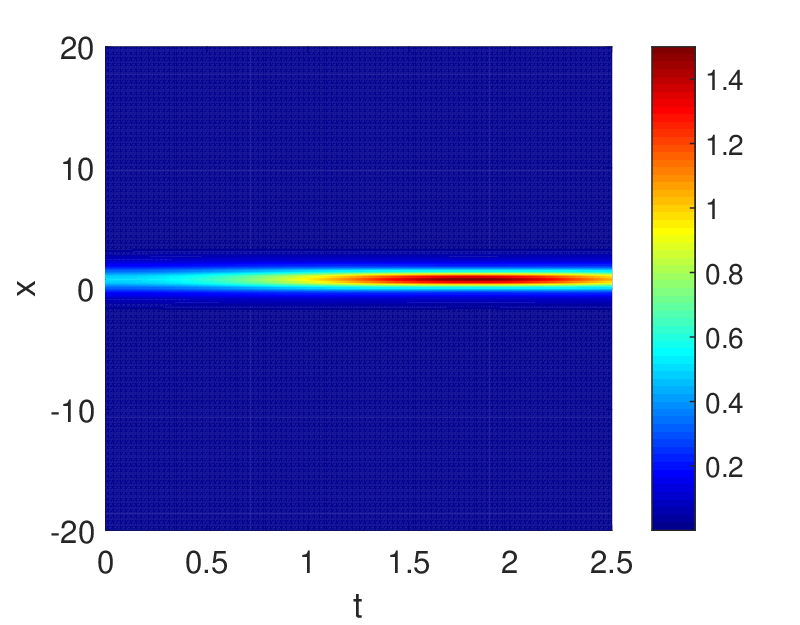}
        \caption{}
    \end{subfigure}
    \hfill
    \begin{subfigure}[b]{0.48\textwidth}
        \centering
        \includegraphics[width=\textwidth, height=4.2cm]{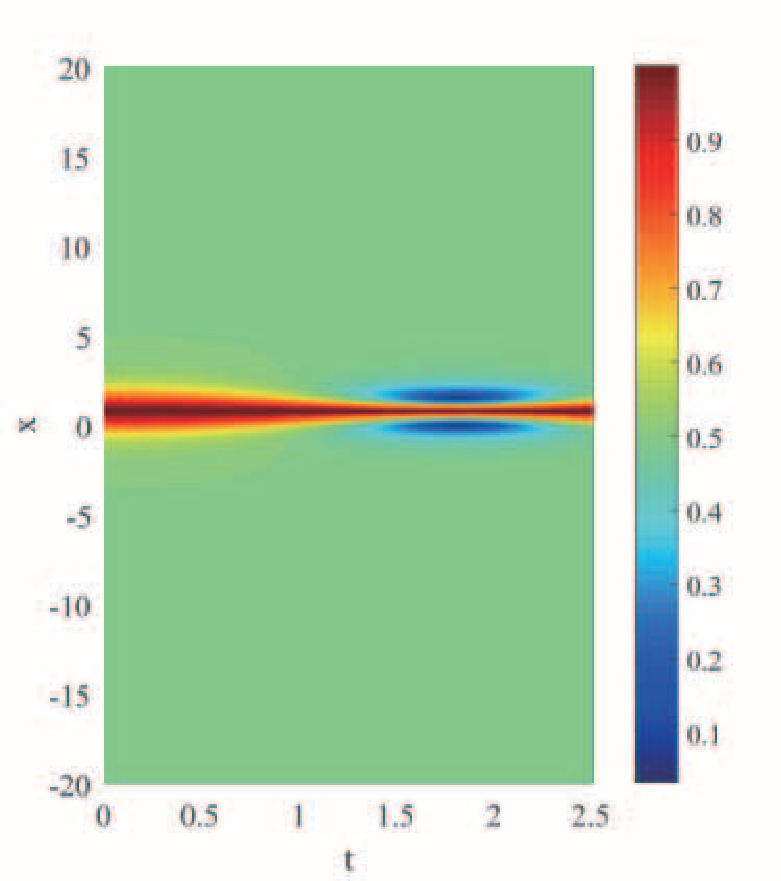}
        \caption{}
    \end{subfigure}
    \caption{Density plots of the Type-I KM solution. (a) $|p(x,t)|$; (b) $|q(x,t)|$. The evolution interval is $t\in[0,2.5]$.}
\end{figure}

Figure~19 shows density plots of $|p(x,t)|$ and $|q(x,t)|$. The $p$ component remains localized on a zero background, whereas the $q$ component oscillates on a finite background. Both components exhibit the temporal periodicity of the KM solution.

\begin{figure}[!tbh]
    \centering
    \begin{subfigure}[b]{0.48\textwidth}
        \centering
        \includegraphics[width=\textwidth, height=4.2cm]{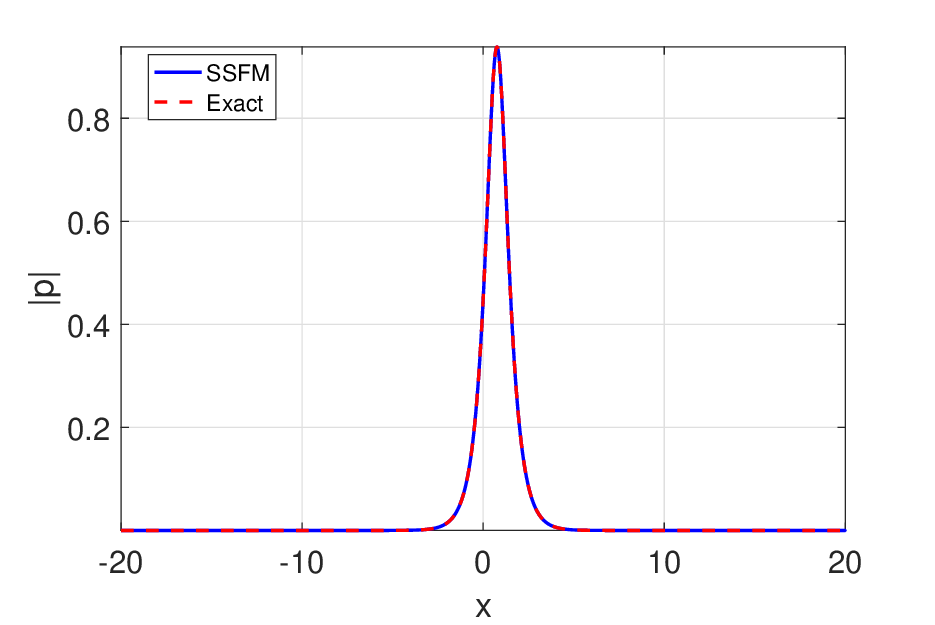}
        \caption{}
    \end{subfigure}
    \hfill
    \begin{subfigure}[b]{0.48\textwidth}
        \centering
        \includegraphics[width=\textwidth, height=4.2cm]{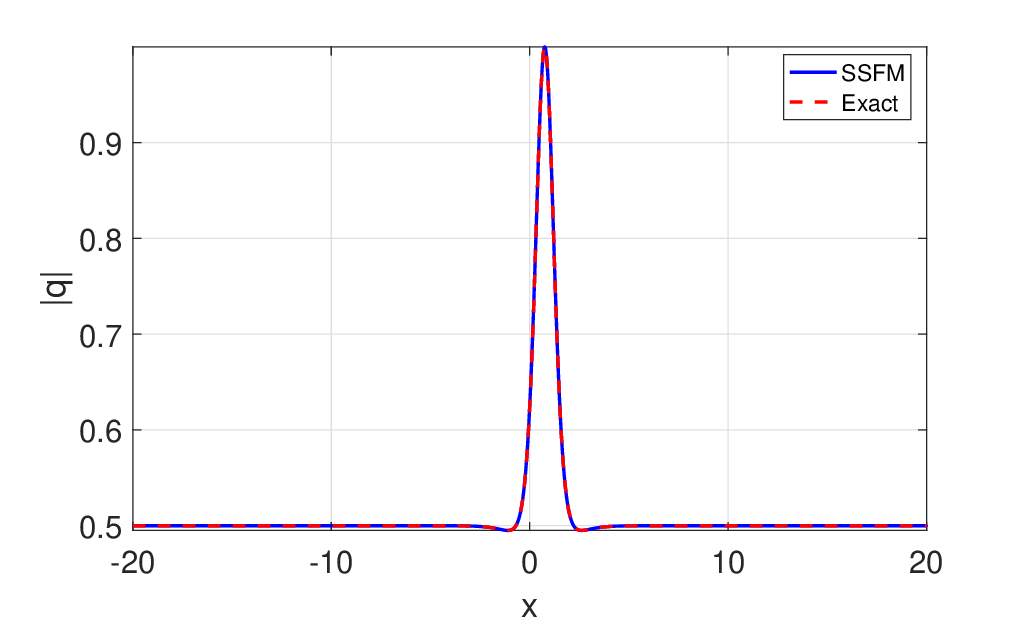}
        \caption{}
    \end{subfigure}
    \caption{Comparison at $t=1.0$ between the SSFM results (solid blue curves) and the exact analytical solutions (dashed red curves). (a) $|p|$; (b) $|q|$. The relative $L^2$ errors are below $10^{-4}$.}
\end{figure}

Figure~20 compares the SSFM results with the exact analytical profiles at $t=1.0$. The numerical and analytical curves agree closely. The relative $L^2$ errors of both components are below $10^{-4}$ over the reported interval. The remaining discrepancy is mainly attributable to spatial truncation and numerical discretization.

To test robustness over the simulated interval, we add complex Gaussian noise to the exact initial fields
\begin{equation}\label{noise-model}
\widetilde p
=p+\varepsilon p_{\max}(\xi_1+\mathrm{i}\xi_2),
\qquad
\widetilde q
=q+\varepsilon q_{\max}(\xi_3+\mathrm{i}\xi_4),
\end{equation}
where $p_{\max}=\max|p|$, $q_{\max}=\max|q|$, and the variables $\xi_j\sim\mathcal{N}(0,1)$ are independent standard normal random variables. The relative noise level is $\varepsilon=2\%$.

\begin{figure}[!tbh]
    \centering
    \begin{subfigure}[b]{0.48\textwidth}
        \centering
        \includegraphics[width=\textwidth, height=4.2cm]{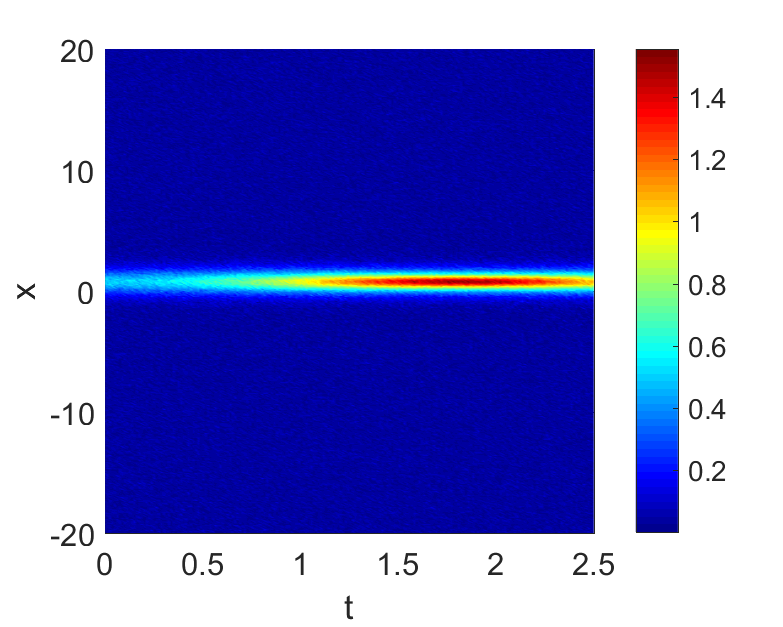}
        \caption{}
    \end{subfigure}
    \hfill
    \begin{subfigure}[b]{0.48\textwidth}
        \centering
        \includegraphics[width=\textwidth, height=4.2cm]{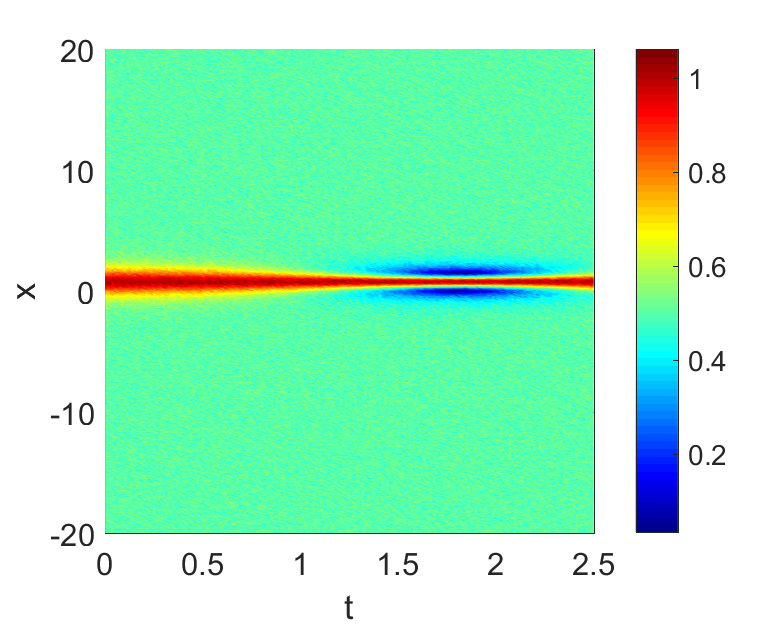}
        \caption{}
    \end{subfigure}
    \caption{Density plots obtained from the Type-I KM initial condition with $2\%$ additive complex Gaussian noise. (a) $|p(x,t)|$; (b) $|q(x,t)|$. The main breathing pattern remains visible over the simulated interval.}
\end{figure}

Figure~21 shows the noisy density plots. Random fluctuations appear across the computational domain, but the central breather structure and its temporal periodicity remain visible. Figure~22 compares one-dimensional profiles at $t=1.0$ before and after the addition of noise. The perturbation produces local fluctuations, while the overall envelopes remain close to the exact profiles. These simulations indicate robustness to small additive noise over the time interval considered; they do not constitute a full spectral-stability analysis.

\begin{figure}[!tbh]
    \centering
    \begin{subfigure}[b]{0.48\textwidth}
        \centering
        \includegraphics[width=\textwidth, height=4.2cm]{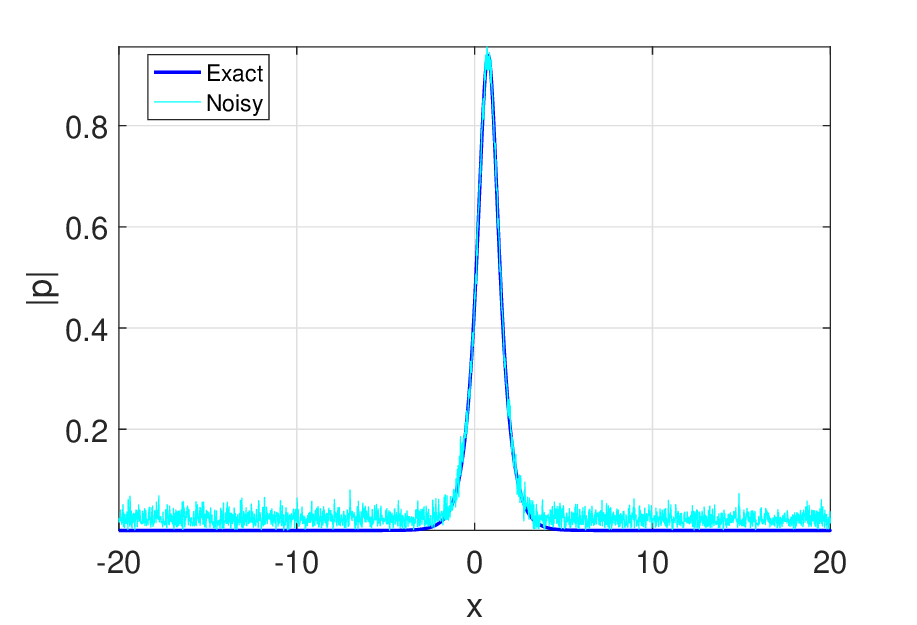}
        \caption{}
    \end{subfigure}
    \hfill
    \begin{subfigure}[b]{0.48\textwidth}
        \centering
        \includegraphics[width=\textwidth, height=4.2cm]{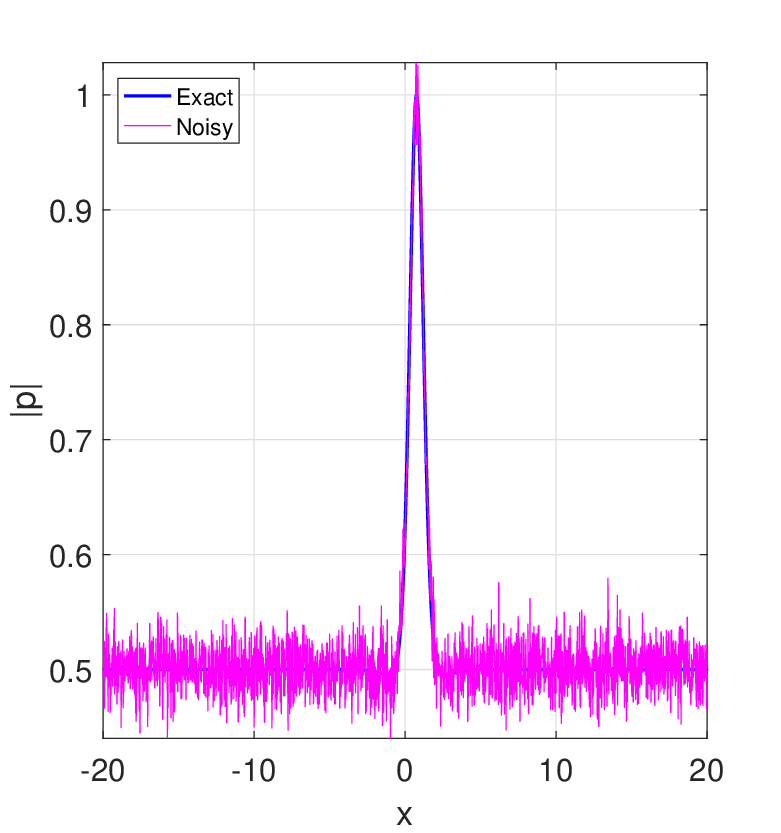}
        \caption{}
    \end{subfigure}
    \caption{One-dimensional profiles at $t=1.0$ with $2\%$ additive Gaussian noise. (a) $|p|$; (b) $|q|$. The exact profiles are compared with the noisy numerical fields. The overall breather envelopes remain close to the unperturbed profiles.}
\end{figure}

\section{Conclusions and discussions}
In this work, we have constructed first- and second-order breather building blocks for the focusing CNLS-p system in the focusing regime by using an $N$-fold Darboux transformation. These building blocks belong to the symmetric normal-mode sector. By assigning two such solutions independently to the half-normal-mode fields, we obtained unshifted Type-I and shifted Type-II configurations. The resulting physical components exhibit background cancellation, overlap, separation, spatial beating, temporal compression, and walk-off-like patterns. These structures arise from normal-mode reconstruction and interference rather than direct nonlinear scattering between the two decoupled normal modes.
For the first-order Type-I KM configuration, we derived the componentwise Fourier decomposition associated with the normal-mode reconstruction. The localized parts of the two physical components share the same continuous spectrum, whereas the retained plane-wave background produces an additional zero-wavenumber Dirac-$\delta$ contribution. Numerical integration based on the SSFM confirms the analytical solution over the simulated interval, and the numerical Fourier transform agrees closely with the exact continuous-spectrum formula.

Future work may extend the Fourier analysis to Akhmediev, Peregrine, and higher-order configurations. It would also be useful to investigate perturbation growth and spectral stability systematically and to relate these results to modulational instability and inverse-scattering spectral data.

\section*{Acknowledgments}
We express our sincere thanks to all the persons who have provided valuable suggestions to this paper.
This work is supported
by the National Natural Science Foundation of China (Grant Nos. 12201622 and 12501590)
and the Project Funded by China Postdoctoral Science Foundation (Grant No. 2024M753506).

\section*{Conflict of Interest}

The authors have no conflicts to disclose.

\section*{Data Availability}
The data that support the findings of this study are available within the article.

\end{document}